\documentclass[epsfig]{article}
\usepackage{makeidx}
\usepackage{graphicx}

\newcommand{\lbar}{\overline}

\newcommand{\reftrans}{2.5}
\newcommand{\refdissol}{2.6} 
\newcommand{\refeos}{2.7}
\newcommand{\refcl}{3.1}
\newcommand{\refryb}{3.3}
\newcommand{\refaccel}{3.4}
\newcommand{\reftestc}{3.4}
\newcommand{\refsuperl}{3.6}
\newcommand{\refgroup}{3.7}
\newcommand{\refformal}{3.8}
\newcommand{\refsyntest}{4.7}
\newcommand{\refgraydisk}{4.2}

\makeindex

\begin{document}

\title{\bf {\sc tlusty}  User's Guide III:\\ Operational Manual}  
\author{I. Hubeny\footnote{University of Arizona, 
Tucson; USA; hubeny{\tt @}as.arizona.edu}~ and 
T. Lanz\footnote{Observatoire de C\^{o}te d'Azur, France; {\tt lanz{\tt @}oca.eu}}}
\date{\today}
\maketitle

\begin{abstract}
This paper presents a detailed operational manual for {\sc tlusty}. It
provides a guide for understanding the essential 
features and the basic modes of operation of the program.
To help the user, it is divided into two parts. The first part describes the most
important input parameters and available numerical options.
The second part covers additional details and a comprehensive description
of all physical and numerical options, and a description of all input parameters, 
many of which needed only in special cases.
\end{abstract}

\tableofcontents

\newpage
\addcontentsline{toc}{section}{PART I}

\begin{center}
        {\Huge \bf Part I}\\ [35mm]
\end{center}

\newpage


\section{Introduction}

The following document is a comprehensive guide to the operation of the program.
It describes in detail the structure of the required input files and resulting
output files, and explains 
the meaning of the individual input parameters. To help the user, the
explanations are supplemented by a number of actual examples of input files
for model atmospheres of various stellar types that may be used as templates
for a construction of similar models.

The program allows for a large number of options, controlling both the physical as
well as numerical setup of the calculations. While some options are very important
or even critical for the numerical performance of the code and the quality of the model,
there are many options and parameters that are important only in special cases.
This document is therefore divided into two parts. The first part deals with the basic 
numerical implementation and provides a guide for understanding the essential 
features and a basic operation of the program. 
Additional details, a comprehensive description
of all physical and numerical options, and a description of all input parameters, 
most of which needed only in special cases, are described in Part II.


\section{Compiling and linking}
\label{compile}

\subsection{Compilation}
\label{comp_comp}

The program is distributed as several files. The largest is {\tt tlusty[nnn].f}
where 
{\tt nnn} represents the current version number. In the following text,
we take 205 for the current version number.
Communication between subprograms is principally carried out through
labeled common blocks. 
To allow for the program to be scaled (re-dimensioned) easily, 
arrays are dimensioned by parameter constants. The arrays and parameters
are defined using INCLUDE files 
\index{INCLUDE files}
\begin{verbatim}
      IMPLIC.FOR 
      BASICS.FOR 
      ATOMIC.FOR 
      MODELQ.FOR 
      ITERAT.FOR
      ARRAY1.FOR 
      ODFPAR.FOR 
      ALIPAR.FOR 
\end{verbatim}

\noindent 
The INCLUDE files have to reside in the same directory as the {\tt tlusty205.f}
file, and, under UNIX/LINUX, their names {\it must} be in capital letters, 
e.g. {\tt BASICS.FOR}.  We stress that the INCLUDE files generally
\index{BASICS.FOR file}
evolve together with {\sc tlusty} (more parameters are being added), so that
the user is advised to copy these files along with any new version of {\sc tlusty}.
In the standard distribution, described in Paper~I, \S\,3.1, the source file as well as the
INCLUDE files are indeed located in the same directory.

The compilation is done as follows:\\ [4pt]
\noindent $\bullet$ Under Mac OSX, or modern versions of Linux
\begin{verbatim}
        gfortran -fno-automatic [-O3] [-o tlusty] tlusty205.f
\end{verbatim}
\noindent where the option {\tt "-fno-automatic"} indicates the static allocation 
of memory, which is a mandatory option. 
The level-3 optimization ({\tt "-O3"}) should be switched on since it
improves the performance of the code considerably. Reaming the executable
as specified by {\tt "-o tlusty"} allows the user to use script file {\tt RTlusty} to
conveniently run {\sc tlusty}. \\ [2pt]
\noindent $\bullet$ Under older versions of LINUX
\begin{verbatim}
        f77 -fno-automatic [-O3] [-o tlusty] tlusty205.f
\end{verbatim}
\noindent $\bullet$ Under generic UNIX
\begin{verbatim}
        f77 [-O4] [-static] [-Nl100] [-o tlusty] tlusty205.f
\end{verbatim}
\noindent where the option {\tt "-Nl100"} is sometimes needed under 
older variants of the SUN operation system (increases the number
of continuation lines to 100); the option {\tt "-static"}, which 
is equivalent to {\tt "`-fno-automatic"}, is needed for some variants of UNIX.
It is a default in most implementations. Similarly,
the optimization (the option {\tt "-O4"}) is a default on most
workstations. If not, the optimization should be switched on since it
improves the performance of the code considerably.\\ [2pt]
\noindent $\bullet$
One may also use (within LINUX) the commercial Portland Group compiler,
pgf77, which provides better optimization (the code is typically
20-30\% faster);
\begin{verbatim}
        pgf77  [-Bstatic] [-fastsse] [-o tlusty] tlusty205.f
\end{verbatim}

\subsection{Reducing the size of the executable file}
\label{reduce}

The strategy adopted in early days of the development of {\sc tlusty},
before FORTRAN90 was introduced,
was to code the maximum dimensions of the important arrays that
determine the overall memory consumption of the code in a few special
files, communicated through the {\tt INCLUDE} statements. This way
it became possible to easily change array dimensions, and thus the
memory needed, by simple changing a few lines in the INCLUDE files.
Although the memory consumption is no longer such a critical issue as
it used to be in the past, it may still be sometimes necessary to change the overall 
memory requirements. For instance, one has to be careful not to exceed 2 GB of
core memory, a typical value of a medium-class Mac or a LINUX box,
when computing NLTE metal line-blanketed models.

The basic parameters defining array dimensions appear in the INCLUDE
file {\tt BASICS.FOR}.
\index{BASICS.FOR file}
The first {\tt PARAMETER} statement there contains the most important 
parameters; which we list below. Their exact meaning is explained through
this paper.
\index{Superlines}
\begin{tabbing}
 MATOM \ \ \ \= maximum number of atomic species \\
 MION    \>     maximum number of explicit ions \\
 MLEVEL  \>     maximum number of explicit levels \\
 MLVEXP  \>     maximum number of explicitly linearized levels \\
 MTRANS  \>     maximum number of all transitions \\
 MDEPTH  \>     maximum number of depth points \\
 MFREQ   \>     maximum number of frequencies \\
 MFREQP \>  maximum dimension of working arrays for setting superlines\\
 MFREQC  \>     maximum number of frequencies in continua \\
 MFREX   \>     maximum number of linearized frequencies \\
 MFREQL  \>     maximum number of frequencies per line \\
 MTOT    \>     maximum number of linearized parameters
                                 (ie. maximum \\ \>dimension of vector $\psi$) \\
 MMU     \>     maximum number of angle points \\
\end{tabbing} 
There are also secondary parameters; their values are rarely changed.
Here is their list:
\begin{tabbing}
 MBF \ \ \ \ \ \ \ \  \=     maximum number of bound-free transitions\\
 MCROSS  \>     maximum number of photoionization
                        cross sections\\ \> plus related quantities\\
 MFIT    \>     maximum number of fit points for the input of 
                bound-free\\ \> cross sections \\
 MITJ    \>     maximum number of lines contributing to the opacity at \\
 \> a given frequency\\
MMCDW   \>     maximum number of levels with pseudocontinuum \\
 MMER    \>     maximum number of ``merged'' levels \\
 MZZ     \>     maximum charge of an ion treated with occupation \\ \> probabilities\\
 MVOIGT  \>     maximum number of lines with Voigt profiles\\
 MSMX    \>     size of largest matrix kept in memory in SOLVE 
\end{tabbing}

The above parameters are universal.
There are also other parameters which specify dimensions of arrays which
are being used only under special circumstances (including using the 
opacity tables, Compton scattering;
tri-diagonal approximate operator, etc.), which are set to 1 for the
bulk of applications. These are the following:
\begin{tabbing}
 MFRTAB\ \ \ \=     maximum number of frequencies in the opacity table;\\
 \> if opacity table is not used, =1\\
MTABT   \>   maximum number of temperatures in the opacity table\\
MTABR   \>   maximum number of densities in the opacity table\\
 MFREQ1   \>    set to MFREQ in case of Compton scattering or DFE transfer\\
                \> solver; otherwise =1\\
 MDEPTC \> set to MDEPTH in case of Compton scattering; otherwise =1\\
 MMUC    \>     maximum number of angle points for Compton scattering\\
 MLEVE3   \>    set to MLEVEL in case of tridiagonal operator;       
                otherwise =1\\
 MLVEX3   \>    set to MLVEXP in case of tridiagonal operator;       
                otherwise =1\\
 MTRAN3   \>    set to MTRANS in case of tridiagonal operator;       
                otherwise =1
\end{tabbing}
In addition, there are several parameters in the INCLUDE file {\tt ODFPAR.FOR}
\index{ODFPAR.FOR file}
that specify array dimensions for a treatment of metal line blanketing.
They are the following:
\begin{tabbing}
 MFODF \ \ \ \= maximum number of frequency points per a superline coss
 section  \\
 \>  (obsolete)\\
 MHOD     \>    maximum number of merged-levels cross sections\\
 MDODF    \>    maximum number of depth points for storing the superline cross \\
 \> sections\\
 MKULEV   \>    maximum number of internal energy levels for an ion 
                treated by \\ \> means of opacity 
              sampling; if opacity sampling 
                is not used, \\ \> it should be set to 1\\
 MLINE    \>    maximum number of internal lines for an ion 
                treated by \\ \> means of opacity 
              sampling; if opacity sampling 
                is not used, \\ \> it should be set to 1\\
 MCFE     \>    maximum total number of internal lines for all ions 
                treated by\\ \> means of opacity 
              sampling; if opacity sampling 
                is not used, \\ \> it should be set to 1
\end{tabbing}
The program checks whether the current values are less than or equal to 
the corresponding maximum dimension, and stops if there is a conflict.
The program then issues a brief message (on the standard output and on the special 
performance and error message file), showing
the corresponding current and maximum values.  The program should then be 
recompiled with the corresponding parameter appropriately modified in the
\index{BASICS.FOR file}
\index{ODFPAR.FOR file}
file {\tt BASICS.FOR} or {\tt ODFPAR.FOR}.

The convention is that the names beginning with {\tt M} designate the
maximum dimension; the analogous names beginning with {\tt N} then denote the
current values; for instance {\tt MATOM} is the dimension of
the arrays containing information about explicit atoms (stored in the
INCLUDE file {\tt ATOMIC.FOR}), while {\tt NATOM} is the current, actual
number of explicit atoms, etc. The exception from this rule is 
{\tt MDEPTH} and the corresponding {\tt ND} for the maximum and actual 
number of discretized depth points.

A reduction of the size of the executable file
is accomplished by reducing some of the above parameters specifying array
dimensions in the INCLUDE files {\tt BASICS.FOR}
\index{BASICS.FOR file}
\index{ODFPAR.FOR file}
and {\tt ODFPAR.FOR}). 
We stress again that when one computes a model without the
Opacity Sampling treatment of line blanketing due to the iron-peak
species, the parameters {\tt MKULEV}, {\tt MLINE}, and {\tt MCFE}
in the file {\tt ODFPAR.FOR} should be set to 2 which leads to a
significant reduction of the size of the executable file.

We stress that the standard distribution of {\sc tlusty}, as specified in Paper~I, \S\,3.1
contains the INCLUDE files {\tt BASICS.FOR} and {\tt ODFPAR.FOR} that
specify the basic array dimensions in such a way that one can run all the
test cases described in Chapter \ref{new_ex}, including a metal line-blanketed model 
atmosphere of a B star (\S\,\ref{examp_bs07}). In other words,
the dimensions are relatively large, and the executable takes about 1.7~GB of
memory. If the user does not have a computer with 2~GB of core memory or more, then
he/she have to modify the files accordingly, for instance by decreasing the values
of the critical parameters in {\tt ODFPAR.FOR} to 1, which would allow to run all tests
cases in Chapter \ref{new_ex} except the one for a B star.

\subsection{Program {\sc pretlus}}
\label{pretlus}
\index{PRETLUS program}

Setting the maximum dimensions of arrays may be a delicate task because
in many instances the actual dimensions depend on input data in a rather
non-trivial manner. In order to assist the user in setting the dimensions
of arrays, and consequently to reduce the size of the executable file, we have developed a utility program {\sc pretlus}, which is distributed along with the the
main {\sc tlusty} program.

Program {\sc pretlus} is easy to run. It accepts exactly the same input
as {\sc tlusty}, and it essentially performs  the initialization part 
of {\sc tlusty} without any
actual calculations, and outputs the list of actual dimensions of all
important arrays. The user has too check files {\tt BASIC.FOR} and
{\tt ODFPAR.FOR} to make sure that the values of the dimension
parameters given there are equal or larger than the actual values 
listed in the standard output from program {\sc pretlus}. An example
of the output produced by {\sc pretlus} is given in \S\,\ref{examp_bs07}.


\section{General scheme of the input}
\label{ginput}

\subsection{Overview}
\label{ginput_over}

The essential feature of the input data format is that there is only a very
\index{Input!general scheme}
short standard input file, which specifies 
(i) the very basic parameters ($T_{\rm eff}, \log g$)  for 
which no reasonable default values can be specified;
(ii) the name of the file where the optional, or keyword, parameters are set up;
and (iii) the names of files where the atomic data for the individual ions are
stored. Keyword parameters are defined as those for which the program assigns
default values, which are optimum for most applications, but could be changed
as required for a particular case. Keyword parameters also allow the user to choose among
several alternative numerical schemes, or to cope with convergence problems.
The most important ones are described in 
Chap.\,\ref{nonst}, the remaining ones  in Chaps.\,\ref{nst2_phys} and
\ref{nst2_num}, and the full list of keyword parameters together with their
default values is presented
in Chap.\,\ref{list}.

Here is a list of input files.
\begin{itemize}
\index{Input!accretion disk switch}
\item{\tt fort.1} --- The basic control file, containing just one single number,
specifying whether one calculates a stellar atmosphere or accretion disk model.
If this number is 0, or if the file is  missing altogether, a
stellar atmosphere model is to be computed.
Otherwise, a disk vertical structure is computed.
\item{Standard input file (unit 5)} --- Main control data. It is a short file with only 
the most important parameters, and filenames of other files.
The structure of the file and the meaning of the individual input parameters
are explained in Chap.\,\ref{new}.
\item{File that specifies non-default values of the keyword parameters}.
Its name is specified in the standard input file.
\item{\tt fort.8} --- A starting model atmosphere (if the calculation does not start
from the scratch, that is, with an LTE-gray model) - see Chap.\,\ref{unit8}.
\end{itemize}
These are universal input files that need to be supplied for any mode of
operation.

In the standard mode, where the opacities are computed on the fly, and
where one introduces the concept of explicit ions and levels (which is
mandatory for NLTE models), there is an important set of input files, namely
\begin{itemize}
\item{Files containing atomic data for the individual ions.}
These files are described in detail in Chap. \ref{ions}. 
A collection of such files is a part of the standard distribution.
It is also available on the {\sc tlusty} website.
\end{itemize}
Further, depending on the mode of application, which is set up by appropriate keyword parameters, there may be additional input files, 
which are also a part of the standard distribution of {\sc tlusty},
namely:
\begin{itemize}
\item{{\tt lemke.dat} or {\tt tremblay.dat} -- special hydrogen line broadening tables} 
-- see also \S\,\ref{nst2_hyd} 
\item{Special tables for hydrogen quasi-molecular data}
-- see also \S\,\ref{nst2_quas} 
\item{\tt tsuji.molec} -- a table with necessary parameters for the molecular state equation -- see also \S\,\ref{examp_optab}, \S\,\ref{nonst_glob} and 
\S\,\ref{nst_exa_optab}
\item{\tt irwin.dat} -- a table of Irwin partition functions
-- see also \S\,\ref{nst2_eos} 
\end{itemize}
In the case where some or all opacities are supplied from the pre-calculated
opacity tables, one needs the following input files: 
\begin{itemize}
\item{\tt absopac.dat} -- opacity table, required if IOPTAB=$-1$ 
-- see also \S\,\ref{examp_optab} and \S\,\ref{nst2_optab} 
\item{Rayleigh scattering opacity table}; if required
-- see also \S\,\ref{nst2_ray} 
\item{State equation tables}, if required. Notice that when using the full opacity
table option, one can still solve the equation of state and compute the
thermodynamic parameters needed for treating convection on the fly, 
so these tables may not be required. 
\end{itemize}

All the input files are ASCII files to enable easy portability.
All the {\tt READ} statements use a free format. Moreover, Unit 5
may contain comment lines; {\sc tlusty} understands a line beginning
with {\tt *} or {\tt !} as comment.
The structure of these files is explained in detail below.  


\subsection{What needs to be specified?}
\label{ginput_what}

There are two types of input parameters; the true {\em physical 
quantities} (such as the effective temperature, $T_{\rm eff}$),
and {\em control parameters} -- typically flags that switch on/off 
various numerical procedures, control the choice of adopted numerical
scheme, etc.

We stress that a NLTE stellar atmosphere code is not a black box.
The setup of a model depends to a large extent of user's judgment
(e.g., determining a degree of sophistication of the model).
The user has to understand the meaning of a number of input parameters.
The philosophy behind introducing several input files is an attempt
to help the user in such a way that important parameters
have to be set up in the standard input file, while other, either less
important, or newly introduced parameters, assume their default values 
unless the user specifically requires to change them. This is done through 
the keyword parameters file. 

As the code evolves, it is often necessary to introduce new input parameters.
These are set through the keyword parameter file, even if they may sometimes
be quite important.
A practical reason for setting newly introduced
parameters in the keyword parameter file instead of the standard input is
that there is a large library of standard input files for various models,
and it would be cumbersome to be forced to change them whenever
a new important input parameter is introduced.

We list the essential categories of the input parameters below.
\begin{itemize}

\item Basic model parameters.\\ 
These are $T_{\rm eff}$ and $\log g$
for stellar atmospheres, and analogous parameters for accretion disks.
These are of course mandatory, and are communicated to {\sc tlusty}
through the standard input -- see \S\,\ref{new1}.

\item Setting LTE or NLTE.\\
This is done in the standard input, see \S\,\ref{new1}.

\item Chemical composition and the choice of explicit species.\\
We recall that
the {\em explicit} atom is defined as such for which a selected set of
energy levels of a selected set of its ionization states are 
considered explicitly, i.e., their populations are determined by 
solving the kinetic equilibrium equations. These species are the only ones
that are allowed to contribute to the total opacity, so their choice
is important because when neglecting an important opacity source the
quality of the resulting model will suffer.
An {\em implicit} atom is not allowed to contribute to the opacity that is
calculated on the fly, but is allowed to contribute to the
total number of particles and to the total charge;
the latter is evaluated assuming LTE ionization
balance, i.e., by solving a set of Saha equations.\\
The choice of explicit and implicit atoms, together with their abundances,
is given in the standard input -- see \S\,\ref{new_at}.

\item Choice of explicit ions.\\ 
Again, the choice has to be made
judiciously, as it significantly influences the overall quality of a model.
The choice is also given in the standard input -- see \S\,\ref{new_ion}.
Notice that if the full opacity table option is adopted, there are no
explicit ions defined.

\item Atomic data for explicit ions.\\ 
This is an important, and sometimes decisive, ingredient of good NLTE models. 
The standard
input file specifies the filenames of the individual atomic data for each
explicit ion; the atomic/ionic data files are separate. This enables us
to store these individual files and distribute them together with the program.
This is actually done on the {\sc tlusty} website, where a collection of 
such files, in varying degree of sophistication and complexity, is presented.
The user can thus use those files without being required to
construct them. The atomic data files also contain a mixture of
physical parameters (such as the level energies, oscillator strengths, 
etc.), and also control parameters for switching particular options.
Therefore, in some cases one may need to modify the atomic data file to
set up a different option if required (for instance, setting a different
choice of level grouping - see \S\,\ref{tricks_num}).
Again, in the case of using full opacity tables, there is no need to have
these files.

\item Auxiliary physical parameters.\\
Here we have for instance convection parameters, microturbulent
velocity, possible external irradiation intensity, etc.
As explained in the next chapters, the user has a choice of
including or not including these physical mechanisms at all,
and if so, to set up corresponding numerical values. These
parameters are communicated through the keyword parameters file,
discussed in detail in Chap.\,\ref{nonst}. 

\item Basic discretization parameters.\\
This category contains the number of depth, frequency, and angle points
used when discretizing the corresponding structural equations.
The number of depth and angle points are set up by corresponding
keyword parameters. The actual values of the individual column masses
(which is the default depth coordinate)
are either given
in the input model, or are constructed in the starting LTE-grey
model atmosphere. The corresponding parameters are also set up by
the keyword parameters - see \S\,\ref{nonst_gr}. If none are
specified, the default values are adopted. The total number of
frequency points is set up by {\sc tlusty}, and depends on the actual
selection of lines and continua that are treated explicitly.
The user can influence some aspects of the selection via the
standard input -- see \S\,\ref{new1}, and through several
optional keyword parameters -- see \S\,\ref{nonst_freq} and \S\,\ref{nonst_blank}.

\item Numerical options\\
There is a large number of flags for controlling many 
numerical aspects, usually communicated as keyword parameters --
see Chaps.\,\ref{nonst} and \ref{tricks}. 
As mentioned above, the atomic data files
also contain some flags of this category.
If the model computation
proceeds well without any special keyword parameters file,
that is if the default values of keyword parameters provide
a satisfactory numerical strategy, the user does not have to set up
this file. However, in a majority of cases one needs to set up
some optional keyword parameters, so one is recommended to obtain at least
a rudimentary knowledge from Chap.\,\ref{nonst}; more dedicated users
are encouraged to study also Chaps.\,\ref{nst2_phys} and \ref{nst2_num}
in detail. A practical advice how to use various options for troubleshooting
is briefly summarized in Chap.\,\ref{tricks}.

\end{itemize}


\section{Standard input file}
\label{new}

We now turn to a detailed description of the individual input parameters.
The standard input file is composed of four basic blocks:

\subsection{First block -- Basic Parameters}
\label{new1}

This block contains only four lines of input.

\subsubsection*{{\em First line:}}

This is the only input record that differs for atmospheres and disks;
all the other input values have the same meaning for both basic
options.

\noindent { $\bullet\bullet$ For stellar atmospheres}:
\begin{description}
\item[TEFF] -- effective temperature [K]
\index{Effective temperature}
\index{Input!effective temperature}
\item[GRAV] -- log $g$ [cm s$^{-2}$]
\index{Surface gravity}
\index{Input!surface gravity}
\end{description}
\noindent { $\bullet\bullet$ For accretion disks}:    
\begin{description}
\item[XMSTAR] -- mass of the central object. There are still three
\index{Input!accretion disk parameters}
possibilities:
\begin{itemize}
\item XMSTAR $>\ 0$ -- in this case the central object is supposed
to be a star; computation of the structure is done in the classical
approximation. The mass can be expressed either in grams, or in solar
masses.
\item XMSTAR $<\ 0$ -- in this case the central object is a black
hole; computation of the structure is done using general relativistic
corrections. The mass of the black hole is then abs(XMSTAR); again,
it is expressed either in grams or in solar units.
\item XMSTAR = 0 -- in this case the meaning of the subsequent
three quantities of the input is different; instead of using basic
parameters $M_\ast$, $\dot M$, $R_\ast$, and  $R/R_\ast$ or  
$R/R_g$, one uses specific parameters for an annulus without
explicit reference to the central object, namely $T_{\rm eff}$, $Q$,
and $m_0$.
\end{itemize}
\end{description}
Specifically, if XMSTAR $>0$, then the subsequent three parameters are:
\begin{description}
\item[XMDOT] -- mass accretion rate. It can be expressed in g$/$s,
or in $M_\odot/{\rm yr}$.
\item[RSTAR] -- again, two different meanings depending whether one
considers a classical or general relativistic central object.\\
$\bullet$ In the classical case, RSTAR is the radius of the central star
(in cm or in $R_\odot$);\\
$\bullet$ in the case of a black hole disk, 
RSTAR has the meaning of the spin (angular momentum) of the black
hole, expressed in geometrized units (i.e. =0 for a Schwarzschild
black hole; =0.998 for a Kerr black hole with maximum stable rotation).

\item[RELDST] -- relative distance of a given annulus with
respect to the stellar radius (classical case), or the gravitational radius
(for the black hole case). Gravitational radius is defined by
$R_g = G M/c^2$. Caution: the Schwarzschild radius, which is
being used by some authors as a reference radius, is given by $R_S = 2 R_g$
\end{description}
In the alternative case, when XMSTAR = 0, the subsequent three
numbers in the first line of input have the following meaning:

\begin{description}

\item[TEFF]  -- effective temperature [K]
\index{Effective temperature!input}
\item[QGRAV] -- the proportionality coefficient, $Q$, to evaluate
the $z$-dependent gravity acceleration,
 $g = Q z$. In the classical
case, $Q = \Omega_{\rm K}^2$, where $\Omega_{\rm K}$ is the Keplerian
velocity, $\Omega_{\rm K} = (G M/R^3)^{1/2}$.

\item[DMTOT] -- column mass at the midplane (=1/2 of the total column mass) [g]
\end{description}

\subsubsection*{{\em Second line:}}

\begin{description}
\item[LTE] -- a logical variable indicating whether an LTE model is
\index{Input!LTE/NLTE switch}
going to be calculated.\\
= .TRUE. -- LTE model is calculated\\
= .FALSE. -- NLTE model is calculated
\item[LTGRAY] -- a logical variable indicating whether an LTE-gray model is 
calculated at the beginning as a starting approximation for the 
linearization iterations.\\
= .TRUE. -- LTE-gray model is calculated as a starting model;\\
= .FALSE. -- LTE-gray model is not calculated; the user has to supply a
starting model atmosphere -- the Unit 8  ({\tt fort.8}) input.
\end{description}

\subsubsection*{{\em Third line:}}
\begin{description}
\item[FINSTD] -- a character variable (up to 20 characters) with the name
\index{Input!keyword parameters filename}
of file containing the values of keyword  parameters. \\
= {\verb|''|} (null string) -- all keyword parameters are taken
with their default values -- see the next section. 
\end{description}

\subsubsection*{{\em Fourth line:}}

\begin{description}
\item[NFREAD] -- an indicator of the number of frequency points and
\index{Frequency points!setup}
their setup:\\
$>$ 0 -- the program sets up the {\em continuum} frequency points. We stress 
that the frequency points in the lines are set up separately based on the input 
data contained in the input atomic data files.
The details of setup are governed by several keyword parameters
(FRCMAX, FRCMIN, CFRMAX, NFTAIL, and DFTAIL -- see Sect.\ref{nonst_freq}).
In the default case, the program sets two frequencies near 
discontinuities corresponding
to the bound-free transitions form all explicit levels, plus approximately
NFREAD frequencies in between, 
plus a number of points in the high- and low-frequency tails
of the spectrum. The endpoints of the high- and low-frequency tails 
are specified by the keyword parameters FRCMAX and FRCMIN, 
respectively; the number of points
in the high-frequency tail are specified by the keyword parameters 
NFTAIL and DFTAIL.
(i.e. they may be changed by a corresponding specification in the
input file FINSTD).\\
$<$ 0 -- the frequency points are set up logarithmically equidistant
between FRCMIN and FRCMAX, with the total number abs(NFREQ).
\end{description}


\subsection{Second block -- Selection of, and basic data for, chemical species}
\label{new_at}

The block contains one record with a value of
\index{Explicit atoms!input}
\index{Input!setting of explicit atoms}
{\rm NATOMS}, and then {\rm NATOMS} analogous records for the individual
species; each containing three parameters: MODE, ABN, MODPF.
The order of individual records {\em must} exactly follow the atomic
number (i.e. H, He, Li, Be, B, C, N, O, etc.).
{\sc tlusty} can potentially treat elements with atomic number 1 -- 99,
but in practice only a smaller number of elements (typically 30)
are taken into account. Another concern is that the values of the
partition functions for species with atomic number higher than 30
are hardwired only for the first two ionization stages, so when treating
higher ions of such species the user has to supply the corresponding
expressions or tables.

The individual input parameters have the following meaning:
\begin{description}
\item[NATOMS] -- the highest atomic number of an element that is
considered (explicitly or non-explicitly).
$<$ 0 -- then abs(NATOMS) has the meaning as above, but all the partition
functions of all species considered by the Opacity Project are evaluated
from the Opacity Project ionization fraction tables, regardless of the
parameter MODPF (see below).\\
= 0 -- no explicit atoms are selected. This option only makes sense
in the case of using complete pre-calculated opacity and state equation
tables (set by the keyword parameter IOPTAB = $-2$).
\item[MODE] -- a specification of the mode of treatment of the given species:\\
= 0  -- the element is not considered at all;\\
= 1  -- he element is treated {\em implicitly}. In this case, 
the species does not contribute to
                the opacity; but it is allowed to contribute to the
                total number of particles and to the total charge;
                the latter is evaluated assuming LTE ionization
                balance, i.e., by solving a set of Saha equations. \\
= 2  -- the element is treated {\em explicitly}, i.e., selected energy levels of 
of the selected ionization states are considered explicitly; i.e., their
populations are determined by solving the corresponding kinetic equilibrium equations.
\item[ABN] -- a specification of the abundance of the given species:\\
= 0 -- the solar abundance is assumed (Grevesse~\& Sauval 1998) -- see 
\index{Abundance!input}
\index{Input!abundances of species}
Paper~II, Table~1;\\
$<$ 0 -- a non-solar abundance is assumed, abs(ABN) has now the meaning of
         the abundance expressed as a multiple of the
         solar abundance (i.e. $-0.1$ means 1/10 of solar, $-5$ means 5 times
         solar abundance, etc.);\\
$>$ 0 -- a non-solar abundance is assumed, expressed as 
         $N({\rm elem})/N({\rm ref})$, i.e. relative by number
	 to the reference species.  The reference atom is H by default,
	 but the reference species can be changed by means of the
	 optional parameter IATREF (see Sect. \ref{nonst_ese})\\
$>10^6$ -- non-homogeneous (depth-dependent) abundance is assumed. In this
  case, the immediately following $N\!D$ lines should be added that
  contain the individual values of the
  abundance (relative to hydrogen by number), for all depth points 
  $d=1,\ldots,N\!D$.
\item[MODPF] -- a flag indicating a mode of evaluation of the partition
functions for the given species. Notice that this may be overwritten by
\index{Partition functions!input}
\index{Input!partition functions setting}
coding NATOMS as negative -- see above.\\
= 0 -- a standard evaluation of the partition functions -- see Paper~II, \S\,\refeos\\
$>$ 0 -- the partition functions evaluated from the
   Opacity Project ionization fraction tables.\\
 $<$ 0 -- non-standard evaluation by a user-supplied formula to be
 implemented in subroutine PFSPEC. For details, refer to \S\,\ref{pfspec}.
\end{description}


\subsection{Third block -- Explicit ions}
\label{new_ion}
\index{Explicit ions!input}
\index{Input!explicit ions setup}

For each 
ion, including the highest ionization degree of a given species, there
must be one input record for each of then relevant ionization stages
containing the following parameters:

\begin{description}

\item[IATII] - the atomic number of the parent species of the ion (i.e.
1 for hydrogen, 2 for all ions of helium, etc.).

\item[IZII] -- the charge of the ion (0 for neutrals, 1 for once ionized, etc.).

\item[NLEVS] -- a number of energy levels considered explicitly.

\item[ILAST] -- an indicator whether the given ion is the highest considered
ionization degree:\\
= 0 -- the ion is not the highest ion of the parent species; the subsequent
input record has to contain parameters for the next higher ion;\\
$>$ 0 -- the ion is the highest ionization degree of the parent species.\\
= 1 -- the program assigns the correct statistical weight of the 
ground state of this ion automatically;\\
$\neq$ 1 -- has the meaning of the statistical weight of the 
ground state of this ion;\\
$<$ 0 -- indicates the last record of the explicit ions input
         block.

\item[ILVLIN] -- an indicator of changing the treatment of a whole group of 
bound-bound transitions, regardless of the input communicated by the atomic
data file - see below.
ILVLIN has the meaning that all lines with the relative index of the lower 
level smaller than ILVLIN are considered in detailed radiative balance
and their opacity is neglected.
The relative index counts the levels within the ion; i.e. the ground state 
of the ion has relative index 1, the last considered level the index NLEVS.
For instance, setting ILVLIN=2 will put all lines originating from the
ground state to the detailed radiative balance, which is often a useful option.
Setting ILVLIN $>$ NLEVS will put all lines of the ion to detailed balance.\\
This option enables one to consider the same atomic input files for LTE, 
NLTE/C (NLTE with continua only), and NLTE/L (NLTE with lines) models
-- see below. 

\item[NONSTD] -- an indicator of an additional input record, to change specific
``non-standard'' parameters for the ion (those having assigned
default values that provide optimum for most applications), 
or to provide necessary filenames for treating ions for which the 
\index{Superlevels}
\index{Superlines}
superlevel and superline formalism is used (typically the iron-peak elements). \\
$= 0$ -- no change of non-standard parameters is required;\\
$> 0$  -- additional record with ``non-standard'' parameters. This is a more
or less outdated option; for completeness see \S\,\ref{new_ion2};\\
$< 0$  -- additional record(s) with filenames  for evaluating cross sections for 
superlines-- see below. 
Moreover, when superlines are treated in the Opacity Sampling option, the
actual numerical value of NONSTD defines the formation of superlines,
\index{Superlines}
namely:\\
$= -1$  -- all internal lines contribute to an appropriate superline;\\
$= - 2$ --  all internal lines  except autoionizing lines (with upper levels 
above the ionization potential) contribute; autoionizing lines are neglected;\\
$< -2$ -- only lines between observed levels (internal levels with measured
energies) are considered.

\item[TYPION] -- a character{\tt *}4 variable containing a descriptive label,
e.g. {\verb|'He 2'|} for He$^+$, etc.

\item[FILEI] -- a character variable containing the filename where the
detailed input of parameters for explicit level, bound-free transitions,
and bound-bound transitions are stored. The structure of this file is described
in detail in Chap. \ref{ions}.

\end{description}

\noindent Note: The number of levels considered for the ion, {\rm NLEVS}, 
{\em must not} exceed the number of levels given in the file {\tt FILEI}.
However, {\rm NLEVS} may be smaller; in such a case the current run
will select {\rm NLEVS} lowest levels from the file {\tt FILEI}. \\ [4pt]
\noindent If the  parameter NONSTD of the standard input (see above)
is coded as negative, the program reads an additional record with the 
following four parameters:

\begin{description}
\index{Superlines}
\item[INODF1, INODF2] -- unit numbers for pre-computed data for superline
cross sections treated through Opacity Distribution Function (ODF). 
This is now an outdated option, these
two numbers should bet set to zero, \\ ${\rm INODF}1={\rm INODF}2=0$.
\item[FIODF1, FIODF2] -- filenames for evaluating superline cross sections:\\
 -- in the standard case of the Opacity Sampling (OS), 
 the first file is the Kurucz level data file (e.g., {\tt gf2601.gam}, for Fe II), 
 and the second file is the line data file ({\tt gf2601.lin}).\\
-- in the (outdated) case of ODF treatment of lines, the
filenames of the corresponding ODF input. 
\item[FIBFCS] -- the name of the file containing the photoionization
cross sections for the individual superlevels. The parameters 
\index{Superlevels}
IFANCY (see Sect. \ref{ion_bf}) has to be set to a value 
between 50 and 99 to switch on reading cross sections from a previously
created file.
\end{description}


\section{General strategy of model construction}
\label{strateg}

As explained in Paper~II, \S\,\refcl\, -- \refryb, any variant of the linearization 
scheme that is used for constructing model atmospheres is quite sensitive 
to a quality of the initial estimate. This feature dictates the strategy of
constructing models, which often consists in a series of consecutive steps.
We shall describe several possible approaches below.

\subsection{LTE models}
\label{strateg_lte}

Constructing an LTE model is usually relatively straightforward. 
The strategy depends on whether an LTE model to be computed is used as
a starting solution for a subsequent NLTE model, or whether it is the final
result on its own right. 
In the former case, the model can be quite simple, for instance without 
taking into account lines. In the latter case, one aims at achieving a
much higher degree of complexity of the model, which usually requires
more refined modeling techniques.

As mentioned earlier, an attractive option to construct an LTE model is using a
pre-calculated opacity table. This can in principle be done at any temperature
range, but at the moment our opacity table only covers temperatures between
4000 and 10,000 K, so it can be used only for G and K star atmospheres.
We plan to extend the opacity table to lower temperatures (say, down to 100 K).
An extension to higher temperatures is possible as well, and in fact is relatively
easy, but it would not have much practical value because LTE model
atmospheres are not accurate enough for higher temperatures (say, for
effective temperatures above 15,000 K).
\smallskip

There are several possibilities to construct an LTE model atmosphere:

(i) The standard way is to start from the scratch; that is, first to construct an 
initial LTE-gray model, and immediately continue to converge the desired
LTE model. As shown in the previous chapter, this is done in one step.
Some examples are shown in \S\,\ref{examp_hhe} (a simple
H-He model), and \S\,\ref{examp_optab} (a complex line-blanketed model,
computed using an opacity table).

(ii) One can use an existing LTE model, either constructed earlier by {\sc tlusty},
or taking a Kurucz model atmosphere, as a starting mode atmosphere,
and converge the desired LTE model from it. Obviously, the basic input
parameters, $T_{\rm eft}$ and $\log g$ for the input model must not be
very different from $T_{\rm eff}$ and $\log g$ of the model to be constructed.
A conservative estimate is to change $T_{\rm eff}$ by no more than 5-10\%,
and $\log g$ by no more than 0.2 dex. Otherwise, the model may still converge,
but could easily take more computer time than when starting from the scratch. 

(iii) When constructing a line-blanketed model without using an opacity table, 
one has to proceed in at least two steps. The first step consists of computing 
a simple
model starting from the scratch (typically without any lines), and the next step
includes all the lines. The reason for this procedure is that the adopted treatment 
of metal line blanketing needs an input starting model in order to construct the
appropriate cross sections for superlines.

(iv) For complex or difficult models, it is possible to construct the desired
model by a sequence of intermediate models. Typically, one first
converges a model with a few or no lines considered explicitly, and then one
adds lines. This approach is analogous to an often used strategy of constructing
NLTE models, which will be described below.

\subsection{NLTE models}
\label{strateg_nlte}

A usual strategy, adopted already in the pioneering study of Auer \& Mihalas (1969)
is to proceed in three steps:
\medskip

LTE $\rightarrow$ NLTE/C $\rightarrow$ NLTE/L, \\ [4pt]
where NLTE/C represents a NLTE model with continua only; i.e., without lines.
In other words, all lines are taken in detailed radiative balance (their radiative
rates are set to 0), and their opacity is neglected. NLTE/L denotes a NLTE with
lines. We stress again that the LTE model used for this purposes does not have to
be a sophisticated LTE model; typically an LTE without lines is sufficient.

The reason for adopting this procedure is the following: 
We know that with increasing depth in the
atmosphere departures from LTE generally decrease, so deep enough the state
of the atmosphere is well described by an LTE model. The continua (bound-free
atomic transitions) are typically less opaque than the bound-bound transitions
(lines), so they are formed deeper in the atmosphere. 
Therefore, including continua only in an NLTE model yields an essentially 
exact atmospheric structure in deep, continuum-forming, layers.
Including lines in the final step then corrects the
structure closer to the surface. By this strategy, one gradually improves the
structure from deep layers to the surface, which is exactly how the energy in 
the atmosphere typically flows. (This also hints why a construction of an 
atmospheric model with a strong external irradiation may be numerically
difficult).

The above reasoning also explains why it is sometimes advantageous to
split the third step into several parts:
\medskip

NLTE/L1 $\rightarrow$ NLTE/L2 $\rightarrow\,\ldots\,\rightarrow$ NLTE/L,\\ [4pt]
where the individual steps consist of adding more and more lines. 
We stress that in some cases this is not
necessary, and a final NLTE/L model can be constructed in one step.
But often this is a good way to proceed. The simplest possibility, always worth
trying in case of convergence problems with an NLTE/L model, is first to set a detailed
balance in all the resonance lines (those with lower levels being the ground states of 
the corresponding ions), and then switch them on, perhaps in several steps starting
with weaker resonance lines (those of less abundant ions), and ending with
the strongest ones.  The reason for this strategy is exactly the same as that
put forward for the general strategy of improving the atmospheric structure
going from deep to upper layers.

In some cases, in particular for models where departures from LTE are
not crucial, say for late B stars and cooler, and also for hot white dwarfs
that contain significant amounts of metals, the following procedure often
proves advantageous:
\medskip

LTE/C $\rightarrow$ LTE/L $\rightarrow$ NLTE/L,\\ [4pt]
that is, constructing first a simple LTE model, than a fully blanketed LTE
model (denoted LTE/L), and from it to proceed directly to fully blanketed
NLTE/L model with all lines. This strategy reflects the physical fact that
when NLTE effects are week, the most important mechanism for determining
the atmospheric structure is the metal line blanketing.

\subsection{Setting a detailed balance in lines}
\label{strateg_db}

One may set up a detailed radiative balance
in a line, and therefore exclude the corresponding transition rate
from the kinetic equilibrium equation, and removing the corresponding
opacity, in several different ways:
 
(i) individually; in which case one
has to modify the corresponding entry in the atomic data file
(see \S\,\ref{ion_bb}). This is the most flexible way, but also the
most cumbersome since one has to modify the atomic data file.

(ii) excluding all lines of an atom/ion that originate from a 
selected set of lower levels. This is driven by the parameter ILVLIN
(see \S\,\ref{new_ion}) for a given atom/ion, which is set in the
\index{ILVLIN positional parameter}
standard input. The meaning of ILVLIN is that all lines originating
at level with index lower than ILVLIN are set to detailed
radiative balance. For instance, ILVLIN=2 excludes all resonance
lines (i.e., originating from the ground state with index 1);
setting ILVLIN=99 (or some other large number) will effectively
set all lines of the given atom/ion to detailed balance.

(iii) The third way to set lines to detailed balance is based on
their frequency. This is driven by keyword parameters
FRLMAX and FRLMIN (see \S\,\ref{nonst_freq}), which put all lines
\index{FRLMIN keyword parameter}
\index{FRLMAX keyword parameter}
with frequency smaller than FRLMIN  and/or larger than FRLMAX to
detailed balance. Setting FRLMIN to values around $10^{13}$ will
not only save computer time because it excludes all far-infrared
lines that typically do not influence the atmospheric structure,
but also remove possible problems with the laser effect in lines, which
may lead to numerical problems.


\section{Examples of standard input; test cases}
\label{new_ex}

\subsection{General remarks about running {\sc tlusty}}
\label{examp_gen}

As already mentioned in Paper~I, \S\,4.4, while
the names of the {\sc tlusty} input and output files can be arbitrary,
it is advantageous to use the following convention: 
Any name of an input or output file is composed of a {\em core name} that may label the
basic physical parameters of a model, with an extension identifying the unit number.
For example, let us take a H-He model for $T_{\rm eff} = 35,000$ K and $\log g = 4$, in
LTE. Let the core name be {\tt hhe35lt}, so the standard input file is then {\tt hhe35lt.5}.
This example is considered below in \S\,\ref{examp_hhe}.

As is shown in Paper~I, \S\,4.4, {\sc tlusty} can be run as 
\begin{verbatim}
      tlusty.exe < [std.input_file] > [std. output_file]
\end{verbatim}
but in this case the executable file {\tt tlusty.exe}
has to be present in (or liked to) the current directory, and also the
necessary atomic data have to be copied or linked to the files specified
by the standard input.

An easier and in fact a safer way is to use the shell script {\tt RTlusty}, 
which is also a part of the standard distribution of {\sc tlusty}. It is called 
with one or two parameters,
\begin{verbatim}
      RTlusty   model_core_name   [core_name_of_starting_model]
\end{verbatim}
The second parameter does not have to be present if the model is calculated
from scratch. 

We stress that the script {\tt RTlusty} is designed to run a model in any directory
provided that the two following requirements are fulfilled:
\begin{itemize}
\item
One sets the environment variable {\tt TLUSTY} that specifies the main 
{\sc tlusty} directory. For instance, using the tar files downloaded from the
Arizona site\footnote{http://aegis.as.arizona.edu/$\,\,\widetilde{\ }\,$hubeny/pub/tlusty205.tar.gz},
and assuming that the tar file is extracted in the home directory, so
the main {\sc tlusty} directory is generated as~ {\tt $\widetilde{}$/tlusty205}, one sets
\begin{verbatim}
setenv TLUSTY ~/tlusty205
\end{verbatim}
\item
The universal directory containing atomic data is a subdirectory
of the main {\sc tlusty} directory, that is \${\tt TLUSTY/data}
\end{itemize}

The script {\tt RTlusty} has the following content:\\ [2pt]
\hrule
\hrule
\begin{verbatim}
#!/bin/bash
#
#       shell script to run tlusty
#
#       print syntax if no parameters specified
#
   if [ $# -lt 1 ]
   then
     echo Usage:
     echo RTlusty  model core name  [core name of starting model]
     exit
   fi
#
# check that the starting model exists
#
   if [ $# -eq 2 ]
   then
     if [ -e $2.7 ] ; then
         echo STARTING MODEL: $2.7;
      else
         echo FILE $2.7 does not exist, therefore quitting ;
         exit;
     fi
#
     rm -f fort.8
     cp $2.7 fort.8
   fi
#
#     link the "data" directory
#
   ln -s -f $TLUSTY/data data
#
#     run tlusty
#
   MOD=$1
   $TLUSTY/tlusty/tlusty.exe  < $MOD.5 > $MOD.6
#
#     save important output files
#
   cp fort.7 $MOD.7
   cp fort.9 $MOD.9
   cp fort.69 $MOD.69
   cp fort.13 $MOD.13
#
   echo "COMPUTED MODEL  $MOD  FINISHED!"
   date
#
\end{verbatim}
\hrule
\hrule
\bigskip

Let us take an
example from the next section, \S\,\ref{examp_hhe}, with the name 
of the input file being {\tt hhe35lt.5}. Then the code can be either as
\begin{verbatim}
     tlusty.exe < hhe35lt.5 > hhe35lt.6
\end{verbatim}
where the standard output is redirected to a file {\tt hhe25lt.6} for
further inspection; or one can use the script {\tt RTlusty}
and to run the test as (this time in the background)
\begin{verbatim}
     RTlusty hhe35lt &
\end{verbatim}
which runs the code and stores not only the standard output (unit 6)
but also other important output files, all with the same core name,
and with the suffix that corresponds to the unit number. They are explained in 
more detail in \S\,\ref{out_bas}. Briefly, {\tt hhe35lt.7} contains a condensed
model atmosphere, {\tt hhe35lt.9} the convergence log, {\tt hhe35lt.13} the
emergent flux, and {\tt hhe35lt.69} the timing log. There are more output files;
if they are intended to be kept for future use, they should be renamed,
perhaps with the same core name.

After a completed run, it is very important to inspect the convergence log
to make sure that all the relative changes of the components of the state
vector are sufficiently small. The standard distribution of {\sc tlusty} also
contains an IDL program {\tt pconv.pro} which plots the contents of the
convergence log. The program needs three output files, units 7, 9, and 69,
and assumes that the file names are constructed using the above convention,
that is the core names are the same. The program It is called as
\begin{verbatim}
     pconv,'hhe35lt'
\end{verbatim}
It can be called  also as
\begin{verbatim}
     pconv
\end{verbatim}
in which case the files are supposed to have generic names {\tt fort.*}.
Several examples of plots generated by {\tt pconv} are presented throughout 
this chapter.

It is also important to inspect the last table of the standard output; in the present
case {\tt hhe35lt.6}, which summarizes the computed model atmosphere.
The last four columns show the computed total flux (radiative plus, if applicable,
convective), and the ratios of the radiative, convective, and computed total
fluxes with respect to the theoretical total flux, $\sigma T_{\rm eft}^4$,
respectively. Therefore, the values of the last column should be very close to
unity (up to at most 1\% departures from it); otherwise the model cannot be 
viewed as sufficiently accurate, even if the convergence log may show that 
the model is formally converged.


\subsection{Simple H-He model atmosphere from scratch}
\label{examp_hhe}

A simple LTE model atmosphere with $T_{\rm eff}$ = 35,000 K, $\log g$ = 4,
composed of H and He (treated as explicit), and C, N, O (implicit; taken only
for particle and charge conservation), is constructed by {\sc tlusty} 
by setting the standard input as follows -- file {\tt hhe35lt.5}:\\ [2pt]

\hrule
\hrule
\begin{verbatim}
 35000. 4.0        ! TEFF, GRAV
 T  T              ! LTE,  LTGRAY
 ''                ! no change of general optional parameters
*-----------------------------------------------------------------
* frequencies
 50                ! NFREAD
*-----------------------------------------------------------------
* data for atoms   
*
 8                 ! NATOMS
* mode abn modpf
    2   0   0
    2   0   0
    0   0   0
    0   0   0
    0   0   0
    1   0   0
    1   0   0
    1   0   0
*-----------------------------------------------------------------
* data for ions
*
*iat    iz   nlevs  ilast ilvlin  nonstd typion  filei
*
   1     0     9      0    100      0    ' H 1' './data/h1.dat'
   1     1     1      1      0      0    ' H 2' ' '
   2     0    14      0    100      0    'He 1' './data/he1.dat'
   2     1    14      0    100      0    'He 2' './data/he2.dat'
   2     2     1      1      0      0    'He 3' ' '
   0     0     0     -1      0      0    '    ' ' '
*
* end
\end{verbatim}
\hrule
\hrule
\bigskip

\noindent 
As mentioned before, one has to make a link from a standard "data" directory
to {\tt ./data} refereed to in the input file. When using the {\tt RTlusty} script,
this is done automatically.

The files {\tt h1.dat, he1.dat, he2.dat} contain necessary 
atomic data for H I, He I, and He II, respectively.
Structure of these, as well as other atomic data files, is explained in detail
in Chap.\,\ref{ions}.
Here we assume that
the model atoms contain 9, 14, and 14 levels of these ions, respectively.
H II and He III are taken as 1-level ions. Notice that there are no
additional input files associated with the highest ions, H II  and He III,
because these ions have no internal structure.
Since the {\tt ILVLIN} parameters are set to 100, higher than the number of
levels, all lines are set to detailed radiative balance, and their opacity is neglected.
This model, and its convergence properties were already shown in 
Paper~1, \S\,\reftestc.

The corresponding NLTE/C model is constructed 
with virtually the same
standard input; the only change is replacing the second record by:
\begin{verbatim}
 F  F              ! LTE,  LTGRAY
\end{verbatim}
The file is stored as {\tt hhe35nc.5}, The model is run, using the script {]\tt RTlusty} as
\begin{verbatim}
     RTlusty hhe35nc hhe35lt 
\end{verbatim}
where the second parameter, {\tt hhe35lt}, stipulates that the starting model
is taken as the result of the previous run, {\tt hhe35lt.7}.

The frequency points in both previous models are set up
with the default values of the highest frequency, 
(such as $\nu_{\rm max}= 10^{11} T_{\rm eff} $ -- see \S\,\ref{nonst_freq}), 
the lowest
frequency $ \nu_{\rm min} = 10^{12}$. The number of frequency points is
approximately 50+2$\times$(9+14+14)+20 = 144 
(i.e. NFREAD + twice the number of
levels from which a bound-free transition can occur + about 20 points at the
short- and long-wavelength tails). The actual number of
points set up by the program is 128 because unnecessary frequency points
for the He II edges that coincide with H edges are removed. 

The final NLTE/L model (NLTE with lines), considering all lines of H and He
explicitly, is calculated by modifying the fourth block of the standard input file
(stored as {\tt hhe35nl.5}) as follows:\\ [2pt]
\hrule
\hrule
\begin{verbatim}
*
*iat    iz   nlevs  ilast ilvlin  nonstd typion  filei
*
   1     0     9      0      0      0    ' H 1' './data/h1.dat'
   1     1     1      1      0      0    ' H 2' ' '
   2     0    14      0      0      0    'He 1' './data/he1.dat'
   2     1    14      0      0      0    'He 2' './data/he2.dat'
   2     2     1      1      0      0    'He 3' ' '
   0     0     0     -1      0      0    '    ' ' '
\end{verbatim}
\hrule
\hrule
\bigskip
The frequency points in all cases are set up
with the default values of the highest frequency, 
(such as $\nu_{\rm max}= 10^{11} T_{\rm eff} $ -- see the next Section), 
the lowest
frequency $ \nu_{\rm min} = 10^{12}$. The number of frequency points is
approximately 50+2$\times$(9+14+14)+20 = 144 
(i.e. NFREAD + twice the number of
levels from which a bound-free transition can occur + about 20 points at the
short- and long-wavelength tails). The actual number of
points set up by the program is 128 because unnecessary frequency points
for the He II edges that coincide with H edges are removed. 

The run is generated analogously as before,
\begin{verbatim}
     RTlusty hhe35nl hhe35nc 
\end{verbatim}
The convergence log, displayed using the IDL program {\tt pconv}, 
is shown in  in Fig. 1.  As in the analogous Fig.~1 of
Paper~I, there are six panels on the figure, the upper three display the relative
changes of temperature (which is usually the most important component of
the state vector), while the lower three panels show the maximum relative change
of all components of the state vector. The leftmost panels show the relative change
as a function of column mass in the absolute scale, and the middle panels the
same in the logarithmic scale. We display both, because the linear scale shows 
the sign of the relative changes but does not properly show those that are small.
These are in turn
clearly seen on the logarithmic scale. The rightmost panels show the maximum
relative change over all depth points. This is an indicator of the global convergence
of the model. Ideally, the relative changes should gradually decrease, which was
clearly the case here. 
The convergence is slower than for the LTE corresponding model, but still
stable. The effects of the Ng acceleration are seen in the 7th iteration, and then
after each 4 iteration steps. The figure also contains a header
that displays the core name and the total execution time. 
In the present case, it is 23 s, longer than
for the LTE model presented in Paper~1, which was less than 1 s.
This, and all other calculations reported below, were
done on MacBook Pro, with 2.2.GHz Intel Core i7.

\begin{figure}[h]
\begin{center}
\includegraphics[width=4in]{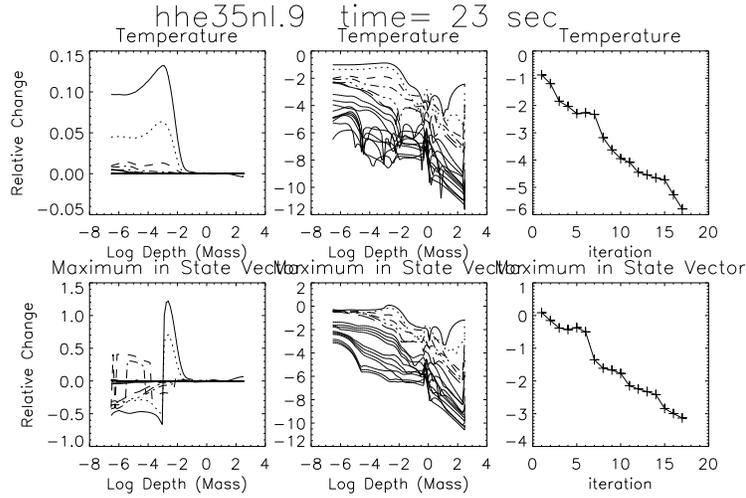}
\caption{Convergence log for the NLTE/L H-He model.}
\end{center}
\vspace{-1em}
\label{fig:f2}
\end{figure}

\begin{figure}
\begin{center}
\includegraphics[width=3.5in]{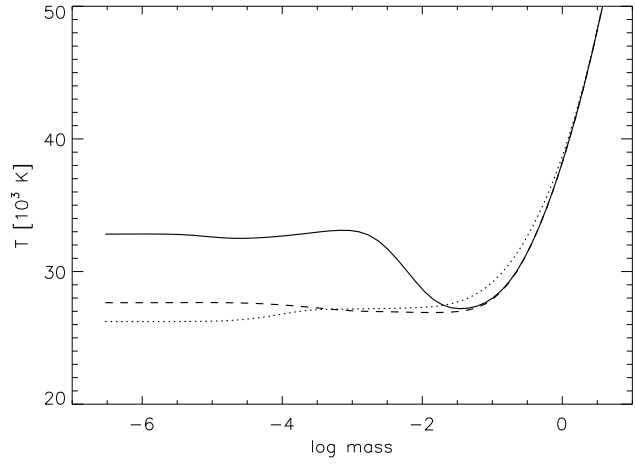}
\caption{Temperature as a function of column mass for the three
H-He modes: NLTE/L (full line), NLTE/C (dashed line), and LTE (dotted line).}
\end{center}
\vspace{-1em}
\label{fig:f3}
\end{figure}

\begin{figure}
\begin{center}
\includegraphics[width=3.5in]{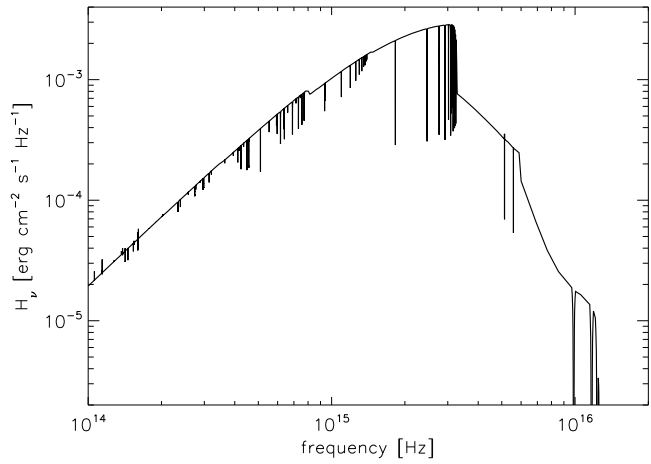}
\caption{Emergent flux for the H-He  NLTE/L model.}
\end{center}
\vspace{-1em}
\label{fig:f4}
\end{figure}

Figure 2 displays the temperature structure for all the three models. The most 
interesting feature is a temperature rise at the surface, first discovered by
Auer \& Mihalas (1969), and explained as an indirect effect of the Lyman~$\alpha$
line on the heating rate in the Lyman continuum. The plot also shows that the
temperature structure for both NLTE models approaches that of the LTE model 
deep in the atmospheres (around $\log m \approx 0.5$ in the present case), and
that the temperature structure for the NLTE/C and NLTE/L models start to
differ around $\log m \approx -1$ where lines become transparent.

Figure 3 shows the emergent flux (better speaking, $H_\nu$) as a function
of frequency, which is given by the first two columns of the file {\tt hhe35nl.13}.
We stress that this is the flux produced directly by {\sc tlusty}, and which is only
a schematic representation of the exact emergent flux because lines of only
H and He are taken into account. One can construct a detailed spectrum, including
all lines, using the associated program {\sc synspec}. 
This was shown in Paper~I, \S\,\refsyntest.


\subsection{NLTE line-blanketed B star model atmosphere}
\label{examp_bs07}

A complex example is provided for a NLTE line-blanketed B star model atmosphere,
with $T_{\rm eff}$ = 20,000 K, $\log g$ = 4, with a solar metallicity.
This example corresponds to the data used for calculating the {\sc bstar2006}
grid (Lanz \& Hubeny 2007), file {\tt BGA20000g400v2.5}, which is called here
{\tt BGA20000g400v2a.5} to stress that one will construct a different model.\\ [2pt]

\hrule
\hrule
\begin{verbatim}
20000. 4.0
 F  F              ! LTE,  LTGRAY
 'nst'             ! keyword parameters filename
*
* frequencies
*
 2000
*
* data for atoms
*
 30                 ! NATOMS
* mode abn modpf
    2   0.      0       ! H
    2   0.      0       ! He
    0   0.      0
    0   0.      0
    0   0.      0
    2   0.      0  ! C
    2   0.      0  ! N
    2   0.      0  ! O
    1   0.      0
    2   0.      0  ! Ne
    1   0.      0
    2   0.      0  ! Mg
    2   0.      0  ! Al
    2   0.      0  ! Si
    1   0.      0
    2   0.      0  ! S
    1   0.      0
    1   0.      0
    1   0.      0
    1   0.      0
    1   0.      0
    1   0.      0
    1   0.      0
    1   0.      0
    1   0.      0
    1   0.      0
    1   0.      0
    2   0.      0  ! Fe
    1   0.      0
    1   0.      0
    1   0.      0
    1   0.      0
*
* data for ions
*
*iat   iz   nlevs  ilast ilvlin  nonstd typion  filei
*
   1    0     9      0      0      0    ' H 1' 'data/h1.dat'
   1    1     1      1      0      0    ' H 2' ' '
   2    0    24      0      0      0    'He 1' 'data/he1.dat'
   2    1    20      0      0      0    'He 2' 'data/he2.dat'
   2    2     1      1      0      0    'He 3' ' '
   6    0    40      0      0      0    ' C 1' 'data/c1.dat'
   6    1    22      0      0      0    ' C 2' 'data/c2.dat'
   6    2    46      0      0      0    ' C 3' 'data/c3_34+12lev.dat'
   6    3    25      0      0      0    ' C 4' 'data/c4.dat'
   6    4     1      1      0      0    ' C 5' ' '
   7    0    34      0      0      0    ' N 1' 'data/n1.dat'
   7    1    42      0      0      0    ' N 2' 'data/n2_32+10lev.dat'
   7    2    32      0      0      0    ' N 3' 'data/n3.dat'
   7    3    48      0      0      0    ' N 4' 'data/n4_34+14lev.dat'
   7    4    16      0      0      0    ' N 5' 'data/n5.dat'
   7    5     1      1      0      0    ' N 6' ' '
   8    0    33      0      0      0    ' O 1' 'data/o1_23+10lev.dat'
   8    1    48      0      0      0    ' O 2' 'data/o2_36+12lev.dat'
   8    2    41      0      0      0    ' O 3' 'data/o3_28+13lev.dat'
   8    3    39      0      0      0    ' O 4' 'data/o4.dat'
   8    4     6      0      0      0    ' O 5' 'data/o5.dat'
   8    5     1      1      0      0    ' O 6' ' '
  10    0    35      0      0      0    'Ne 1' 'data/ne1_23+12lev.dat'
  10    1    32      0      0      0    'Ne 2' 'data/ne2_23+9lev.dat'
  10    2    34      0      0      0    'Ne 3' 'data/ne3_22+12lev.dat'
  10    3    12      0      0      0    'Ne 4' 'data/ne4.dat'
  10    4     1      1      0      0    'Ne 5' ' '
  12    1    25      0      0      0    'Mg 2' 'data/mg2.dat'
  12    2     1      1      0      0    'Mg 3' ' '
  13    1    29      0      0      0    'Al 2' 'data/al2_20+9lev.dat'
  13    2    23      0      0      0    'Al 3' 'data/al3_19+4lev.dat'
  13    3     1      1      0      0    'Al 4' ' '
  14    1    40      0      0      0    'Si 2' 'data/si2_36+4lev.dat'
  14    2    30      0      0      0    'Si 3' 'data/si3.dat'
  14    3    23      0      0      0    'Si 4' 'data/si4.dat'
  14    4     1      1      0      0    'Si 5' ' '
  16    1    33      0      0      0    ' S 2' 'data/s2_23+10lev.dat'
  16    2    41      0      0      0    ' S 3' 'data/s3_29+12lev.dat'
  16    3    38      0      0      0    ' S 4' 'data/s4_33+5lev.dat'
  16    4    25      0      0      0    ' S 5' 'data/s5_20+5lev.dat'
  16    5     1      1      0      0    ' S 6' ' '
  26    1    36      0      0     -1    'Fe 2' 'data/fe2va.dat'
   0    0                                      'data/gf2601.gam'
                                               'data/gf2601.lin'
                                               'data/fe2p_14+11lev.rap'
  26    2    50      0      0     -1    'Fe 3' 'data/fe3va.dat'
   0    0                                      'data/gf2602.gam'
                                               'data/gf2602.lin'
                                               'data/fe3p_22+7lev.rap'
  26    3    43      0      0     -1    'Fe 4' 'data/fe4va.dat'
   0    0                                      'data/gf2603.gam'
                                               'data/gf2603.lin'
                                               'data/fe4p_21+11lev.rap'
  26    4    42      0      0     -1    'Fe 5' 'data/fe5va.dat'
   0    0                                      'data/gf2604.gam'
                                               'data/gf2604.lin'
                                               'data/fe5p_19+11lev.rap'
  26    5     1      1      0      0    'Fe 6' ' '
   0    0     0     -1      0      0    '    ' ' '
*
* end
\end{verbatim}
\hrule
\hrule
\bigskip

File {\tt nst}, described in detail in \S\,\ref{nst_exa_bs07}, contains
the necessary keyword parameters: 
\begin{verbatim}
ND=50,NLAMBD=3,VTB=2.,ISPODF=1,DDNU=50.,CNU1=6.,NITER=0
\end{verbatim}
For the testing purposes, this model is not identical to the corresponding model
from the{\sc bstar2006} grid. It is calculated with a lower sampling step in frequency,
50 fiducial Doppler widths, and no iteration ({\tt NITER=0}) of the hybrid CL/ALI
scheme is done. This run therefore only initializes the superline cross sections
and performs a global formal solution (i.e., a simultaneous solution of the radiative 
transfer and the kinetic equilibrium equations to obtain new values of mean 
intensities of radiation and atomic level populations),  
so it tests most of the features specific to NLTE metal line blanketed models.
The user can obviously increase the number of iterations by skipping the
{\tt NITER=0} statement, in which case the run would however take several hours.

The first 30 elements, except Li, Be, and B, are included, all
with the solar abundances.
The model atmosphere includes 11 explicit atoms (H, He, C, N, O, Ne,
Mg, Al, Si, S, and Fe), 35 explicit ions,
and 1127 explicit NLTE (super)levels. Opacity Sampling is used
(ISPODF=1). Unlike the models from the {\sc bstar2006} grid that were calculated
with a sampling step of 0.75 fiducial Doppler width for
iron, here we use for simplicity a lower resolution of 50 fiducial Doppler widths. 
The whole spectrum is represented by 123,396 frequencies.
The mean intensity of radiation is linearized at 16 frequencies.

Again, a subdirectory {\tt "./data"}, in which all the atomic data files referred to 
in the standard input are located, must exist.
When using the standard directory tree and the script {\tt RTlusty},
the link to the standard "{\tt data}" directory is done automatically.
The model is then run simply as
\begin{verbatim}
RTlusty  BGA20000g400v2a  BGA20000g400v2 
\end{verbatim}
where {\tt BGA2000g400v2a} is the core name of the new model,
and the second parameter, {\tt BG20000g400v2} sets the
file {\tt BG20000g400v2.7} as a starting model for the
present run.

Here we also demonstrate the use of the program {\sc pretlus}. It is run as
\begin{verbatim}
cd $TLUSTY/examples/bstar
ls -s -f $TLUSTY/data  data
$TLUSTY/pretlus/pretlus.exe <BGA20000g40v2a.5
\end{verbatim}
The standard output from the program is the following: \\ [2pt]
\hrule
\hrule
\begin{verbatim}
 MATOM  =      99
 MION   =      35
 MLEVEL =    1127
 MLVEXP =     222
 MTRANS =   20476
 MDEPTH =      50
 MFREQ  =  123396
 MFREQP =  215637
 MFREQC =    4015
 MFREX  =      16
 MFREQL =    4935
 MTOT   =     241
 MMU    =       3
 MFIT   =     157
 MITJ   =     199
 MMCDW  =       6
 MMER   =       1
 MVOIGT =      79
 MZZ    =       6
 NLMX   =      80
 MSMX   =       1
 MFREQ1 =       1
 MFRTAB =       0
 MTABT  =       0
 MTABR  =       0
 MCROSS =    1132
 MBF    =    1127
 MDEPTC =       0
 MMUC   =       0

parameters in ODFPAR.FOR:

 MDODF  =          3
 MKULEV =       6870
 MLINE  =    1133840
 MCFE   =    7823728
\end{verbatim}
\hrule
\hrule
\bigskip
This is to be compared to the content of the files {\tt BASICS.FOR} and {\tt ODFPAR.FOR}
to check that the actual values of the parameters are greater than or equal to the values listed above.
If not, the files {\tt BASICS.FOR} and/or {\tt ODFPAR.FOR} have to be modified accordingly.


\subsection{LTE model atmosphere of a K star}
\label{examp_optab}

Here is an example of using the pre-calculated opacity table, while solving the
equation of state on the fly. This example also shows how to deal with
convection. The standard input file, called {\tt s4545.5} is as follows: \\ [2pt]

\hrule
\hrule
\begin{verbatim}
4500. 4.5            ! TEFF, GRAV
 T  T                ! LTE,  LTGRAY
 's45.param'         ! keyword parameters filename
*
* frequencies
*
 -15000
*
* data for atoms   
*
 30                     ! NATOMS
* mode abn modpf
    1   0.      0       ! H
    1   0.      0       ! He
    0   0.      0
    0   0.      0
    0   0.      0
    1   0.      0       ! C
    1   0.      0       ! N
    1   0.      0       ! O
    1   0.      0
    1   0.      0       ! Ne
    1   0.      0
    1   0.      0       ! Mg
    1   0.      0       ! Al
    1   0.      0       ! Si
    1   0.      0
    1   0.      0       ! S
    1   0.      0
    1   0.      0
    1   0.      0
    1   0.      0
    1   0.      0
    1   0.      0
    1   0.      0
    1   0.      0
    1   0.      0
    1   0.      0       ! Fe
    1   0.      0
    1   0.      0       ! Ni
    1   0.      0
    1   0.      0
*
* data for ions
*
*iat   iz   nlevs  ilast ilvlin  nonstd typion  filei
*
   0    0     0     -1      0      0    '    ' ' '
\end{verbatim}
\hrule
\hrule
\bigskip
\begin{figure}
\begin{center}
\label{fig5}
\includegraphics[width=4in]{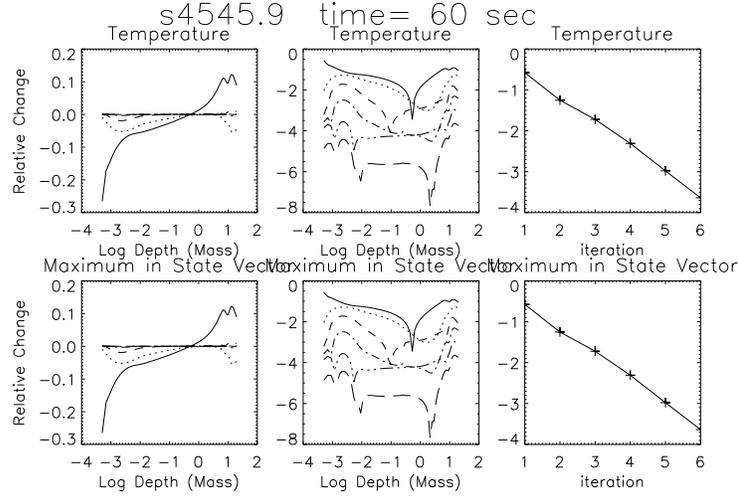}
\caption{Convergence log for the {\sc tlusty} LTE test-case model
for $T_{\rm eff}=4500$ K and $\log g=4.5$. Since the Rybicki scheme is
used, the maximum relative change of all quantities is equal to the relative 
change of temperature, so the lower
panels are redundant because they are identical to the upper ones.}
\end{center}
\vspace{-1em}
\end{figure}

The file {\tt s45.param}, described in detail in \S\,\ref{nst_exa_optab}, contains
the necessary keyword parameters, most importantly {\tt IOPTAB=-1}
which sets the mode of evaluation to the pre-calculated opacity table,
but solves the equation of state and evaluates the thermodynamic parameters
on the fly. 
The file looks like this:
\begin{verbatim}
IOPTAB=-1,IFRYB=1,IFMOL=1,IDLST=0
FRCMAX=3.2e15,FRCMIN=1.5e13,IBINOP=0,
ITEK=50,IACC=50
TAUFIR=1.e-7,TAULAS=1.0e2
HMIX0=1,NDCGAP=10,ICONRE=0,IDEEPC=3,CRFLIM=-10
\end{verbatim}

%
\begin{figure}
\begin{center}
\label{fig6}
\includegraphics[width=4in]{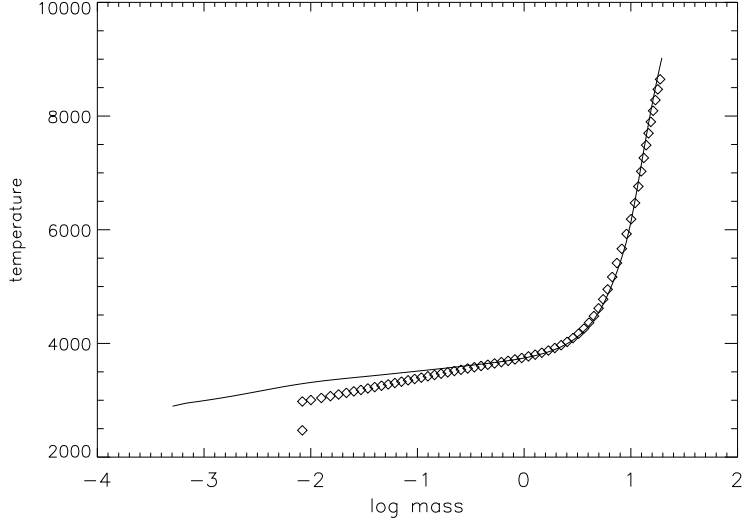}
\caption{Temperature as a function of column mass for the 
{\sc tlusty} test-case LTE model for $T_{\rm eff}=4500$ K and $\log g=4.5$
(solid line), together with the Kurucz model for the same parameters
(diamonds).}
\end{center}
\vspace{-1em} 
\end{figure}
\begin{figure}
\begin{center}
\label{fig7}
\includegraphics[width=4in]{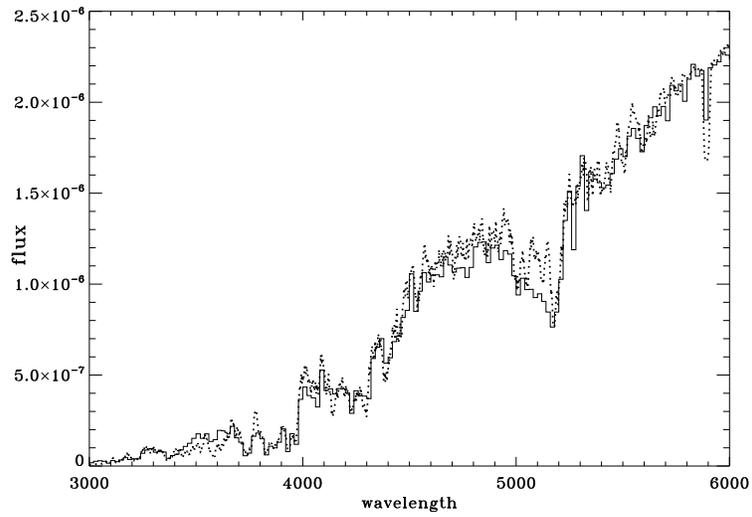}
\caption{Emergent flux for the {\sc tlusty} test-case model
with $T_{\rm eff}=4500$ K and $\log g=4.5$, smoothed to about 
1\AA \  resolution (dotted line), 
together with the Kurucz model for the same parameters 
(solid line histogram).}
\end{center}
\vspace{-1em}
\end{figure}
The frequency number parameter NFREAD is set to $-$15000, which sets
logarithmically equidistant frequencies between FRCMIN and FRCMAX,
which are set in the {\tt s45.param} file.
The first 30 elements, except Li, Be, and B,  are considered in the 
equation of state with solar abundances. In this mode, there are no explicit
atoms and ions; the input block for ions thus contains only the final line
indicating its end.

Due to the specification of the keyword parameters, several
additional input files are needed:
\begin{tabbing}
 {\tt data/absopac.dat} -- \= opacity table (because 
IOPTAB  is set to IOPTAB=$-1$); \\
{\tt data/tsuji.molec} -- \>
molecular state equation parameters (because  IFMOL=1);\\
{\tt data/irwin.dat} -- \> Irwin partition functions (by default if IFMOL $>0$).
\end {tabbing}
Using the standard {\sc tlusty} directory tree, these files are readily available.

The convergence of the model is excellent -- see Fig.\,4. It can
hardly be ever better; this is essentially thanks to the use of the Rybicki scheme
which linearizes the mean intensities of radiation in all 15,000 frequency
points, so the adopted iteration scheme is essentially an exact Newton-Raphson method.
The execution was also quite fast; the complete model was calculated in 60 s on 
a 2.2 GHz Mac. Actually, we used here a binary form of the opacity table;
its reading took 20 out of 60 s.
When using an ASCII table, the time for its reading increases to  about 50s.

In Fig.\,5, we plot
the temperature structure, together with the temperature structure of the
Kurucz model with the same $T_{\rm eff}$ and $\log g$. Overall temperature
structure is quite similar, but there are small
differences, in particular close to the surface. They are most likely caused
by our neglecting the contribution of molecules in the opacity table.
We stress that the current opacity table is not meant to provide the exact
opacity, but rather to be used for testing purposes.

Figure 6 shows an analogous comparison of the emergent flux. Again, an
agreement between the two models is quite good, in particular in view of the
fact that our opacity table is just a beta-version of a more complete opacity
table that is currently under construction.


\subsection{Moderately cool DA white dwarf}
\label{examp_cwd}

We take an example of a DA (pure-hydrogen) white dwarf with 
$T_{\rm eff}=10,000$~K and $\log g =8$. The atmosphere is
convective; we assume the ML2-type convection with the mixing length equal
to 0.6 pressure scale height. Hydrogen line broadening is treated using the
Tremblay  tables (see \S\,\ref{nst2_hyd}). The standard input 
for an LTE model is as follows (file {\tt t100g80l.5}): \\ [2pt]
\hrule
\hrule
\begin{verbatim}
  10000.0      8.0
  T    T
  'cwd.flag'
  200
*
* data for atoms
*
       1
* mode abn modpf
    2   0   0
*
*iat iz  nlv  ilst ilvln nonst typion  filei
*
 1    0   16    0    0    0   ' H 1' './data/h1s16.dat'
 1    1    1    1    0    0   ' H 2' ' '
 0    0    0   -1    0    0   '    ' ' '
 \end{verbatim}
 \vskip-6pt
\hrule
\hrule
\bigskip
The keyword parameter file {\tt cwd.flag} given as
\begin{verbatim}
IFRYB=1,IHYDPR=2
IPRINT=3,ITEK=40,IACC=40
TAUDIV=1.e-2,IDLST=0
HMIX0=0.6,MLTYPE=2,IMUCON=40,ICONRS=5
NDCGAP=5,ICONRE=0,IDEEPC=3,CRFLIM=-10.
 \end{verbatim}
Due to the specification of the keyword parameters, one needs
the additional file {\tt data/tremblay.dat}. 
Again, the model is run as
\begin{verbatim}
     RTlusty t10080l
\end{verbatim}
The convergence pattern is displayed in Fig.\, 7. The convergence is very
smooth; this is essentially due to to using the Rybicki scheme, and because 
of employing special procedures to treat convection in
the global formal solution as described in Paper~II, Appendix B2. 

\begin{figure}
\begin{center}
\label{fig8}
\includegraphics[width=4in]{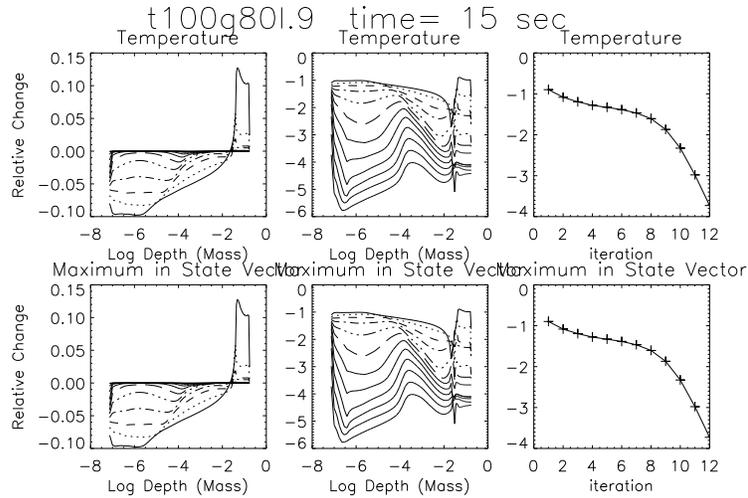}
\caption{Convergence log for the {\sc tlusty} test-case LTE model
of a DA white dwarf with $T_{\rm eff}=10,000$ K and $\log g=8$.}
\end{center}
\vspace{-1em}
\end{figure}

NLTE effects are not expected to be crucial, but to verify this statement it is necessary to
compute the corresponding NLTE model. The standard input (file {\tt t100g80n.5})
is very similar to the LTE input file, with the first three lines being now
\begin{verbatim}
      10000.0      8.0
  F    F
'cwdn.flag'
\end{verbatim}
with the keyword parameters file {\tt cwdn.flag} is a slightly modified {\tt cwd.flag},
namely
\begin{verbatim}
IFRYB=1,IHYDPR=2
IPRINT=3,ITEK=20,IACC=14
TAUDIV=1.e-2,IDLST=0
HMIX0=0.6,MLTYPE=2
NDCGAP=5,ICONRE=0,IDEEPC=3,CRFLIM=-10.
IMUCON=40,ICONRS=5
 \end{verbatim}
The most interesting feature is that one is able to converge the model while
still using the Rybicki scheme. Not only it works well in this case,
but in fact it works significantly better than the original hybrid CL/ALI scheme,
which is surprisingly difficult to converge. The reason for one can use the
Rybicki scheme at all is that departures from LTE are not very large,
and the reason for its stability is that it represents an essentially exact
Newton-Raphson scheme without an ALI treatment of the radiation
field in any frequency. The reason why the original CL/ALI scheme has problems
is that the energy balance, represented here by the radiative/convective equilibrium,
leads to a steep temperature gradient, which is hard to converge if some 
quantities are lagged behind.

The model is run simply by 
\begin{verbatim}
     RTlusty t100g80n t100g80l
\end{verbatim}

\begin{figure}
\begin{center}
\label{fig9}
\includegraphics[width=4in]{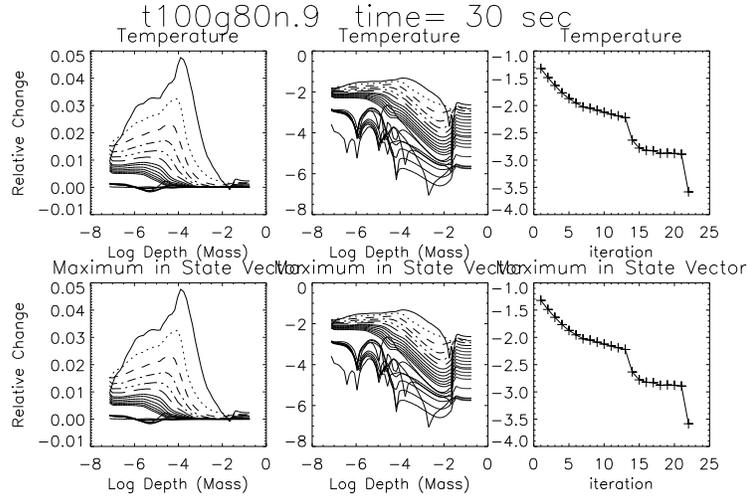}
\caption{Convergence log for the {\sc tlusty} test-case NLTE model
of a DA white dwarf with $T_{\rm eff}=10,000$ K and $\log g=8$.}
\end{center}
\vspace{-1em}
\end{figure}
\noindent The convergence pattern is shown in Fig. 8. The convergence is relatively slow,
but stable. We see a significant effect of the Ng acceleration, which is first
applied in the 14th iteration, and then  after each 4 iterations. Usually, one
does not apply Ng acceleration for models with convection, but in this case
its use is warranted because the atmospheric structure in deep layers, where
convection is present, is not expected to be influenced by NLTE effects to any
significant degree.

\subsection{Accretion disk around a white dwarf}
\label{examp_accr}

The standard input for a simple H-He model of the vertical structure of 
an accretion disk around a white dwarf, analogous to the simple model 
atmosphere computed in \S\,\ref{examp_hhe}, is as follows (file {\tt d1lt.5}): \\ [2pt]

\hrule
\hrule
\begin{verbatim}
 0.8 1.e-9  0.01 2        ! XMSTAR,XMDOT,RSTAR,RELDST
 T  T                     ! LTE, LTGRAY
 'param'                  ! keyword parameters file
*------------------
* frequencies
 50                ! NFREAD
*------------------
* data for atoms   
*
 8                 ! NATOMS
* mode abn modpf
    2   0   0
    2   0   0
    0   0   0
    0   0   0
    0   0   0
    1   0   0
    1   0   0
    1   0   0
*-----------------
* data for ions
*
*iat    iz   nlevs  ilast ilvlin  nonstd typion  filei
*
   1     0     9      0    100      0    ' H 1' './data/h1.dat'
   1     1     1      1      0      0    ' H 2' ' '
   2     0    14      0    100      0    'He 1' '/data/he1.dat'
   2     1    14      0    100      0    'He 2' '/data/he2.dat'
   2     2     1      1      0      0    'He 3' ' '
   0     0     0     -1      0      0    '    ' ' '
*
* end
\end{verbatim}
\hrule
\hrule
\bigskip
and the keyword parameter file {\tt param} is simply
\begin{verbatim}
ALPHAV=0.3
\end{verbatim}

{\bf Important note}: To tell {\sc tlusty} that an accretion disk model is computed,
one has to create a file {\tt fort.1} with just one number, 1.

This run computes an LTE model (from scratch) of a ring
of an accretion disk around a star with mass 0.8 solar masses and
radius $0.01 R_\odot$  (a typical white dwarf),  and with a mass accretion rate
$\dot M = 10^{-9}$ $M_{\odot}$/year, and at a distance 2 stellar
radii. The viscosity parameter is set to $\alpha=0.3$.

An essentially identical model is obtained replacing the first line 
of the standard input  file with
\begin{verbatim}
0.  35165.  3.934e-2  5.742e1       ! TEFF, QGRAV, DMTOT  
\end{verbatim}
Here, as explained in \S\,\ref{new1}, the central star mass is
set to 0, in which case the following three parameters have the
meaning of the effective temperature $T_{\rm eft}$ (TEFF), gravity
acceleration parameter $Q$ (QGRAV), and the total column mass
$m_0$ (DMTOT). These parameters can be found at the beginning
of the standard output file of the previous run. In this case the
parameter $\alpha$ is not needed (because it is used only to evaluate
the total column mass), so that the keyword parameter file {\tt param}
can be empty. However, since $\alpha$ is not used, its value set up
in {\tt param} is inconsequential, and therefore the run may be done
with the file {\tt param} unchanged. 

One can also continue on and construct NLTE/C and NLTE/L models,
exactly as in the case of a simple H-He atmosphere of
\S\,\ref{examp_hhe}.  When this is done, one finds that the departures
from LTE are much smaller now because of much higher effective
gravity acceleration (that is, $g$ is large everywhere except very close
to the midplane), and the corresponding high density of the material.


\section{Keyword parameters}
\label{nonst}
\index{Keyword parameters}

\subsection{General description}
\label{key_gen}

There are over 210 individual keyword parameters. We stress that a
keyword, or optional, parameter is defined as such for which the program assigns
a default value that is meant to provide an optimum value for  most applications,
or the most reasonable value. A small number of keyword parameters are
actual physical parameters, like the convective mixing length or the
turbulent velocity, which are set to zero for most applications. 
However, the majority
of keyword parameters are various computational flags and switches
that specify a detailed setup of the numerical method.
Notice that the overall degree of sophistication of the resulting model
atmosphere is determined by the atomic input files, described in Chap.\,\ref{ions}.
The keyword parameters discussed in this section,
with the exception of the physical parameters mentioned above, will mostly
influence the numerical performance, such as the rate of convergence (or a lack 
thereof!), the total computer time, the numerical accuracy, the degree of auxiliary
output, etc.

The file, whose name is specified by the standard parameter FINSTD (see 
\S\,\ref{new1}), contains a list of the keyword parameters to be changed
from their default values. The input format is very simple, namely
\begin{verbatim}
   PARAM1=VALUE1,PARAM2=VALUE2,...
\end{verbatim}
where {\tt PARAM1}, etc., are the names of the parameters as specified below,
and {\tt VALUE1}, etc, are the corresponding numerical values of the 
parameters. The individual  entries are delimited by one of the following
special characters: comma ({\tt ,}),  space, left or right bracket ({\tt (} or
{\tt )}), asterisk ({\tt *}), slash ({\tt /}), or carriage return.
The parameters may appear in any order. If a parameter does not appear in the
list, its default value is adopted by the program. 
The name of a parameter has to be coded with capital letters. 
The numerical values 
{\tt VALUE1}, etc., may be up to 6 character long. 
This system of input was adopted from Carlsson's program 
{\sc multi} (Carlsson 1986);
we have used and modified his routines GETWRD and RINPUT.

{\bf Hint}: Since parameters to be changed have to be specified by capital letters.
changing one letter to a lower case may be used as a way of ``commenting
out'' a given parameter, which can then easily be reinstated later.
For instance, coding\\ [2pt]
{\tt ITEK=8}\\ [2pt]
has the effect of starting the Kantorovich acceleration 
(see Sect.\ref{nonst_acc}) at the 8th iteration; while setting\\  [2pt]
{\tt ITE=8} \ \ {\rm or} \ \ {\tt itek=8}  \ \ {\rm or} \ \ {\tt iTEK=8}\\ [2pt]
has no effect, and ITEK will assume its default value (ITEK=4).

In this section, we will describe only the most important parameters. 
The less important, very specific, or rarely used parameters are described 
in Part II in Chaps.\,\ref{nst2_phys} and \ref{nst2_num}.


\subsection{Input physical parameters}
\label{nonst_phys}

Most basic physical parameters ($T_{\rm eff}$, $\log g$, chemical
abundances; and basic parameters for a ring in an accretion disk) 
are set up by the standard input file. However, there are some 
physical parameters which
are being set by corresponding keyword parameters. We list them below.

\subsubsection{Universal parameters}

\begin{description}

\item[HMIX0] -- the mixing length 
\index{HMIX0 keyword parameter}
\index{Mixing length!input}
parameter:\\
$\bullet\,>$ 0 -- convection is considered; HMIX0 has the meaning of 
mixing length;\\
$\bullet\,= 0$ -- convection is suppressed, but the adiabatic and radiative gradients
are calculated and printed (essentially, to check for the convective stability of
the resulting model);\\
$\bullet\,< 0$  -- convection is suppressed and no gradients are calculated.\\
DEFAULT: HMIX0=-1.

\item[MLTYPE] -- switch for the type of the mixing length 
\index{MLYTPE keyword parameter}
prescription:\\
$\bullet\,= 0$ or 1 -- normal mixing length (type 1);\\
$\bullet\,= 2$ -- the so-called ML2 prescription (Fontaine et al 1981; Bergeron et al.
1992)\\
DEFAULT: MLTYPE=0.

\item[VTB] -- microturbulent velocity; in km s$^{-1}$ (if $<10^3$)
\index{VTB keyword parameter}
\index{Turbulent velocity!input}
or cm s$^{-1}$ (if  $\geq 10^3$).\\
DEFAULT: VTURB=0.

\item[IPTURB] -- switch for setting the effects of microturbulent 
\index{IPTURB keyword parameter}
velocity:\\
$\bullet\,= 0$ -- the turbulent velocity is considered only for modifying the 
Doppler widths, but not for turbulent pressure 
\index{Turbulent pressure!setup}
(which is set to zero);\\
$\bullet\,= 1$ -- the turbulent velocity is considered both for Doppler widths as well
as for determining the turbulent pressure via 
$P_{\rm turb}= (1/2) \rho v_{\rm turb}^2$\\
DEFAULT: IPTURB=1.

\end{description}

\subsubsection{White dwarf--specific parameters}
\label{nonst_wd}

They account for equilibrium diffusion of helium and thus enable to
compute models with
helium stratification, by specifying the total mass of the layer of
\index{Helium stratification}
hydrogen that floats on a helium layer. The formalism is taken from 
Vennes et al. (1988).

\begin{description} 

\item[HCMASS] -- if set to a non-zero value, the total mass of the 
\index{HCMASS keyword parameter}
hydrogen layer (in g or in $M_\odot$) 
for stratified H-He models with equilibrium helium diffusion.\\
$\bullet\,=0$ -- no helium stratification\\
DEFAULT: HCMASS=0.

\item[RADSTR] -- stellar radius (in units of the solar radius) of the
\index{RADSTR keyword parameter}
star (for stratified models with equilibrium helium diffusion)\\
DEFAULT: RADSTR=0.

\end{description}

\subsubsection{Accretion disk--specific parameters}
\label{nonst_accr}

\begin{description}

\item[ALPHAV] -- the  viscosity 
\index{ALPHAV keyword parameter}
\index{Viscosity!parameter $\alpha$}
parameter $\alpha$. \\
$\bullet\,>0$ -- the numerical value of $\alpha$;\\
$\bullet\,\leq 0$ -- the viscosity is set through the Reynolds number.\\
DEFAULT: ALPHAV=0.1

\item[REYNUM] -- the Reynolds number.  It takes an effect,  and
\index{REYNUM keyword parameter}
\index{Reynolds number}
the viscosity is evaluated through the Reynolds number, only 
if the keyword parameter ALPHAV is set to zero or a negative value.\\
DEFAULT: REYNUM=0.

\item[IVISC] -- a mode of treating depth-dependent 
\index{IVISC keyword parameter}
\index{Viscosity!setup}
viscosity:\\
$\bullet\,\geq 0$ -- the depth-averaged viscosity is determined through
the ``averaged'' $\alpha$, given by the parameters ALPHAV. By default, the
local kinematic viscosity is constant, but can also be treated as a two-step
power law as described by Hubeny \& Hubeny (1998). In this case, one
needs to employ four more keyword parameters,
as described in \S\,\ref{nst2_disk}.\\
$\bullet\,< 0$ -- the local viscous stress is given by 
$t_{\phi r} = \alpha P_{\rm gas}$;
i.e. $\alpha$ determines the {\em local} viscosity.\\
DEFAULT: IVISC=0

\end{description}


\subsection{External irradiation}
\label{exter}
\index{External irradiation}

External irradiation brings complications into the methodology
of computing model stellar atmospheres because 
it breaks several fundamental assumptions of the adopted modeling 
procedure. First, it generates departures from horizontal homogeneity. 
Second, it involves a dependence of specific intensity not only on the 
azimuthal angle, but also on the polar angle.

There is a special variant of {\sc tlusty}, briefly described in Sudarsky et al. 
(2005), that is able to deal with a full angular dependence of the external irradiation.
The corresponding calculations are, however, quite time-consuming even if one employs
a powerful ALI-based numerical scheme for solving the radiative transfer equation.
Experience shows that such detailed calculations are necessary only in special
cases, for instance for strongly  irradiated extrasolar
giant planets. 

The mainstream {\sc tlusty} therefore adopts a simpler approach.
One replaces a highly anisotropic irradiation by an isotropic radiation intensity
that gives the same total incoming energy. The irradiated intensity is taken either
as hoc, as a diluted black-body spectrum, or through a theoretical flux from
an actual model atmosphere. In that case, the dilution factor is either set up
to a specific value (either an empirical one, or that obtained by independent 
calculations), or is given through the radius and distance of the irradiating object.
In the present version, the irradiating object is assumed to be a spherical star.

To deal with departures from the horizontally homogeneous approximation,
one can either construct several model atmospheres for varying distances 
from the substellar point (the point on the surface of the irradiated star at which 
the external irradiation comes in the direction of the normal to the surface),
or to assume that the irradiated intensity is spread over the surface of the
irradiated star. One then introduces a factor $f$ that represents the ratio between
the specific intensity at the substellar point and the specific intensity that is
characteristic of the averaged irradiation over the whole irradiated hemisphere.
Assuming a homogeneous distribution of energy gives $f=1/2$, but it turned 
out (see Burrows et al. 2008) that the most appropriate value is $f=2/3$.

The external specific intensity is given by
\begin{equation}
\label{extint}
I_\nu^{\rm irrad} = W H_\nu^\ast
\end{equation}
where $H_\nu^\ast$
is the Eddington flux (the first moment of the specific intensity) at the surface
of the irradiating object (star), in erg cm${}^{-2}$ s${}^{-1}$ Hz${}^{-1}$.
The reason for using the Eddington flux is that most stellar atmosphere programs 
and model grids ({\sc tlusty}; Kurucz) provide $H_\nu$ to represent the emergent 
spectrum. The dilution factor in this case is given by
\index{Dilution factor}
\begin{equation}
\label{wdil}
W= f (1/2) (R^\ast/d)^2,
\end{equation}
where the factor $1/2$ comes from the the conversion of $H$ to $I$, see 
Eq. (7) of Paper~II.

Here is a list of keyword parameters that control the treatment of
irradiation:

\begin{description}
\item[TRAD] -- a switch for treating the external
\index{TRAD keyword parameter}
\index{External irradiation!setup}
irradiation:\\
$\bullet\,= 0$ -- no external irradiation is considered;\\
$\bullet\,> 0$  -- the external irradiation is considered; the irradiated
intensity of radiation is assumed to be given by Eq, (\ref{extint}),
$I_\nu^{\rm irrad} = W  B_\nu(T^\ast)$,
where $W$ is the dilution factor, $B$ the Planck function, and $T^\ast$ the
characteristic temperature of incoming radiation. The parameter TRAD
has then the meaning of $T^\ast$. This is a general option, which
may mimic for instance an irradiation from a distributed source.
In the special case 
of an irradiation by a spherical star, $T^\ast$
is given by the effective
temperature of the irradiating star. \\
$\bullet\,< 0$  -- the irradiated intensity at the top of the atmosphere
is given by the emergent spectrum of the source of irradiation, Eq. (\ref{extint}), 
where the Eddington flux $H_\nu$, in erg cm${}^{-2}$ s${}^{-1}$ Hz${}^{-1}$,
is read from a special input file {\tt source.flux}.\\ 
The quantity abs(TRAD) then represents the number of frequency points in the
file {\tt source.flux}. These frequencies  can be arbitrary; the program then interpolates to the
frequency points used by {\sc tlusty}.\\
The dilution factor $W$ is given either through the
parameter WDIL, or is computed using the radius of the irradiating star,
$R^\ast$ (given by keyword parameter RSOURC),
the distance between the modeled object and the irradiation source, $d$
(parameter ADIST), and the spreading factor $f$ (parameter SPRFAC) -- see below.\\
DEFAULT: TRAD=0.

\item[WDIL] -- the dilution factor $W$, used either for TRAD $>0$,
\index{WDIL keyword parameter}
or for TRAD $<0$ if the parameters RSOURC and ADIST are not set.\\
DEFAULT: WDIL=1.

\item[RSOURC] -- the radius if the irradiating star.
\index{RCOURC keyword parameter}
Can be expressed either in cm, or in the units of solar radii. It takes effect
only for TRAD $<0$.\\
DEFAULT: RSOURC=0.

\item[ADIST] -- distance between the center of the irradiating star
\index{ADIST keyword parameter}
and the top of the irradiated atmosphere. It can be expressed either in 
cm or in Astronomical Units. It takes effect only for  TRAD $<0$.\\
DEFAULT: ADIST=0.

\item[SPRFAC]  -- spreading factor $f$ used in Eq. (\ref{wdil}). It takes effect 
\index{SPRFAC keyword parameter}
only for  TRAD $<0$.\\
DEFAULT: SPRFAC=0.667

\end{description}


\subsection{Basic numerical setup parameters}
\label{nonst_bas}

\subsubsection{Global setup}
\label{nonst_glob}

These are potentially very important parameters that in fact determine the
overall mode of calculation of a model atmosphere.

\begin{description}
\item[IFRYB] -- a switch for invoking the Rybicki 
\index{Rybicki scheme!setup}
\index{IFRYB keyword parameter}
scheme:\\
$\bullet\,= 0$ -- original hybrid CL/ALI scheme is used\\
$\bullet\,\geq 1$ -- Rybicki scheme is used\\
DEFAULT: IFRYB=0

\item[ISPODF] -- the basic mode of treating 
\index{ISPODF keyword parameter}
\index{Opacity Distribution Function}
line blanketing:\\
$\bullet\,= 0$ --  a classical option -- line blanketing is either not considered,
or treated through the (obsolete) Opacity Distribution Function (ODF)
option. The frequency points are 
set up for each line
(or ODF) separately. Lines are assumed to have a Doppler or Voigt profile.
The regions between lines are represented by a small number of
frequency points, set up through input parameter NFREAD, and 
additional keyword parameters -- see \S\,\ref{nonst_freq}.\\
$\bullet\,\geq 1$ - the Opacity Sampling (OS) mode. Frequencies are set up 
\index{Opacity Sampling}
systematically with a variable step depending on the kind of opacity 
source to be represented.
Small frequency steps are adopted to sample adequately lines of light 
elements. The regions in between these lines and the iron-peak lines are 
sampled with a step DDNU times the fiducial Doppler width.
For details refer to \S\,\ref{nonst_blank}.

\item[IOPTAB] -- a switch specifying the use of pre-calculated 
\index{IOPTAB keyword parameter}
opacity tables:\\
$\bullet\,= 0$ -- classical option -- no pre-calculated opacity table is used;\\
$\bullet\,<$ 0 -- all opacities are calculated by means of a pre-calculated opacity
table. In this case, one does not specify any explicit atoms, ions, and levels.
Only LTE models can be computed with this option.\\
$\bullet\,=-1$ -- in addition, equation of state and thermodynamic parameters are
computed on the fly.\\
$\bullet\,=-2$ -- in addition, also equation of state and thermodynamic parameters
are given by pre-calculated tables.\\
$\bullet\,>$ 0 -- hybrid option -- one still selects explicit atoms, ions, and levels, for
which the opacity is computed on the fly, while for remaining species one
uses an appropriate  pre-calculated opacity table.\\
DEFAULT: IOPTAB=0 

\item[IFMOL] -- a switch for including molecular formation in the equation 
\index{IFMOL keyword parameter}
of state.\\
$\bullet\,= 0$ -- molecular formation neglected\\
$\bullet\,> 0$ -- molecules are included in the equation of state, as described 
in Paper~II, \S\,\refeos.\\
DEFAULT: IFMOL=0

\item[ICOMPT] -- a switch for including Compton scattering. In this case,
\index{ICOMPT keyword parameter}
several other keyword parameters take effect -- see 
\S\,\ref{nst2_compt}.\\
$\bullet\,= 0$ -- Compton scattering is not included; electron scattering is 
treated as Thomson (coherent) scattering;\\
$\bullet\,>0$ -- Compton scattering is included.\\
DEFAULT: ICOMPT=0

\end{description}

\subsubsection{Setup of frequency points}
\label{nonst_freq}

\begin{description}

\index{FRCMAX keyword parameter}
\item[FRCMAX] --  the maximum frequency $\nu_{\rm max}$.\\
$\bullet\,= 0$ -- maximum frequency is set up such as: \\
(a) for atmospheres: $\nu_{\rm max} = 8 \times 10^{11} T_{\rm eff}$;
i.e. $h\nu_{\rm max}/kT_{\rm eff} \approx 38$;\\
(b) for accretion disks:
$h\nu_{\rm max}/kT_{\rm mid} = 17$, where $T_{\rm mid}$ is the
local temperature at the midplane, evaluated approximately through
the effective temperature and the total column density $\Sigma$ as
$T_{\rm mid} = 2.83 \times 10^{11}\, T_{\rm eff} \, (0.2\, \Sigma)^{1/4}$
(see Hubeny et al. 2001).\\
$\bullet\,>$ 0 -- the value of maximum frequency.\\
DEFAULT: FRCMAX=0.

\item[CFRMAX] -- an auxiliary parameter for setting the maximum
\index{CFRMAX keyword parameter}
frequency. \\
$\bullet\,=0$ -- the value of $\nu_{\rm max}$ given by FRCMAX is unchanged;\\
$\bullet\,>1$ --  the value of $\nu_{\rm max}$ is set to the maximum of the one given
by FRCMAX and CFRMAX times the frequency of the explicit
photoionization edge with the highest-frequency. \\
DEFAULT: CFRMAX=2 for atmospheres; =0 for disks

\index{FRCMIN keyword parameter}
\item[FRCMIN] --  the minimum frequency.\\
$\bullet\,= 0$ -- the minimum frequency is set to $10^{12}$ s$^{-1}$.\\
$\bullet\,> 0$ -- the value of the minimum frequency.\\
DEFAULT: FRCMIN=$10^{12}$

\index{FRLMAX keyword parameter}
\item[FRLMAX] --  the maximum frequency in the line transitions.
If the central frequency of a line is larger than FRLMAX, the line is neglected.\\
$\bullet\,= 0$ -- the maximum frequency in line transitions is set to FRCMAX.\\
$\bullet\,> 0$ -- the value of maximum frequency in line transitions.\\
DEFAULT: FRLMAX=FRCMAX

\index{FRLMIN keyword parameter}
\item[FRLMIN] --  the minimum frequency in the line transitions.\\
$\bullet\,= 0$ -- the minimum frequency in the line transitions is set to 
$10^{13}{\rm s}^{-1}$.\\
$\bullet\,> 0$ -- the value of minimum frequency in the line transitions\\
DEFAULT: FRLMIN=$10^{13}$

\item[NFTAIL] -- one of parameters that determine the 
\index{NFTAIL keyword parameter}
setting of frequency points in the high-frequency tail.\\
$\bullet\,>$ 0 --  it has the meaning of the number of frequency points
between the highest-frequency continuum edge and the maximum frequency
given by parameter {\rm FRCMAX}. 
The integration is done by the Simpson formula, 
so that NFTAIL must be an odd number. Specifically, the integration is done 
by two Simpson integrations, dividing the total tail region into two parts,
each part is done by a (NFTAIL/2+1)-point Simpson integration, and the part 
nearer the photoionization edge is DFTAIL times the total interval. This allows
one to consider denser grid of frequency points just blueward of the
highest-frequency discontinuity, which yields more accurate evaluation of the
photoionization rate of the corresponding transition.\\
$\bullet\,<$ 0 --   the frequencies are set as equidistant in $\log \nu$
between the minimum and maximum frequency. The total number of
frequency points is now exactly {\rm NFREAD}. This option is very useful
if there are many explicit levels which would generate an
unnecessarily large number of frequencies around completely unimportant
edges.\\
DEFAULT: NFTAIL=21

\index{DFTAIL keyword parameter}
\item[DFTAIL] -- see above.\\
DEFAULT: DFTAIL=0.25

\end{description}

\subsubsection{Setup of the radiative transfer equation and ALI}
\label{nonst_rte}
\begin{description}

\item[ISPLIN] -- Mode of numerical representation of the radiative
\index{ISPLIN keyword parameter}
\index{Radiative transfer equation!setup}
transfer equation (for details see Paper~II, \S\,3.8.1)\\
$\bullet\,= 0$ -- standard second-order Feautrier scheme \\
$\bullet\,= 1$ -- spline collocation scheme \\
$\bullet\,= 2$ -- Auer's fourth-order Hermitian scheme\\
$\bullet\,= 3$ -- Rybicki--Hummer improved Feautrier scheme \\
$\bullet\,= 5$ -- Discontinuous Finite Element (DFE) scheme  \\  
DEFAULT: ISPLIN = 0

\index{IFALI keyword parameter}
\index{Accelerated Lambda Iteration (ALI)}
\item[IFALI] -- basic switch for treating the radiative transfer equation for
frequency points that are treated with ALI in the hybrid CL/ALI method:\\
$\bullet\,= 0$  -- no ALI option, i.e. all frequency points are explicitly linearized. All
transition that are not selected for linearization are treated using the
''fixed-rates''  option (essentially an ordinary Lambda iteration) as in original
{\sc tlusty} -- Hubeny (1988);\\
$\bullet\,= 1 - 4$ -- ALI scheme, with some limitations. It has only a historical
significance; it was used for testing purposes.\\
$\bullet\,= 5$ -- ALI scheme with diagonal $\Lambda^\ast$ is used for frequency
points treated within the ALI  method;\\
$\bullet\,= 6$ -- ALI scheme with tri-diagonal $\Lambda^\ast$. This option was
not fully tested in version 205; and does not work properly for some
setups. It is not recommended.\\
DEFAULT: IFALI=5

\index{JALI keyword parameter}
\item[JALI] -- switch determining the type of evaluation of the 
$\Lambda^\ast$ operator.\\
$\bullet\,= 1$ -- $\Lambda^\ast$ evaluated by the Rybicki-Hummer (1991) algorithm;\\
$\bullet\,= 2$ -- Olson-Kunasz (1987) algorithm.\\
DEFAULT: JALI=1

\item[IFRALI] --  a switch for a global change of the ALI mode of a whole
\index{IFRALI keyword parameter}
group of frequency points\\
$\bullet\,= 0$ -- the ALI mode for all frequencies is determined by the input for the
individual frequency points and/or for individual transitions;\\
$\bullet\,= 1$ -- all frequency points in lines are set to ALI mode (thus overwriting
a specific input for line transitions in an atomic data file);\\
$\bullet\,= 2$ -- all frequency points altogether are set to the ALI mode (i.e.
the fully ALI scheme is forced regardless of other input).\\
$\bullet\,= -1$ -- all points are in the linearized mode, regardless of other
input (i.e., the original complete linearization method).\\
DEFAULT: IFRALI=0

\item[RADZER]  -- a parameter for zeroing the mean intensity.
\index{RADZER keyword parameter}
This option avoids numerical problems connected with extremely small
intensity at the highest frequencies, in particular for accretion disks.
When the mean intensity decreases below RADZER times the maximum relative
intensity, it is set to zero, and the radiative transfer equation is written
and linearized as $J_\nu =0$.\\
DEFAULT: RADZER=1.e-20

\end{description}

\subsubsection{Setup of the kinetic equilibrium equations}
\label{nonst_ese}
\index{Kinetic equilibrium equation}
\begin{description}
\index{IATREF keyword parameter}
\item[IATREF] -- a flag for setting up the reference atom.\\
{\bf Reference atom} is the species to which all abundances
\index{IATREF keyword parameter}
 are related (usually, but not necessarily, hydrogen).
 IATREF refers to the {\em index} of the atomic species in the numbering
 of explicit species. Therefore, for instance, if H is not considered
 explicitly at all (mode 0 or 1) and He is explicit, then the index 
 IAT for He is set to 1, and the IATREF should be set to 1 (which is
 the default anyway).\\
$\bullet\,= 0$ -- IATREF is set to 1 (i.e. the first explicit species).\\
DEFAULT: IATREF=1

\item[IDLTE] --  a depth point below which all the explicit levels are forced
\index{IDLTE keyword parameter}
to have LTE populations (even for NLTE models);\\
DEFAULT: IDLTE=1000 (no LTE populations are forced)

\item[POPZER] --  a value of the ratio of a level population over the population
\index{POPZER keyword parameter}
of the most populated level of a given species, below which the population
is declared to be ``too small'' and is set to 0. This option allows to consider
many ionization degrees of an atom without running into numerical problems
connected with too small/large numbers.\\
DEFAULT: POPZER=1.e-20

\item[NITZER] --  iteration number till which the population can be
\index{NITZER keyword parameter}
zeroed or un-zeroed. After NITZER-th iteration, any new zeroing is 
switched off, so the populations that were already zeroed at this 
iteration stay zeroed till the end of the run.\\
DEFAULT: NITZER=1

\end{description}

\subsubsection{Setup  of the radiative equilibrium equation}
\label{nonst_re}
\index{Radiative equilibrium equation}

\begin{description}

\item[NDRE] -- a parameter that defines the treatment of the radiative
\index{NDRE keyword parameter}
equilibrium equation, namely a form of the superposition of the integral 
and the differential equation representations:\\
$\bullet\,= 0$ -- a linear combination of both forms is used, after Hubeny \& Lanz
(1995). In this case, the form of the linear combination is given by
parameters TAUDIV and IDLST, in such a way that:\\ [2pt]
   -- the integral form is used for depth points $d$, 
 $d=1,\ldots,$ND$-$IDLST \\ [2pt]
   -- the differential form is used for depth points where the Rosseland optical 
 depth is greater than TAUDIV.\\  [2pt]
The coefficients $\alpha$ and $\beta$ in Eqs. (16) or (19) of Paper~II
are thus given by
\begin{eqnarray}
\alpha_d=1\quad {\rm for}\quad d \leq {\rm IDLST}, \quad
\alpha_d=0\quad {\rm elsewhere}, \nonumber \\ 
\beta_d =1 \quad {\rm for}\quad \tau_{\rm Ross} > \tau_{\rm div}, \quad
\beta_d=0 \quad {\rm elsewhere}, \nonumber
\end{eqnarray}
$\bullet\,>$ 0 -- the coefficients are step functions with discontinuity at
$d$ = NDRE, i.e.:\\ [2pt]
-- for depth points  ($d=1,\ldots,$NDRE$-$1)  -- purely integral form ;\\ [2pt]
-- for depth points  ($d=$NDRE,$\ldots$,ND) -- purely differential form.\\ [2pt]
This option is outdated and is not recommended. Much better option is
the previous one, NDRE=0.\\ [2pt]
DEFAULT: NDRE=0

\item[TAUDIV] --  see above (effective only if NDRE=0)\\
\index{TAUDIV keyword parameter}
DEFAULT: TAUDIV=0.5

\item[IDLST] -- see above (effective only if NDRE=0)\\
\index{IDLST keyword parameter}
DEFAULT: IDLST=5

\item[NRETC] -- a switch for setting a fixed temperature
\index{NRETC keyword parameter}
at the upper layers\\
$\bullet\,= 0$ -- no fixed temperature; radiative equilibrium is solved for
all depths\\
$\bullet\,>0$ -- the temperature at the first NRETC depth points is held fixed
(given by the input model); radiative equilibrium at those depths
is not solved for the temperature.\\
DEFAULT: NRETC=0

\end{description}

\subsubsection{Discretization and linearization loop control}
\label{nonst_discr}
\begin{description}

\index{ND keyword parameter}
\item[ND] -- number of depth points\\
DEFAULT: ND=70

\item[NMU] -- number of angle points for the formal solution of the
\index{NMU keyword parameter}
transfer equation; using the double-Gaussian quadrature in angles.\\
DEFAULT: NMU=3

\index{NITER keyword parameter}
\item[NITER] -- maximum number of global iterations of the linearization
scheme.\\
DEFAULT: NITER=30

\item[CHMAX] -- maximum relative change of the state vector. If all the
\index{CHMAX keyword parameter}
relative changes of all state parameters at all depth points are below
this value, the model is declared converged and the execution stops
after a finished formal solution.\\
DEFAULT: CHMAX= $10^{-3}$

\item[NLAMBD] -- number of  iterations for the  global formal solution; which
typically consists of solving the coupled radiative
\index{NLAMBD keyword parameter}
transfer and kinetic equilibrium equation.
Historically, these were ordinary Lambda iterations, hence the name NLAMBD.
Currently the default procedure is instead the ALI scheme with preconditioning.\\
DEFAULT: NLAMBD=2 for NLTE; NLAMBD=1 for LTE models

\item[CHMAXT] --  a parameter which enables to change the number of
\index{CHMAXT keyword parameter}
iterations of the global formal solution (NLAMBD)  when the model 
is almost converged.
If the maximum of the absolute values of the relative changes 
of temperature at all depths decreases below CHMAXT,
 the number of lambda iterations is set to NLAMT. \\
DEFAULT: CHMAXT=0.01

\index{NLAMT keyword parameter}
\item[NLAMT] --  the reset number of iterations of the global formal
solution -- see above.\\
DEFAULT: NLAMT=1

\end{description}

\subsubsection{Setup of the linearization matrices}
\label{nonst_lin}

Each of the following parameters: INHE, INRE, INPC, INSE, INMP, INDL, INZD, 
corresponds to one equation and one model parameter, as described below.

 If INxx = 0, then the corresponding equation is not solved, and the
              corresponding quantity is not a component of the state vector.
              Instead, it is held fixed to the values from the starting model.
              
 If INxx $>$ 0, the corresponding equation is solved, and the
              corresponding quantity is the  (NFREQE+INxx)-th component
              of the vector $\psi$ of unknown model parameters
              (the first NFREQE components are mean intensities of
              radiation in the explicitly linearized frequency points).
              In the following, by ``the position'' we mean the position of the
              corresponding quantity in the state vector after the first NFREQE
              quantities.
\begin{description}

\index{INHE keyword parameter}
\item[INHE] -- the position of $N$; the index of the hydrostatic equilibrium
equation.\\
DEFAULT: INHE=1

\item[INRE] -- the position of $T$; the index of the radiative equilibrium
\index{INRE keyword parameter}
equation.\\
DEFAULT: INRE=2

\item[INPC] -- the position of $n_{\rm e}$; the index of the charge conservation 
\index{INPC keyword parameter}
equation (or the number conservation equation, depending on parameter
ICHC). In any case, it is the equation which determines the electron 
density. \\
DEFAULT: INPC=3 (or INPC=4 for convective models)

\item[INSE] -- the position of $n_1$, i.e. the first population; index of the 
\index{INSE keyword parameter}
first kinetic equilibrium equation.\\
DEFAULT: INSE=4 (or INSE=5 for convective models or disks)

\item[INMP] -- the position of $n_{\rm m}$ -- massive particle number
\index{INMP keyword parameter}
density; after Auer \& Mihalas (1969). This option is included for historical 
reasons only.\\
DEFAULT: INMP=0

\item[INDL] -- the position of $\nabla$ -- the logarithmic gradient of 
\index{INDL keyword parameter}
temperature. It is used only for convective models.\\ 
DEFAULT: INDL=0; or INDL=3 for convective models

\item[INZD] -- a position of $z$ -- vertical distance from the
\index{INZD keyword parameter}
midplane. Used for disk models only.\\
DEFAULT: INZD=0 for atmospheres; INZD=4 for disks.

\item[IFIXMO] -- a shortcut for setting all parameters INHE, INRE,
INPC and (for disks) INZD, to zero. This is done if IFIXMO is set to a non-zero
\index{IFIXMO keyword parameter}
value.\\
DEFAULT: IFIXMO=0

\end{description}

\subsubsection{Acceleration parameters}
\label{nonst_acc}

\begin{description}

\index{IACC keyword parameter}
\index{Ng acceleration!setup}
\item[IACC] -- a switch for the Ng acceleration procedure:\\
$\bullet\,\leq$ 4 -- Ng acceleration is done  in the 7th, 11th, etc, iteration;\\
$\bullet\,\geq$ 5 -- Ng acceleration is done in the iterations ITER=IACC, IACC+IACD, 
            IACC+2$\times$IACD, etc.\\
$\bullet\,\leq$ 0 -- no Ng acceleration.\\
DEFAULT: IACC=7

\index{IACD keyword parameter}
\item[IACD] -- the step for the Ng acceleration -- see above.\\
DEFAULT: IACD=4

\item[KSNG] -- if set to 1, one excludes populations of all
\index{KSNG keyword parameter}
levels except ground states and the highest ions from contributing
to evaluating the acceleration parameters. In other words, these
populations have weights $W_{di}=0$ in Eq. (140) of Paper~II.\\
DEFAULT: KSNG=0

\index{ITEK keyword parameter}
\index{Kantorovich acceleration!setup}
\item[ITEK] -- the iteration after which the Kantorovich method is set up.\\
DEFAULT: ITEK=4

\index{ORELAX keyword parameter}
\index{Successive over-relaxation!setup}
\item[ORELAX] -- the over-relaxation coefficient.\\
DEFAULT: ORELAX=1.

\end{description}


\subsection{LTE-gray model input}
\label{nonst_gr}

\subsubsection{Stellar atmospheres}
\label{nonst_gr_atmos}

\begin{description}

\index{TAUFIR keyword parameter}
\item[TAUFIR] -- the Rosseland optical depth in the first depth point.\\
DEFAULT: TAUFIR $=10^{-7}$

\index{TAULAS keyword parameter}
\item[TAULAS] -- the Rosseland optical depth in the last depth point.\\
DEFAULT: TAULAS $=3.16\times 10^2$

\item[ABROS0] -- the initial estimate of the Rosseland opacity (per gram) 
\index{ABROS0 keyword parameter}
at the first depth point.\\
DEFAULT: ABROS0 = 0.4

\item[IPRING] -- a flag that controls the diagnostic output (to the standard
output file, unit 6) of the LTE-gray model
\index{IPRING keyword parameter}
calculations:\\
$\bullet\,= 0$ -- no output;\\
$\bullet\,= 1$ -- only final LTE-gray model is printed;\\
$\bullet\,= 2$ -- results of all internal iterations are printed;\\
DEFAULT: IPRING=0

\item[NCONIT] -- a number of internal iterations for calculating the
\index{NCONIT keyword parameter}
 gray model with convection (for convective models).\\
DEFAULT: NCONIT=10

\item[ICHANM] -- a switch to setting the mass scale of the LTE-gray model
\index{ICHANM keyword parameter}
in the convection zone.\\
$\bullet\,= 0$ -- the column mass scale that follows from the first estimate 
of the model without convection is held fixed;\\
$\bullet\,> 0$ --  the column mass scale is recalculated from $\tau_{\rm ross}$
and the current $\chi_{\rm ross}$ that already takes into account a change
of the structural parameters ($T$, $n_{\rm e}$) due to the presence of
convection.\\
DEFAULT: ICHANM=1

\end{description}

\subsubsection{Accretion disks}
\label{nonst_gr_disk}

For details of the physical and the mathematical formulation, see Paper~II, \S\,\refgraydisk.

\begin{description}

\index{DM1 keyword parameter}
\item[DM1] -- the column mass in the first depth point.\\
DEFAULT: DM1 $=10^{-3}$

\item[ABPLA0] -- the initial estimate of the Planck mean opacity (per gram)
\index{ABPLA0 keyword parameter}
at the first depth point.\\
DEFAULT: ABPLA0 = 0.3
 
\index{ABPMIN keyword parameter}
\item[ABPMIN] -- the floor value of the Planck mean opacity. That is, the Planck
mean opacity is set to the maximum of computed Planck mean and the value of ABPMIN.
It is introduced to avoid spuriously low values of the Planck mean opacity at the
initial stages of computation.\\
DEFAULT: ABPMIN $=10^{-5}$

\item[ABROS0] -- the initial estimate of the Rosseland opacity (per gram) 
\index{ABROS0 keyword parameter}
at the first depth point -- the same as in the case of stellar atmospheres.\\
DEFAULT: ABROS0 = 0.4

\item[IDMFIX] -- a switch for setting the characteristic temperature for
\index{IDMFIX keyword parameter}
evaluating the sound speed and the gas and radiation pressure scale heights
when computing the starting LTE-gray model.\\
$\bullet\,=1$ -- the characteristic temperature is set to the effective temperature;\\
$\bullet\,\not= 1$ the characteristic temperature is set to $T_1$, the temperature
at the first depth point, which is determined iteratively -- see \S\,\ref{nst2_disk}.\\
DEFAULT: IDMFIX=1

\item[ITGMAX] -- the number of internal iterations in computing the
\index{ITGMAX keyword parameter}
LTE-gray model.\\
DEFAULT: ITGMAX=10

\index{NNEWD keyword parameter}
\item[NNEWD] -- indicator of changing the depth grid in the
LTE-gray model.\\
$\bullet\,= 0$ -- the grid is set up once and for all and is not changed;\\
$\bullet\,> 0$ -- the grid is updated NNEWD times.\\
DEFAULT: NNEWD=0

\item[TDISK] -- if set to positive value, it has the meaning
\index{TDISK keyword parameter}
of the temperature of the disk, assumed constant with height.
This provides a means to compute the hydrostatic structure of an isothermal
disk. Otherwise, the temperature structure is determined by solving
the energy balance equation.\\
DEFAULT: TDISK=0 

\end{description}


\subsection{Parameters for modifying the starting input model atmosphere}
\label{nonst_mod}

\begin{description}
\item[INTRPL] -- a switch indicating that the input model atmosphere
\index{INTRPL keyword parameter}
has to be interpolated to a new depth scale to serve as a starting
model for the current run. For a detailed discussion, see Chap. \ref{unit8}.\\
$\bullet\,= 0$ -- no interpolation; i.e. the same depth grids are used  in the input and 
the current models;\\
$\bullet\,> 0$ -- the input model is interpolated to a new depth grid.
The actual interpolation scheme is the polynomial
interpolation of the (INTRPL$-$1)th order;\\
$\bullet\,< 0$ -- the starting model atmosphere is a Kurucz model.\\
DEFAULT: INTRPL=0

\item[ICHANG] -- switch indicating a change of the explicit level structure
\index{ICHANG keyword parameter}
between the input model atmosphere and the current run. It has an effect
only for NLTE models. For a detailed discussion, see Chap.\,\ref{unit8}.\\
$\bullet\,= 0$ -- no change of level structure (i.e. the same explicit levels, with the
same overall indices, are considered in the input model and in the
current model to be computed);\\
$\bullet\,> 0$ -- a ``simple'' change of the level structure -- levels are only added,
and only for new species. Starting populations of the new levels are computed in
LTE;\\
$\bullet\,< 0$ -- change of the structure; a detailed additional input for each level
of the present run is required. \\
DEFAULT: ICHANG=0

\end{description}


\subsection{Line blanketing}
\label{nonst_blank}

\begin{description}

\index{ISPODF keyword parameter}
\item[ISPODF] -- the basic mode of treating line blanketing:\\
$\bullet\,=0$ -- line blanketing, described through the concept of superlevels and
superlines, is either not considered, or is treated in the Opacity Distribution Function (ODF) 
\index{Superlevels}
\index{Superlines}
\index{Opacity Distributton Function}
mode (which is still offered by {\sc tlusty}, but is obsolete). 
Here, frequencies are set up for each line or ODF. The corresponding
cross sections must be pre-tabulated; the filenames of the corresponding
tables are specified in the standard input. In this case, there are 
no other optional parameters to be specified.\\
$\bullet\,\ge 1$ -- the Opacity Sampling (OS) mode. Frequencies are set up 
\index{Opacity Sampling!setup}
systematically, with a variable
step depending on lines and continua. Small frequency steps are
adopted to sample the lines of light elements adequately. In between
these lines, iron-peak lines are sampled with a step DDNU times
the fiducial Doppler width for Fe. Line cross sections are calculated
by {\sc tlusty} based on Kurucz data files. Lines are selected dynamically
based on a criterion including ionization, excitation and the $g\!f$-value.
Parameter STRLX sets the selection criterion (the smaller STRLX the 
more lines are selected; 
values $10^{-6} \leq {\rm STRLX} \leq 10^{-10}$ are generally appropriate).
The cross sections are recalculated after the first 3 Ng accelerations if
the maximum relative temperature change is larger than CHMAXT.
Cross sections are calculated by default (JIDS=0) at three depth points
(top, bottom, and the layer where $T \approx T_{\rm eff}$) and are
logarithmically interpolated
in between. They may be calculated at more depth points, 
in which case JIDS $>0$ gives the number of layers.\\
{\bf Note}: The default is set to 0 to help most new users to start with
simple models (say, H-He only), with no lines or with a few lines, in which case
ISPODF=0 is appropriate. However, we stress that for computing
fully blanketed models the option ISPODF=1 (the Opacity Sampling approach) is preferable.\\
DEFAULT: ISPODF=0

\item[DDNU] -- the step for the Opacity Sampling expressed in fiducial
\index{DDNU keyword parameter}
Doppler widths for iron (see above).\\
DEFAULT: DDNU=0.75

\item[CNU1] -- Controls the highest frequency at which lines are taken
\index{CNU1 keyword parameter}
             into account (lines at higher frequencies are omitted).
             This frequency is defined as
             $\nu_{\rm max} = {\tt CNU1} \times 10^{11} \times T_{\rm eff}$.
             Frequencies for Opacity Sampling are set up from $\nu_{\rm max}$
             to lower frequencies, or from the highest bound-free limit
             if this limit exceeds $\nu_{\rm max}$. \\
             DEFAULT: CNU1=4.5

\item[CNU2] -- Defines the lowest frequency for Opacity Sampling, lower 
\index{CNU2 keyword parameter}
frequencies
             are only included to represent the continuum and lines from
             light elements down to frequency {\tt FRLMIN}. This lowest frequency
             is defined as \\ $\nu_{\rm min} = 3.28805\,10^{15}/{\tt CNU2}^2$. \\
             DEFAULT: CNU2=3.

\item[STRLX] -- selection criterion for iron-peak element lines
\index{STRLX keyword parameter}
in the Opacity Sampling mode (see above)\\
DEFAULT: STRLX=1.e-10

\item[STRL1, STRL2] -- Define groups of intermediate and weaker lines that will be
\index{STRL1 keyword parameter}
\index{STRL2 keyword parameter}
represented by a limited number of frequencies, hence optimizing
the number of frequencies in Opacity Sampling mode. \\
DEFAULT: STRL1=0.001; STRL2=0.02

\item[JIDS] -- number of depths at which the cross sections for
\index{JIDS keyword parameter}
\index{Superlines}
superlines are computed and stored.\\
$\bullet\,=0$ -- sets the default of 3 depths (with depth indices $I\!D=1, N\!D$, 
and the depth where $T \approx T_{\rm eff}$);\\
$\bullet\,>0$ -- number of depths. Their indices are set equidistant between 1 and 
$N\!D$ (including these).\\
DEFAULT: JIDS=0

\end{description}


\subsection{Treatment of convection}
\label{nonst_conv}
\index{Convection}

\begin{description}

\index{ICONV keyword parameter}
\item[ICONV] --  a flag to switch on convection: \\
$\bullet\,= 0$ -- convection is neglected. However, if the keyword parameter
HMIX0 is set to a positive value, ICONV is reset to ICONV=1.\\
$\bullet\,>$ 0 -- convection is considered, and is linearized. There are several
numerical options, but they are obsolete. The user
is recommended to use the value ICONV=1 (which is set automatically if
the parameter HMIX0 is set).\\
Note: if ICONV is set to ICONV $>$0, but HMIX0=0, then all
convective routines are called, but the convective flux will
always be zero, so effectively no convection is allowed for.\\
$\bullet\,<$ 0 -- convection is taken into account, but it is not linearized.\\
DEFAULT: ICONV=0

\item[ICONRE] -- if set to a positive value, subroutine CONREF is called
\index{ICONRE keyword parameter}
that recalculates temperature in the convection zone between the
individual iterations of the global linearization scheme, using the
procedure specified in Paper~II, Appendix B2.
ICONRE has the meaning of the  iteration number till which the
correction procedure is being performed.  However, the correction is done
only at depth points where the convective flux exceeds a certain fraction
of the total flux, given by the parameter CRFLIM. Usually, this
routine helps significantly; there may however be cases when it does not
help. The symptom of this problem is that the program produces essentially identical 
corrections of temperature
in the linearization (that is, CONREF changes the temperature and the linearization
changes it back). In those cases, routine CONREF should be disabled by setting
ICONRE=0.\\
DEFAULT: ICONRE=1

\item[ICONRS] -- a similar switch for the starting iteration number in which
\index{ICONRS keyword parameter}
the correction procedure CONREF is performed.\\
DEFAULT: ICONRS=10

\item[CRFLIM] -- minimum local value of $F_{\rm conv}/F_{\rm tot}$ for
\index{CRFLIM keyword parameter}
which routine CONREF corrects the temperature in the convection zone.\\
DEFAULT: CRFLIM=0.7

\item[IMUCON] -- a switch for invoking an additional temperature
correction procedure,
\index{IMUCON keyword parameter}
defined by Eq. (343) of Paper~II, Appendix B2,
on top of the standard one that is set up through the keyword parameters
ICONRE and ICONRS. It takes effect only if ICONRE $>0$ (that is,
if a correction is done at all).\\
$\bullet\,=0$ -- no additional correction procedure;\\
$\bullet\,>0$ -- an additional correction procedure is performed at the global
iteration starting with iteration number IMUCON.\\
DEFAULT: IMUCON=10

\end{description}

If ICONRE is set to a positive value, then there are two additional
parameters that control the work of routine CONREF, namely:
\begin{description}
\item[IDEEPC] -- if set to a positive value, then the correction to the current
\index{IDEEPC keyword parameter}
temperature is performed even in the so-called "convection gaps",
which are defined as depth zones that are located between zones in which
there is a convection ($\nabla > \nabla_{\rm ad}$), but for which the
current values lead to  $\nabla < \nabla_{\rm ad}$ and thus are viewed
as convectively stable (this may happen during a global iteration process
because in the convection zone $\nabla$ is always very close to $\nabla_{\rm ad}$,
and the linearization may easily lead to $\nabla$ slightly smaller than
$\nabla_{\rm ad}$). Moreover, the actual value of IDEEPC controls 
the treatment of the deepest layers of the atmosphere:\\
$\bullet\,= 0$ -- convection gaps are not considered; the temperature
correction procedure is performed only in the layers with  
$\nabla > \nabla_{\rm ad}$.\\
$\bullet\,= 1$ -- convection gaps are considered as described above,
but the position of the convection zone in the deepest layers is left 
as specified by the current values of $\nabla$ and  $\nabla_{\rm ad}$.\\
$\bullet\,\geq 2$ -- if the convection zone ends at ND-1 depth zone,
then it is reset to ND (ND being the total number of depth zones). Again, this
may happen in the iteration process, and is essentially always spurious.\\
$\bullet\,=3$ -- the end of convection zone is reset to ND regardless of
where it currently ends.\\
$\bullet\,=4$ -- if in the current iteration
the end of convection zone appears above the beginning of the convection zone
in the previous iteration, the current beginning and end are reset to the values
from the previous iteration.\\
DEFAULT: IDEEPC=2

\index{NDCGAP keyword parameter}
\index{Convective gap}
\item[NDCGAP] -- specifies the number of the depth zones that define the
``convection gap". \\
DEFAULT: NDCGAP=2

\end{description}

It is best to explain these parameters and corresponding computational strategies
on an example. Let the convection instability condition $\nabla > \nabla_{\rm ad}$
be currently satisfied for depth zones 10, 35-38, and 41-50. The convection at depth 10
is completely spurious, while the convection zone should extend from 35 to 50.

If IDEEPC is set to 0, then the convection zone is assumed to extend from depth
10 to 50, and the temperature correction is performed there. Since the algorithm
forces the gradient to be close to the adiabatic one, this clearly leads to
wrong results, and although linearization will try to correct temperature back,
the routine CONREF with IDEEPC=0 would do more harm than help in this case.

If IDEEPC=1 or 2, and NDCGAP is set between 3 and 24, then the gap 
between 10 and 35
is too large to be considered a convection gap, and the extent of the convection
zone is set correctly between depths 35 and 50. If NDCGAP is set to 1 or 2, then
the gap between 38 and 41 is too large, and the temperature correction would be
performed only between 41 and 50, so the convection zone would be spuriously narrow and the correction procedure would not work properly. This demonstrates that while NDCGAP should be set up to a reasonably large value, one should not set it to be too large. When the procedure does not converge, one should inspect the standard output file where some details about temperature correction in the convection zone are provided, and which may give a clue as to what values of the convection switches to choose.


\subsection{Artificial truncation of the radiation pressure}
\label{nonst_misc}
\index{Radiation pressure}
The reason for considering a truncation of radiation pressure is to avoid
numerical as well as physical instability on the surface layers of hot stars.
In this case the true radiation pressure may exceed the gas pressure, leading
to an onset of stellar wind, which cannot be treated by {\sc tlusty}. However,
it is still quite meaningful to construct hydrostatic model photospheres for the
deeper layers of such stars, for which the hydrostatic equilibrium is an excellent approximation, and where most of the observed spectral features
(except strongest lines) do originate. A more detailed discussion of this
topic is presented in Lanz \& Hubeny (2003).
\begin{description}
\item[XGRAD] -- a switch controlling an artificial lowering of radiation 
\index{XGRAD keyword parameter}
acceleration at surface layers.\\
$\bullet\,= 0$ -- allows to auto-limit the radiative acceleration to some 
fraction of the gravity in the 10 uppermost depth points. The fraction
of the gravity acceleration to which the radiation acceleration is 
limited to was found empirically and is hard-wired in the program; 
it is given as
$(0.1,0.3,0.5,0.7,0.9,0.92,0.94,0.96,0.98,0.99)$ times $g$, in the
first 10 depths, respectively.\\
$\bullet\,= -1$ or $-2$ -- allows to impose a more stringent cut-off in the 20 
uppermost layers.\\
$\bullet\,> 0$ -- auto-limits the radiative acceleration to fraction XGRAD 
of the gravity acceleration $g$ {\em everywhere}
in the atmosphere. \\
DEFAULT: XGRAD=0.

\index{IFPRAD keyword parameter}
\item[IFPRAD] -- a switch for turning off the radiation pressure completely.\\
$\bullet\,=0$ -- radiation pressure is set to zero everywhere;\\
$\bullet\,>0$ -- radiation pressure is considered (but still may be limited by means
of the keyword parameter XGRAD).\\
DEFAULT: IFPRAD=1

\end{description}


\subsection{Examples}
\label{nst_exa}

Now we take the examples of files with keyword parameters that were considered
in Chap.\,\ref{new_ex} and explain in more detail their contents and the reasons for
the adopted settings.

\subsubsection {Simple H-He model atmosphere in \S\,\ref{examp_hhe}}
\label{nst_exa_hhe}

In this case, the standard input file sets the name of the keyword parameter file
FINSTD to a null string. in which case no keyword parameters were set, and
thus all assume their default values. Recapitulating the basic features of
the standard, default setup:\\
$\bullet\,$ model is computed by the hybrid CL/ALI method, with Kantorovich
acceleration started in the 4th and Ng acceleration in the 7th global iteration;\\
$\bullet\,$ no superlines, no convection, no molecules, and no additional opacities.

\subsubsection{NLTE line-blanketed model of a B star in \S\,\ref{examp_bs07}}
\label{nst_exa_bs07}

The keyword parameter file {\tt nst} (with the name specified in the standard input)
has the following content:
\begin{verbatim}
ND=50,NLAMBD=3,VTB=2.,ISPODF=1,DDNU=50.,CNU1=6.,NITER=0
\end{verbatim}
\noindent which have the following meaning:\\
$\bullet\,$ ND=50 -- sets the number of depth points to 50. It is lower than the
\index{ND keyword parameter}
default value ND=70, and was used in our {\tt ostar2002} and {\tt bstar2006}
\index{OSTAR2002 model grid}
\index{BSTAR2006 model grid}
model grids because: (i) the overall accuracy of the model is only insignificantly
lower than for 70 depth points; (ii) the convergence of the ALI (and thus the
hybrid CL/ALI) scheme is somewhat faster; and (iii) for practical reasons, namely
that setting ND=50 (and also MDEPTH=50 in the ``INCLUDE'' file {\tt BASICS.FOR})
decreases the memory of consumption significantly, considering that line-blanketed
models need a large amount of memory for computing superline cross sections
due to the required large values of parameters {\tt MKULEV}, {\tt MLINE}, and, 
in particular {\tt MCFE} in the "INCLUDE" file {\tt ODFPAR.FOR}. If the core memory
is no longer an issue, one could calculate models with default ND=70
or with an even higher value of ND.\\ [2pt]
$\bullet\,$ NLAMBD=3 -- sets the number of internal iterations in the global
\index{NLAMBD keyword parameter}
formal solution to 3. 
This choice leads to a more stable overall iteration process.
Having more iterations would  improve the global convergence 
properties even more, but the formal solution would be more time-consuming. 
The adopted value of 3 represents a reasonable compromise.\\ [2pt]
$\bullet\,$ VTB=2. -- set the turbulent velocity to 2 km/s \\ [2pt]
$\bullet\,$ ISPODF=1 -- the basic parameter that sets the calculation of a 
line-blanketed
\index{ISPODF keyword parameter}
model with treating the superline cross sections in the OS (Opacity 
Sampling) approach.\\ [2pt]
$\bullet\,$ DDNU=50. -- sets the sampling step to 50 fiducial Doppler width. This
is a lower resolution that was used in {\sc bstar2006} grid, where DDNU=0.75
\index{DDNU keyword parameter} 
was used. \\ [2pt]
$\bullet\,$ CNU1=6.  -- specifies the highest frequency for which are the
superlines consdired, as explained
\index{CNU1 keyword parameter}
in \S\,\ref{nonst_blank}. \\ [2pt]
$\bullet\,$  NITER=0 -- sets the number of the linearization iterations to zero; 
that is, only an initialization and the first formal solution is performed.

\subsubsection{LTE model atmosphere of a K star in \S\,\ref{examp_optab}}
\label{nst_exa_optab}

The content of the keyword parameter file {\tt s45.param} is as follows
\begin{verbatim}
IOPTAB=-1,IFRYB=1,IFMOL=1,IDLST=0
FRCMAX=3.2e15,FRCMIN=1.5e13,IBINOP=0,
ITEK=50,IACC=50
TAUFIR=1.e-7,TAULAS=1.0e2
HMIX0=1,NDCGAP=10,ICONRE=0,IDEEPC=3,CRFLIM=-10
\end{verbatim}

\noindent which has the following meaning:\\ [2pt]
$\bullet\,$ IOPTAB=$-1$ -- sets the mode of evaluation of opacities to using 
the pre-calculated opacity table, but with
\index{IOPTAB keyword parameter}
solving the equation of state and evaluating the thermodynamic parameters
on the fly.\\ [2pt]
$\bullet\,$ IFRYB=1 -- sets the global iteration method to the Rybicki scheme. 
For LTE models of cool stars, and in particular when convection is present, 
this scheme works much better than the standard hybrid CL/ALI scheme as it is more
\index{Rybicki scheme}
\index{IFRYB keyword parameter}
stable, and the convergence rate is much faster.\\ [2pt]
$\bullet\,$ IFMOL=1 -- stipulates that molecules are included in the equation of state (as is of \index{IFMOL keyword parameter}
course mandatory for such low temperature models).\\ [2pt]
$\bullet\,$ IDLST=0  -- sets the proper treatment of the energy equation at the
\index{IDLST keyword parameter}
lower boundary. This option has to be used in conjunction with IFRYB=1.\\ [2pt]
$\bullet\,$ FRCMAX, FRCMIN -- sets the minimum and maximum 
\index{FRCMAX keyword parameter}
\index{FRCMIN keyword parameter}
frequency.\\ [2pt]
$\bullet\,$ IBINOP=0 -- signifies that the opacity table is an ASCII file. 
The default value
\index{IBINOP keyword parameter}
of the parameter, IBINOP=1 would accept the opacity table as a binary file, which
works faster, but it is less portable.\\ [2pt]
$\bullet\,$ ITEK=50, IACC=50 -- inhibits both Kantorovich and Ng acceleration
(that is, they would start at 50th iteration, but the total defaults number of CL
iterations is 30). This option is much safer when dealing with convective 
models.\\ [2pt]
$\bullet\,$ TAUFIR=1.e-7,TAULAS=1.0e2 -- sets the minimum and maximum
\index{TAUFIR keyword parameter}
\index{TAULAS keyword parameter}
Rosseland optical depth when computing the starting LTE-gray model. In fact,
TAUFIR does not have to be set because 1.e-7 is the default anyway. The
model can easily be computed also with the default TAULAS (3.16e2), 
but one would
go to unnecessary large depth in the atmosphere, which is inconsequential
for the bulk of atmospheric structure and emergent radiation.\\ [2pt]
$\bullet\ $ HMIX0=1 -- switches on the convection, with the mixing length equal
\index{HMIX0 keyword parameter}
to 1 pressure scale height.\\ [2pt]
$\bullet\ $ NDCGAP=10,ICONRE=0,IDEEPC=3,CRFLIM=-10 -- this was
found, by some experimentation, to be the a good set of parameters 
for the numerical
treatment of convection, namely how the convective flux and temperature are
being recalculated in the global formal solution step to lead to a stable convergence of the CL iterations. For details, refer to \S\,\ref{nonst_conv}.
\index{NDCGAP keyword parameter}
\index{ICONRE keyword parameter}
\index{IDEEPC keyword parameter}
\index{CRFLIM keyword parameter}

\subsubsection{Cool white dwarf model in \S\,\ref{examp_cwd}}
\label{nst_exa_cwd}

The content of the keyword parameter file {\tt cwd.flag} is as follows:
\begin{verbatim}
IFRYB=1,IHYDPR=2
IPRINT=3,ITEK=40,IACC=40
TAUDIV=1.e-2,IDLST=0
HMIX0=0.6,MLTYPE=2
NDCGAP=5,ICONRE=0,IDEEPC=3,CRFLIM=-10.
IMUCON=40,ICONRS=5
\end{verbatim}
\noindent 
Here, many keyword parameters are set similarly as in the previous
case of a K star; here we point out only different or otherwise
interesting  parameters 
which have the following meaning:\\ [2pt]
$\bullet\,$ IFRYB=1 -- sets the global iteration method to the Rybicki scheme. 
As pointed
out in \S\,\ref{examp_cwd}, this scheme
\index{Rybicki scheme}
works much better than the standard hybrid CL/ALI scheme as it is
\index{IFRYB keyword parameter}
more stable, and the convergence rate is much faster. In the present case,
this setup works well even for NLTE models (file {\tt cwdn.flag}).\\ [2pt]
$\bullet\,$ IHYDPR=2 -- sets the evaluation of the hydrogen line profiles
to using the Tremblay tables, which are most suitable for white dwarf models --
see \S\,\ref{nst2_hyd}.\\ [2pt]
$\bullet\,$ IPRINT=3 -- this setup produces more output on Unit 6
(standard output), namely a table of convection parameters printed after each
completed iteration of the global linearization scheme. This is useful if 
something goes wrong with a treatment of convection, or with the
temperature correction procedures.\\ [2pt]
$\bullet\,$ ITEK=40,IACC=40 -- as is prudent for models with convection,
both accelerations (Ng and Kantorovich) are turned off.\\ [2pt]
$\bullet\,$ TAUDIV=0.01 -- setting the coefficients of the
linear combination of the differential and integral form of the energy balance
equation -- see \S\,\ref{nonst_re}. 
The value of 0.01, which yields accurate results (that is, a better conservation
of the total flux) than the default value
of 0.5, can only be set that small if the Rybicki scheme is used.\\ [2pt]
$\bullet\,$ HMIX0=0.6  -- convection is switched on; the mixing length set to
0.6 pressure scale height, which is often considered as a typical value for 
white dwarfs.\\ [2pt]
$\bullet\,$ MLTYPE=2 -- a variant of the mixing-length description, ML2,
is considered. Again, it is often used for white dwarfs.\\ [2pt]
$\bullet\,$ NDCGAP=5,ICONRE=0,IDEEPC=3,CRFLIM=-10. -- analogous
as before.\\ [2pt]
$\bullet\,$ IMUCON=40,ICONRS=5 -- additional convection parameters.\\ 

The keyword parameter file for a NLTE models, {\tt cwdn.flag}, is very similar,
the only difference is:\\ [2pt]
$\bullet\,$ ITEK=20,IACC=14 -- the Ng acceleration now starts at the 14th iteration,
and the Kantorovich acceleration at the 20th. As mentioned in \S\,\ref{examp_cwd}.
one can afford to switch on the Ng acceleration because NLTE effects are not
supposed to be influencing the temperature structure in any significant way.
Moreover, the temperature structure, as well as the position of the convection 
zone, were well established in the previously computed LTE model. Although the model would converge without the Ng acceleration (the user may easily verify 
that by setting IACC=40), the Ng acceleration improves the convergence behavior significantly, as is clearly seen in Fig. 8.

\subsubsection{Accretion disk around a white dwarf in \S\,\ref{examp_accr}}
\label{nst_exa_accr}

In this case, the keyword parameter file {\tt param} is simple:
\begin{verbatim}
ALPHAV=0.3
\end{verbatim}
\index{ALPHAV keyword parameter}
which sets the viscosity parameter $\alpha=0.3$. Without setting it,
\index{Viscosity!parameter $\alpha$}
$\alpha$ would be set to 0.1, which in fact is a typical value used for
accretion disk studies. The present value of 0.3 has no special meaning;
it was chosen just for the purposes of showing how to set $\alpha$.


\section{Starting model atmosphere}
\label{unit8}

Except when computing an LTE model from scratch (i.e. LTE-gray model, 
by setting {\tt LTGRAY = .TRUE.}) (for details, refer to Paper~II, Chap.\,4),
a starting model atmosphere is needed.  
It is transmitted by input Unit 8, i.e. as a file with name {\tt fort.8}.
In most applications, the user does not have to care about the structure of
the file because it is usually created by a previous run of {\sc tlusty}. 

However, the program also accepts model atmospheres created by {\sc tlusty} with 
different choice of explicit atoms, ions, and energy levels than that
which is specified for the current run. This is straightforward if 
the starting model is an LTE one. However, if 
the starting model is a NLTE one, the input also contains all
the explicit level populations, and therefore would be 
incompatible with the current indexing of levels. 
Consequently, a special input is needed in these
cases, which is signaled by coding the parameter {\rm ICHANG} 
to be non-zero (see \S\,\ref{inp_chan}). Some examples were also presented
in Paper~I, \S\,5.6 and 5.7.

Finally, the program accepts as a starting model a Kurucz model atmosphere
in Kurucz's standard format.
We will now describe the relevant input parameters in detail.


\subsection{{\sc tlusty} input model atmosphere}
\label{inp_tlus}

As mentioned earlier, the input file {\tt fort.8} is an output file {\tt fort.7} from
a past run of {\sc tlusty}. Its structure is as follows:\\

\smallskip
\noindent $\bullet$ {\bf 1st line}:

\begin{description}
\item[NDEPTH] -- number of depth points in which the initial model is
given. Usually, but not necessarily, it is equal to the number of depth
points, ND, considered for the current run. If it is not equal to ND,  one
has to interpolate the structure from the input model to the current
column mass scale -- see below.
\item[NUMPAR] -- number of input model parameters in each depth:\\
= 3 (or 4 for disks) -- for an LTE model ($T$, $n_{\rm e}$, $\rho$; plus $z$ 
for disks);\\
$>$ NLEVEL+3 (or 4 for disks) --  for a NLTE model (as above, plus 
populations of all explicit levels of the input model).
\end{description}

\medskip
\noindent $\bullet$ {\bf Next block}:

\begin{description}
\item[DEPTH] ($d=1,\ldots,$NDEPTH) -- depth grid for the input model,
i.e. the column mass [g cm$^{-2}$] for all depth points. If INTRPL=0
and NDEPTH=ND, this depth grid will be used for the current model as
well (DM $\equiv$ DEPTH). If INTRPL=0 and NDEPTH $\not=$ ND,
the program stops.
\end{description}

\medskip
\noindent $\bullet$ {\bf For each depth point}:

\begin{description}
\item[T]   -- temperature, $T$ [K]
\item[ANE] -- electron density, $n_{\rm e}$ [cm$^{-3}$]
\item[RHO] -- mass density, $\rho$ [g cm$^{-3}$]
\item[ZD] -- geometrical distance from the midplane [cm] -- for disks only
\item[level populations] -- $n_i$ [cm${}^{-3}$] for $i=1,\ldots,$NLEVEL  of the input
model. They are needed only for a NLTE input model, but
can be present for an LTE model as well, although they will
be immediately recomputed.
The number of input level populations does not have to be
equal to the current NLEVEL; in that case the keyword ICHANG should be 
coded as non-zero, and the appropriate initial populations of the current
levels are computed -- see below.
\end{description}

\medskip
\noindent $\bullet$ If INTRPL $>$ 0. This signifies that the current depth 
scale is going to be different from that for the inout model (DEPTH -- see above). 
If so, there is an additional input from unit 8, directly after the input of
model parameters, which sets the
new depth scale DM (i.e. the column mass in g cm$^{-2}$) which 
will be used in the current run. All the model quantities  are interpolated
from the initial to the new column mass grid. The interpolation is a log-log
interpolation; the numerical value of INTRPL sets the order of interpolation
(INTRPL=1 signifies a linear interpolation in logarithms, INTRPL=2 the
quadratic one, etc.).


\subsection{Change of the input level structure}
\label{inp_chan}

The change of the explicit level structure is controlled by parameter
ICHANG -- see \S~\ref{nonst_mod}. If this parameter is coded negative, then
an additional input, which specifies the correspondence of the
``old'' (i.e. the input model) level populations and the  `new'' ones
(i.e. those which are to be computed in the current run), is required.
Generally, this option is useful, for instance, for adding more
explicit levels to an already converged model, without the necessity to
start again from the scratch.

For each explicit level in the ``new'' level system, II=1,$\ldots,$NLEVEL, 
the following parameters are required:
\begin{description}
\item[IOLD]  -- basic correspondence indicator:\\
$\bullet\,>$ 0  -- means that population of this level is
                 contained in the set of input populations;
                 IOLD is then its index in the "old" (i.e. input)
                 numbering.
                 All the subsequent parameters have no meaning
                 in this case.\\
$\bullet\,= 0$  -- means that this level has no equivalent in the
                 set of "old" levels. An initial estimate of the 
                 population of this level has thus to be computed,
                 following the specifications based on the following 
                 parameters:
\item[MODE]   --    indicates how the population is evaluated:\\
$\bullet\,= 0$  -- population is equal to the population of the "old"
              level with index ISIOLD, multiplied by REL;\\
$\bullet\,= 1$  -- the level is assumed to be in LTE with respect to the
              first state of the next ionization degree whose
              population must be contained in the set of "old"
              (ie. input) populations, with index NXTOLD in the
              "old" numbering.
              The population determined of this way may further
              be multiplied by REL.\\
$\bullet\, = 2$  -- population is determined assuming that the b-factor
              (defined as the ratio between the NLTE and
              LTE population) is the same as the b-factor of
              the level ISINEW (in the present numbering). The
              level ISINEW must have the equivalent in the "old"
              set; its index in the "old" set is ISIOLD, and the
              index of the first state of the next ionization
              degree, in the "old" numbering, is NXTSIO.
              The population determined of this way may further
              be multiplied by REL.\\
$\bullet\,= 3$  -- a level corresponds to an ion or atom which was not
              explicit in the old system; population is assumed
              to be an LTE one.
\item[NXTOLD]  --  see above
\item[ISINEW]  --  see above
\item[ISIOLD]  --  see above
\item[NXTSIO]  --  see above
\item[REL]     -- a population multiplier -- see above\\
$\bullet\, = 0$  --  the program sets REL=1
\end{description}
In our experience, it is usually sufficient to use MODE=1 for the ``new'' levels.
Recall that even if this sets the LTE populations for such levels, they will be
immediately recalculated to yield an (approximate) NLTE population in the first formal solution even before entering the first linearization step.

Recall that when the keyword parameter ICHANG is set to 1 
(see \S\,\ref{nonst_mod}), one specifies the so-called ``simplified change'',
which can be used only if (a) one leaves the structure of the input energy
levels unchanged, while (b) one only adds levels of new species, which are
now indexed with I=NLEVEL0+1,$\ldots$,NLEVEL, where NLEVEL0 is
the number of levels in the input model. Their initial level populations 
are set to the LTE values. In this case, no other additional input parameters 
are needed.


\subsection{Kurucz input model atmosphere}
\label{inp_kur}

{\sc tlusty} also accepts Kurucz (1979, and later Kurucz CD-ROMs)
models as starting models. The format of
\index{Kurucz model atmospheres}
the file is the standard Kurucz output file. The Kurucz model is read
if the parameter INTRPL is set to $-1$.

One may still interpolate to a different depth grid than that used by
Kurucz. Since parameter INTRPL cannot be used for this purpose, the user
has to append  the end of the Kurucz model file by the parameter
INTRPL followed by the values of the column mass (in g cm$^{-2}$).

\noindent Note:
The first depth point of Kurucz models is usually incorrect, 
and is therefore skipped. If the user does not want to interpolate 
in depth, the keyword parameter ND should be set to 63 (i.e. {\tt ND=63}) 
should appear in the keyword parameter file (see \S~\ref{nonst_glob}).


\section{Output}
\label{out}

There are several output files. We divide them into two groups as listed
and described below.
By default, all the output files are generated as ASCII files for portability.
{\sc tlusty} does not contain any explicit {\tt OPEN} statements for the
output files, so the files are generated with names {\tt fort.}{\it nn},
where {\it nn} is the corresponding unit number.

\begin{enumerate}
\item Basic output, generated always 
 \begin{itemize}
  \item Unit 6 -- Standard output
  \item Unit 7 -- Condensed model atmosphere
  \item Unit 9 -- Convergence log
  \item Unit 10 -- Performance and error log
  \item Unit 11 -- Mean opacities for the resulting model
  \item Unit 12 -- $b$-factors (NLTE departure coefficients)              
  \item Unit 13 -- Emergent flux in all frequency points (spectral energy
  distribution)
  \item Unit 18 -- Convergence log for the formal solution
  \item Unit 69 -- Timing log        
 \end{itemize}
\item Auxiliary output, generated only if required (by setting the corresponding
keyword parameters -- see Sect. \ref{nonst_print})
 \begin{itemize}
  \item Unit 14 -- Emergent (angle-dependent) specific intensities 
   in all frequency points (at present it is implemented only in the case of
   Compton scattering)
  \item Unit 16 -- A check of the accuracy of the numerical solution of the kinetic 
  equilibrium equation
  \item Unit 17 -- Condensed model atmosphere in every iteration
  \item Unit 85 -- Opacities for all continuum frequencies
  \item Unit 86 -- Total and net cooling rates
  \item Unit 87 -- Net cooling rates separately for all ions
 \end{itemize}
\end{enumerate}

\subsection{Basic standard output}
\label{out_bas}

\subsubsection*{Unit 6: Standard output.}

This a general log of the model construction procedure, It contains
\index{Output!standard}
tables displaying the input data, some performance (and possibly error)
messages, and prints several tables of the output model. In most cases,
these tables are self-explanatory.

The amount of output on Unit 6 depends upon input parameters.
For instance, an important
portion of the output are tables containing various quantities
produced if convection is taken into account.

Note: Unit 6, if accidentally or intentionally deleted, may to
a large extent be re-created by a simple run of {\sc tlusty} using unit 7
output as an input (unit 8), and with the same unit 5; one only needs
to specify NITER=0 in the keyword parameter file.

\subsubsection*{Unit 7: Condensed model atmosphere.}

This is the basic output in a machine-oriented form, i.e. without any
\index{Output!condensed model atmosphere}
table headers, etc. 
This file may serve as the input model file for another run of
{\sc tlusty} as Unit 8, or for {\sc synspec} and various interface and utility programs.
Its structure is described in detail in \S\,\ref{inp_tlus}. 

The file is generated after each set of formal solutions before entering
a next iteration of the complete linearization, and is always overwritten
so that only the quantities at the last completed iteration are stored.

\subsubsection*{Unit 12: $b$-factors.}

The file is exactly analogous to the Unit 7 output file; only instead 
\index{Output!$b$-factors}
of populations the file contains the $b$-factors (NLTE departure coefficients).

\index{Convergence log}
\index{Output!convergence log}
\subsubsection*{Unit 9: Convergence log.}
This is a very important output file, and the user is strongly
encouraged to inspect it carefully after each run. As mentioned in
\S\,\ref{examp_gen} ,there is also
an IDL routine {\tt pconv} for plotting its contents. 
The individual columns have the following meaning:
\begin{enumerate}
\item iteration number
\item depth index ($d$)
\item relative change in $T$, i.e. $(T_d^{\rm new}-T_d^{\rm old})/T_d^{\rm new}$
\item relative change in $n_{\rm e}$, (defined analogously)
\item maximum relative change of all explicit level populations
\item maximum relative change of linearized mean intensities
\item maximum relative changes of all component of the state vector
\item index of the level population that exhibits the maximum relative change
\item index of the frequency point that exhibits the maximum relative change
\end{enumerate}
Notice that if the Rybicki scheme is used, all the relative changes except that
of temperature are equal to zero.

\subsubsection*{Unit 10: Performance and error log.}
This file, created by several different subroutines, contains
\index{Output!performance and error log}
messages about performance (log of accelerations, recalculation of
the radiative equilibrium equation division optical depths), 
and all messages printed by the program when an error status 
occurs. These errors are either fatal (occurred when checking actual dimensions
against maximum allowed dimensions; calling various routines with inconsistent parameters;
divergence of complete linearization); or warnings
(slow convergence of subroutine {\tt ELCOR} - a solution of the non-linear system
of statistical equilibrium + charge conservation; negative opacities,
etc.). All messages are more or less self-explanatory.

\subsubsection*{Unit 11: Mean opacities for the resulting model.}
This file is generated by subroutine {\tt ROSSTD} in the final iteration. 
It is composed of ND records, each containing: 
\begin{enumerate}
\item depth index 
\item column mass $m$ [g cm${}^{-2}$]
\item Rosseland optical depth 
\item Rosseland opacity per unit mass [cm${}^{2}$ g ${}^{-1}$]
\item temperature $T$ [K] 
\item electron density $n_{\rm e}$ [cm${}^{-3}$] 
\item mass density $\rho$ [g cm${}^{-3}$] 
\item numerically integrated Planck function, $\sum_{i} w_i B(\nu_i,T)$,
\item analytically integrated Planck function ($\pi\sigma_R T^4$),
\item Planck mean opacity [cm${}^{2}$ g ${}^{-1}$]
\end{enumerate}
This table, which shows the basic structural
parameters ($T$, $n_{\rm e}$, $\rho$) as a function of the Rosseland
optical depth is useful for comparisons with other models; a comparison of the
two integrated Planck functions to check the accuracy of frequency coverage and
integrations.

\subsubsection*{Unit 13 -- Emergent flux.}
Generated by subroutine {\tt OUTPRI} in the final iteration. It prints the
\index{Output!emergent flux}
following quantities for all the frequency points, sorted by wavelength:
\begin{enumerate}
\item  frequency [s$^{-1}$]
\item  emergent flux, precisely the first moment of the specific 
intensity of radiation, $H_\nu$, at the surface,
[erg cm$^{-2}$ s$^{-1}$Hz$^{-1}$]
\index{Eddington factor}
\item   the surface Eddington factor $f_H$, where 
$f_H= H_\nu(0)/J_\nu(0)$, $J_\nu(0)$ 
being the mean intensity of radiation at the surface.
\end{enumerate}                
Note: In versions prior to 200, additional quantities were printed
out in this file.

\subsubsection*{Unit 18 -- Convergence log of the formal solution.}
Generated by subroutine {\tt RESOLV}.
The individual columns are: 
\begin{enumerate}
\item iteration number of the linearization procedure
\item iteration number of the formal  solution 
\item depth index 
\item maximum relative change of all populations 
\item index of level for which the maximum relative
change occurs.
\end{enumerate}
This file is seldom used. It may be useful for diagnosing problems
that may occur in the formal solution.

\subsubsection*{Unit 69: Timing.}
As pointed out above, this file is produced only under Unix or Linux. 
\index{Timing log}
\index{Output!timing log}
The individual columns represent:
\begin{enumerate}
\item{iteration number}
\item a label for distinguishing the two parts of the given iteration step: 
1 for the global formal solution part (subroutine {\tt RESOLV}),
and 2 for the linearization part 1 (subroutine {\tt SOLVE});
\item total elapsed time up to the given step [s]
\item time spent in the given step [s]
\item a descriptive label corresponding to the second column.
\end{enumerate}
%


\subsection{Auxiliary output}
\label{out_aux}

\subsubsection*{Unit 14 -- Angle-dependent emergent specific intensities.}
\index{Output!specific intensities}
This file is generated only in the case of Comptonization, i.e. if ICOMPT
$>$ 0. 

\index{Output!check of kinetic equilibrium}
\index{ICHCKP keyword parameter}
\subsubsection*{Unit 16 -- Check of the kinetic equilibrium equations.}

This output is generated only if the keyword parameter ICHCKP is
set to a non-zero value.

For each explicit level, it prints the total rate in and out, and their 
difference divided by the rate in, for each depth. The last column should
thus contain much lower values than the previous two columns.
Generated by subroutine {\tt CHCKSE} in the final iteration.

\index{IPRIND keyword parameter}
\subsubsection*{Units 17 and 20 -- condensed models in each iteration.}

This output is generated only if the keyword parameter IPRIND is
set to a non-zero value.

These files are completely analogous to the Unit 7 and 12 outputs,
respectively; the only difference being that the values for
all completed iterations are stored. These files can be used to
recover partially converged models in case the final iteration crashed
or diverged due to numerical reasons (e.g., an improper use of
acceleration techniques, etc.).

\index{IPOPAC keyword parameter}
\subsubsection*{Units 85 and 86 - opacities and other quantities for the continuum frequencies.}
This output is generated only if the keyword parameter IPOPAC is
set to a non-zero value.\\
$\bullet\ = 1$ --
The file contains NFREQC (the number of frequency points in the continuum)
blocks; each block contains data for one frequency point. The block
contains first the frequency index and the value of frequency (Hz),
then ND values of opacity (per gram) for all depth points, ID$=1,\ldots,$ND.
Frequencies go from the highest to the lowest.\\
$\bullet\ = 2$ -- a more extended table with the frequency index and the value
of frequency, followed by ND records for each depth that contains 5 numbers: 
the monochromatic optical depth, the true
absorption coefficient, scattering coefficient, emission coefficient, and the
mean intensity of radiation.

\subsubsection*{Units 87 and 88 - cooling rates.}
\index{Cooling rates}
\index{Output!cooling rates}
\index{ICOOLP keyword parameter}

This output is generated only if the keyword parameter ICOOLP 
is set to a non-zero value.

Unit 87 contains four columns; the depth index, the total net cooling rate
without an influence of scattering ($C_{\rm net}$), 
the total net cooling rate including scattering ($C_{\rm net}^\prime$), 
and the total cooling rate (i.e., the true radiative cooling
without a balancing effect of radiation heating, $C_{\rm tot}$). 
The two net cooling rates are identical in the case of 
coherent scattering (e.g. Thompson limit of electron scattering);
but are generally different for non-coherent scattering (Compton).
$$C_{\rm net} = 4\pi \int_0^\infty (\eta_\nu - 
\kappa_\nu J_\nu) d\nu\, ,$$ 
$$C_{\rm net}^\prime = 4\pi \int_0^\infty (\eta_\nu + \sigma_\nu J_\nu - 
\chi_\nu J_\nu) d\nu\, ,$$ 
$$C_{\rm tot} = 4\pi \int_0^\infty \eta_\nu  d\nu\, ,$$ 
where $\eta_\nu$ is the thermal emission coefficient; $\kappa_\nu$ is
the (thermal) absorption coefficient, and $\chi_\nu$ is the total extinction
(absorption + scattering) coefficient, all at frequency $\nu$.

 If IPRINP is set to a value larger than 10, an additional output
is generated -- Unit 88 -- that contains the cooling rates 
separately for all explicit ions separately.


\section{Basic troubleshooting}
\label{trouble}

Here we describe the most common errors when running {\sc tlusty}, and the 
ways to fix them. Advanced troubleshooting, and a summary of useful
numerical and physical tricks, will be covered later in \S\,\ref{tricks} in Part II.

\subsection{Program stops immediately}
\label{trouble_immed}

On several occasions, the program stops either immediately, or
after a few seconds. There are several possible errors:
\begin{itemize}
\item No standard output created, and the message 
\begin{verbatim}
   Wrong architecture 
\end{verbatim}
or 
\begin{verbatim}
   dyld: Library not loaded: /usr/local/lib/libg2c.0.dylib
   Referenced from: ...  
   Reason: image not found
   Trace/BPT trap
\end{verbatim}
is generated. This indicated that
the code was compiled on a different machine and the executable was
accidentally copied. The fix: recompile the code on the current machine.
\item Program stops with the message: 
\begin{verbatim}
    Bus error
\end{verbatim}
Most likely, the core memory is not
sufficient to run the current executable. \\
The fix: run the {\sc pretlus} program
with the standard input for the current model, and update the {\tt BASICS.FOR}
and {\tt ODFPAR.FOR} files accordingly, and recompile {\sc tlusty}. If the
problem persists, the core memory of the computer is not sufficient
to run the code with the current input data. One has to modify the input
data to make the model less demanding on the computer memory, for instance, by 
decreasing the number of explicit ions, levels, frequencies, depths, etc.
(Obviously, a better option is to use a computer with a larger memory.)
\item Program stop with a message 
\begin{verbatim}
    Segmentation fault
\end{verbatim}
Most likely, the code was not compiled with the specification
{\tt -fno-automatic}.\\
The fix: recompile the code with {\tt -fno-automatic}
(or {\tt -static} for the generic Unix), and run again.
\item The program runs for a short time, and generates some content to 
the standard output
file (the header, plus possibly a few tables), and then stops. In this case 
inspect the standard output, in particular the
end of it. The last line will most likely contain a message that a dimension of 
some array is larger than the maximum available dimension.
For instance, the last line of the standard input may read
\begin{verbatim}
    ilev.gt.mlevel         164       134
\end{verbatim}
which means that the level index {\tt ilev} (=164) surpassed the
maximum number of explicit levels (MLEVEL=134), set up in the ``INCLUDE''
file BASICS.FOR (see \S\,\ref{reduce}).

The fix: run the {\sc pretlus} program with the same input data as the current
mode, and check whether all the dimension parameters in BASICS.FOR and
ODFPAR.FOR are larger than or equal to the values specified by the 
{\sc pretlus} output. It is much
preferable to run {\sc pretlus}, because the above message does not mean
that it is sufficient to reset MLEVEL to, say 165, because {\sc tlusty} 
stopped in the first occurrence of {\tt ilev} $>$ MLEVEL, but had it continued 
it could have reached much
higher values of {\tt ilev}. In such a case, the user would have to recompile and
rerun the code many times.
\item Program may also stop because some file, whose need is stipulated by
the input data, is missing. A message is issued to this effect. The obvious fix is
to provide the needed file, or, if the file is not available, to run the code with the
input parameters that do not require such a file.
\end{itemize}


\subsection{Program runs, but does not converge}
\label{trouble_conv}

This is of course the most common problem, and often the most difficult to
solve. There is no general recipe for fixing it. 
However, it is our belief that for any model, regardless how complex it may be, 
there is essentially always a way to converge it within the means provided
by {\sc tlusty}, but sometimes it is very difficult to find the proper way. 
Over the time, we have developed many strategies and procedures
to help with convergence problems, but while one strategy may work perfectly
in some cases, it could rather cause problems in others. Therefore, finding a
good way to converge a model is sometimes a true art.

Here we will provide a guidance about possible strategies that are
worth trying in difficult cases. We will consider separately several classes
of models, starting with some general strategies applicable for all models.

\subsubsection{General strategies}
\label{trouble_conv_gen}

The most important part of the diagnosis of the problem is to
inspect the convergence log, output file {\tt fort.9}, preferably by plotting
its contents using our IDL routine {\tt pconv}, or, if IDL is not available,
to write an analogous routine in a graphical package that is available.
Sometimes it is also worthwhile to inspect the standard output, in particular
for convective models.

Here we list several general symptoms, and corresponding strategies to
fix the problem.
\begin{itemize}
\item If the relative changes, in particular the maximum change over all
depths (the rightmost panels in Figs, 1, 4, 7, 8) exhibits a sudden increase
in the 7th iteration while the keyword IACC that specifies the Ng acceleration
is not set, thus assuming its default value IACC=7 (or if IACC is specifically set
differently, and an increase of the maximum relative change occurs at the
iteration number IACC), then there is a good probability that the culprit is the 
Ng acceleration.

The fix: increase IACC to start the Ng acceleration later, or disable it completely
(by setting IACC larger than NITER).

\item An analogous problems may happen if the Kantorovich acceleration
is switched on too early. Again, a symptom would be that the relative changes
exhibit strange increase at the iteration at which the Kantorovich scheme is
set (controlled by the keyword ITEK, with the default equal to 4). 
This scheme usually
does not cause divergences but rather only a slower convergence, but it may
cause problems if a model structure changes dramatically at iteration steps
after the ITEK-th iteration. For instance, the convection zone
may easily change position, so the Jacobi matrix evaluated in an earlier iteration
step would be inconsistent with the present atmospheric structure.

\item The convergence pattern exhibits a spike, or a sudden increase, at the
depth point which corresponds to TAUDIV, the point at which the coefficients
of the linear combination of the integral and differential form of the radiative
equilibrium equation suddenly change (called the division point). This depth is 
found in the standard output in the table entitled:
\begin{verbatim}
  id        redif          reint
\end{verbatim}
where {\tt id} represent the depth index, {\tt reint} and {\tt redif} are the parameters
$\alpha$ and $\beta$ in Eq. (16) of Paper~II, respectively. The division point is the
depth where {\tt redif} suddenly changes value from 0 to 1.

A possible fix: try to change the value of keyword TAUDIV, or, perhaps much
better, to set the basic mode of calculation to the Rybicki scheme (IFRYB=1).
There are some other possibilities; we will mention some of them below
when dealing with specific types of models.

\end{itemize} 

\subsubsection{LTE models without convection}
\label{trouble_conv_lte}

Non-convective LTE models are usually the easiest to converge, but even in
this case one may sometimes face problems. Here are several typical situations:
\begin{itemize}
\item If one faces generic convergence problems or a divergence not related
to the cases described above, the first possible fix is again
to set the basic mode of model calculation to the Rybicki scheme. In most
cases it helps significantly.
\item If there are problems even with the Rybicki scheme being set, and in particular
if the relative changes are large in the upper layers, the problems may be 
connected to spectral lines. In that case, try first to converge a model without
any lines (as is done, for instance, in the first H-He model in 
\S\,\ref{examp_hhe}), and then
to add lines. If there are still problems, one may try to converge first a model
with fewer lines and consecutively adding more, lines, as explained in 
\S\,\ref{strateg_lte}.
\item If the convergence pattern exhibits large changes in deep layers 
(possibly oscillating and increasing in absolute value), and the Rybicki 
scheme does not help, then the problem 
may be that  the model is in fact convective, but the convection was not set up
(HMIX0 $\leq 0$). The obvious fix is to set up the convection, HMIX0 $> 0$.
\end{itemize}

\subsubsection{LTE convective models}
\label{trouble_conv_ltec}

The first action one needs to take in the case of convergence problems with
LTE convective model is to rerun the model with IPRINT=3, in which case
the standard output will contain the table of interesting model quantities
that enter the evaluation of the convective flux, together with the corrected
temperature in the case of one or both of the correction procedures described in
Paper~II, Appendix B2 and here in \S\,\ref{nonst_conv} are applied. One can use
the test cases considered in \S\,\ref{examp_optab} and \S\,\ref{examp_cwd} as
a template, but if that does not work, one has to experiment with changing
the individual parameters to see whether that helps. The hints how to set
up some parameters (e.g., NDCGAP, IDEEPC) are given in 
\S\,\ref{nonst_conv}; the parameters ICONRE, ICONRS, and IMUCON
that set the first and the last iteration step at which the correction procedures
are performed have to be found by trial and error. Unfortunately,
we do not know of any general recipe that would be applicable in
all cases.

Interestingly, the correction procedures work best for a strong convection
(where the dominant part of the flux is transported by convection), in
which case the temperature gradient is close to the adiabatic gradient.
Convection is more difficult to treat, and the correction procedures may not
be helping, if the convection is weak but still present. But in those
cases the Rybicki scheme, if not already set, may help significantly.

\subsubsection{NLTE models}
\label{trouble_conv_nlte}

Again, there is no general recipe to fix the convergence problems
for NLTE models. All the strategies described in \S\,\ref{trouble_conv_gen}
and \ref{trouble_conv_lte} are of course applicable here as well. Some
additional possibilities are described below:
\begin{itemize}
\item One may try to increase NLAMBD, the number of global formal
solutions between two consecutive iterations of the linearization scheme,
setting for instance NLAMBD=4.
\item If the strategy of converging first a model without lines and then adding
consecutively more and more lines does not work, one can resort to
set IFIXMO=1, in which case the temperature, electron density, and mass
density remain fixed to the values of the input model, and only the level populations
and the radiation intensities are computed (the so-called restricted NLTE problem).
This model, when converged, can be used as the input model for the 
subsequent run where all the 
state parameters are being updated. If this still does not work, one may try 
to first converge a model with 
INRE=0 (that is, keeping the temperature fixed and updating everything else),
and only after that to continue to the full model.
\item As demonstrated in \S\,\ref{examp_cwd}, it is sometimes helpful to apply
the Rybicki scheme even for NLTE models. If it converges very slowly, and
can still use the resulting model as a new starting model for the full NLTE
solution using the default hybrid CL/ALI method.
\end{itemize}
As an example, we take a NLTE/L H-He model considered in \S\,\ref{examp_hhe}.
First we leave the standard input data {\tt hhe35nl.5} unchanged, but instead 
of starting with the  NLTE/C (model {\tt hhe35nc.7}), we take as a starting model
the LTE one, {\tt hhe35lt.7}. Although we know that it is better to start with the 
NLTE/C model, we take this case as an example of finding the way to converge 
a difficult model, or, in other words, to converge a model with inappropriately 
chosen initial model.

%
\begin{figure}
\begin{center}
\label{fig10}
\includegraphics[width=4in]{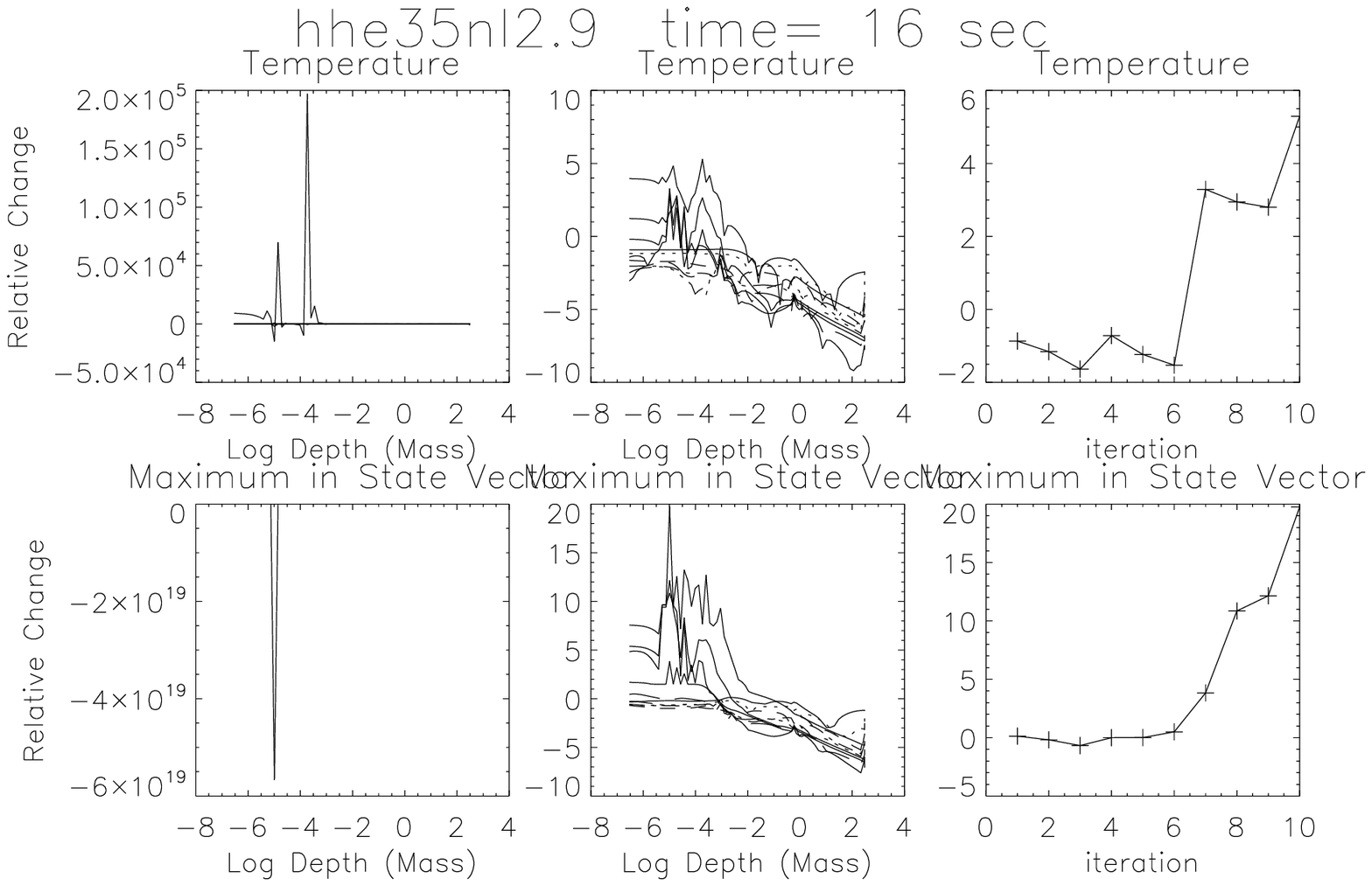}
\caption{Convergence log for the {\tt hhe35nl2} model.}
\end{center}
\vspace{-1em}
\end{figure}
%
\begin{figure}
\begin{center}
\label{fig11}
\includegraphics[width=4in]{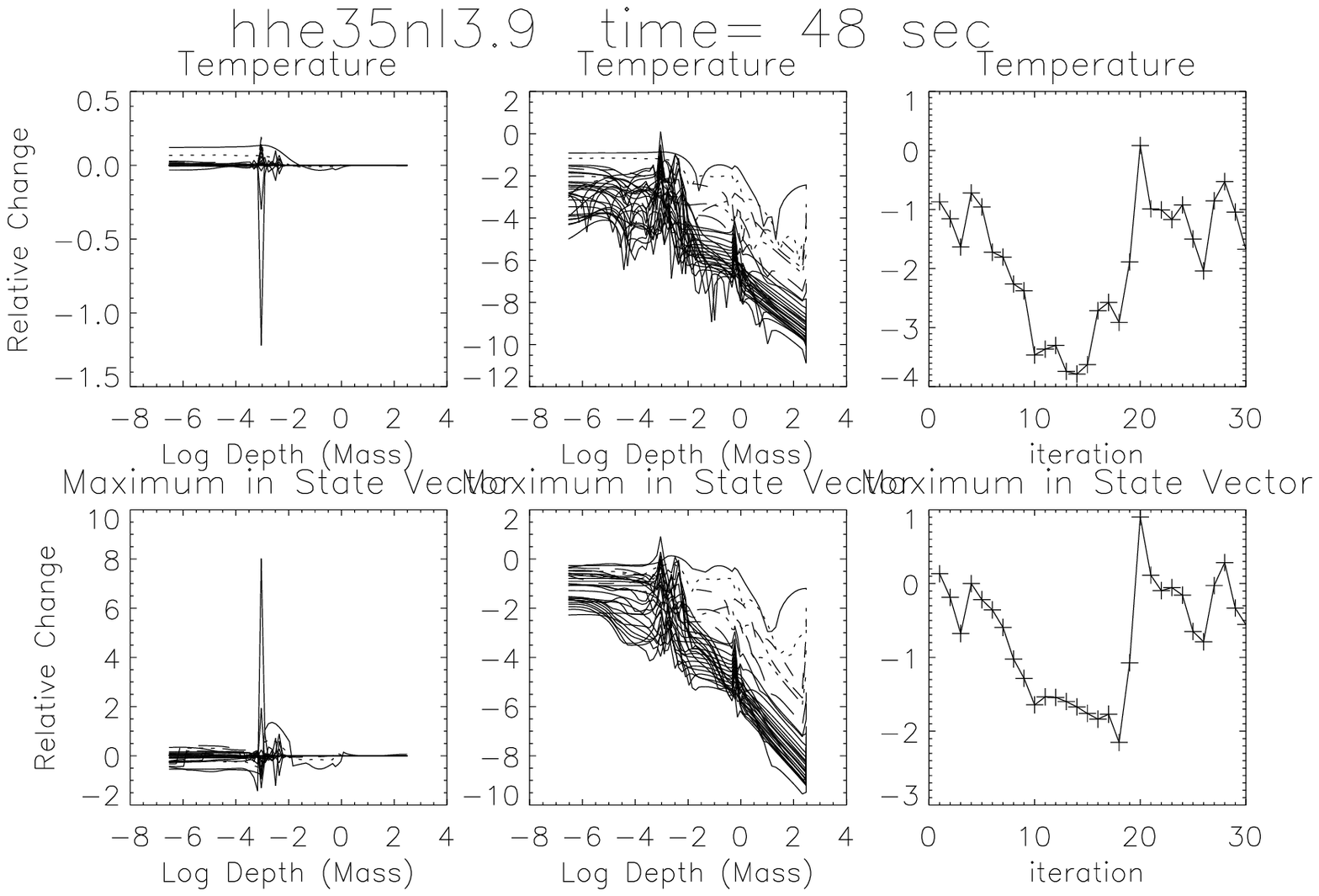}
\caption{Convergence log for the {\tt hhe35nl3} model.}
\end{center}
\vspace{-1em}
\end{figure}
%
\begin{figure}
\begin{center}
\label{fig12}
\includegraphics[width=4in]{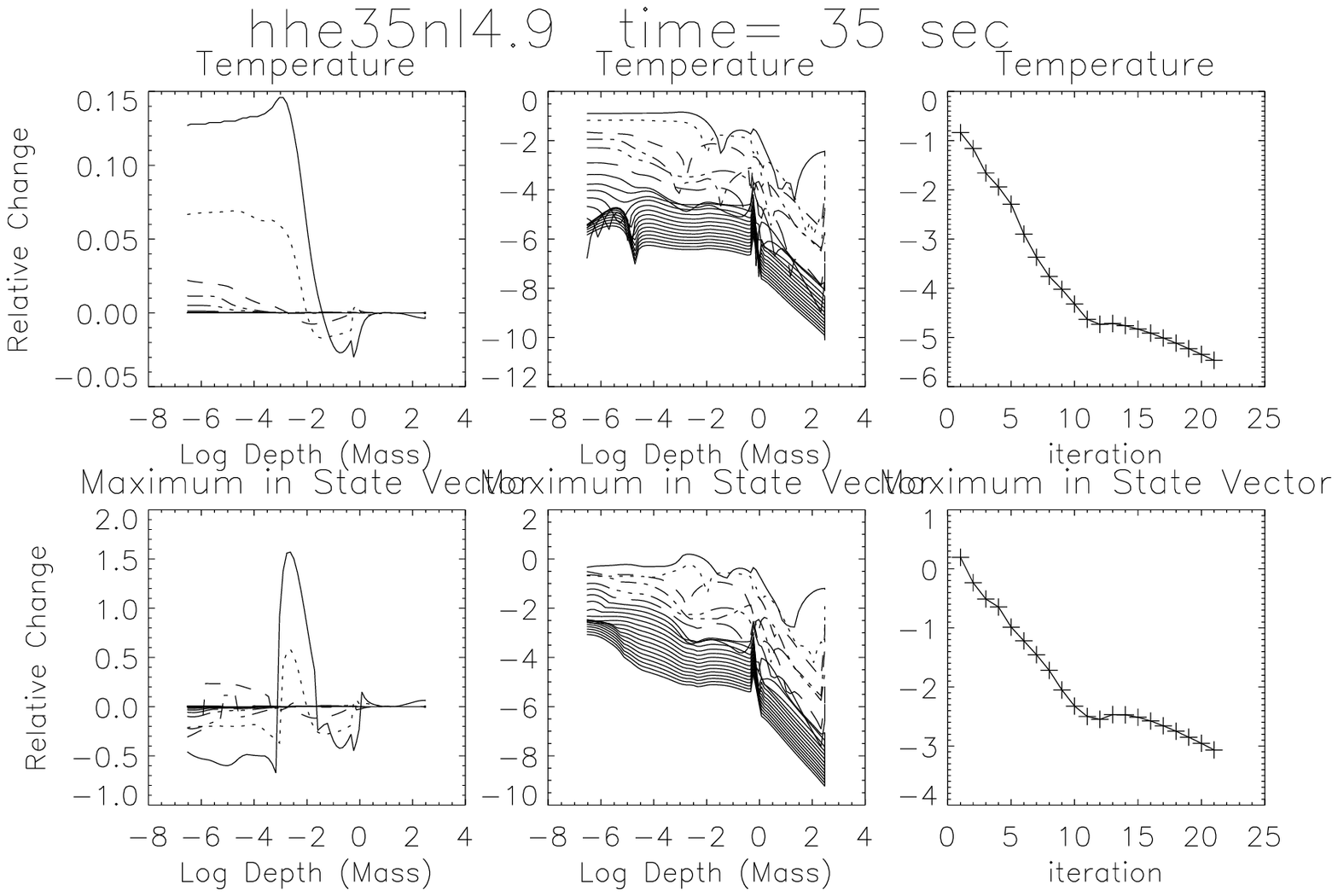}
\caption{Convergence log for the {\tt hhe35nl4} model.}
\end{center}
\vspace{1em}
\end{figure}
 
The model run is submitted as
\begin{verbatim}
   cp hhe35nl.5 hhe35nl2.5
   RTlusty hhe35nl2 hhe35lt
\end{verbatim}
so that the new model files have a core name {\tt hhe35nl2}.
The convergence pattern is displayed in Fig. 9.
It is seen that the maximum relative change exhibits the first large jump at the 7th
iteration, which, as explained in \S\,\ref{trouble_conv_gen}, points to the 
Ng acceleration as the likely source of problem. One can therefore try to 
disable the Ng acceleration, and to be on the safe side, also the 
Kantorovich acceleration. One then creates a standard input file
{\tt hhe35nl3.5}, which is identical to {\tt hhe35nl2.5}, except the 3rd line
which now specifies the name of the keyword parameter file, {\tt 'f3'}.
This file contains just one record:
\smallskip

{\tt IACC=40,ITEK=40}\\ [2pt]
which effectively disables both accelerations. The convergence pattern
of this model is shown in Fig. 10.
The convergence behavior is clearly improved, but the model still cannot be
viewed as converged. As suggested above, one may try to increase the
number of iterations of the global formal solutions. This model, called {\tt hhe35nl4},
has the keyword parameter file, named {\tt 'f4'} which is:
\smallskip

{\tt IACC=40,ITEK=40,NLAMBD=4} \\ [2pt]
The convergence pattern of this model is shown in Fig. 11.
Indeed, the convergence is now much better. Comparing this model
and the original {\tt hhe35nl} model considered in \S\,\ref{examp_hhe}
reveals that both models are essentially identical, as one should expect
from two converged models computed with the same physical parameters.

There are several more advanced procedures, often applicable to
special types of objects. They will be discussed in Chap.\,\ref{tricks}.


\newpage
\addcontentsline{toc}{section}{PART II}

\begin{center} 
        {\Huge \bf Part II}\\ [35mm]
\end{center}

\newpage


In the second part of this operational manual, we describe the topics that
are not absolutely necessary for a casual user of {\sc tlusty}, in a sense
that one can construct basic model atmospheres and accretion disks
using information from Part I only. However, the material presented in
Part II is quite important for a dedicated user. It provides a more detailed 
understanding of the employed numerical procedures, the treatment of input
atomic data, the calculation of specific models, as well as the description of 
additional flags that help cope with convergence problems.


\section{Input files for individual model atoms}
\label{ions}

Each explicit ion has an individual file which
contains information about three or four basic types 
of atomic data (the fourth type is only required in the Opacity Sampling mode), 
namely:

\begin{enumerate}
\item Energy levels (i.e. level energies, statistical weights, etc.).
\item Bound-free transitions (modes of evaluation of the photoionization
cross section, collisional ionization rates, etc.).
\item Bound-bound transitions (necessary data for lines, such as the oscillator
strengths, line broadening parameters, a mode of evaluation of the
collisionakl excitation rates, etc.).
\item Energy bands for setting up superlevels of iron-peak elements
(only required in the Opacity Sampling mode)
\end{enumerate}

The overall system of input enables the user to set up a library
of more or less universal data sets for all the astrophysically important ions,
and to select a desired degree of sophistication of a model atmosphere easily 
by pointing to those filenames in the standard input (unit 5). 
In fact, such a library is available through the {\sc tlusty} 
website.\footnote{{\tt http://tlusty.oca.eu}}

Moreover, there is an IDL--based program called MODION, written by F. Varosi
\index{MODION program}
(NASA/GSFC), which is designed to construct the individual atomic data 
files directly from the Opacity Project (OP) database TOPbase. The program displays
the Grotrian diagram of a selected ion and the user selects explicit levels,
and build superlevels, simply by a mouse. Program MODION then builds
an array of bound-bound and bound-free transitions. For the latter, the
user may form an approximate photoionization cross section graphically
from the detailed OP cross sections. 

{\bf Important note}: Unlike the standard input (Unit 5), the atomic data files
{\em must not} contain ``comment lines'' beginning with {\tt *} or {\tt !}.
Instead, there is one mandatory record beginning with {\tt *} immediately
preceding each block, i.e the structure of the file looks for instance like:
\begin{verbatim}
****** Levels
\end{verbatim}
followed by data for energy levels, without any comment line,
\begin{verbatim}
****** Continuum transitions
\end{verbatim}
followed by data for bound--free transitions, without any comment line,
\begin{verbatim}
****** Line transitions
\end{verbatim}
followed  by data for bound--bound transitions, 
without any comment line.\\
This is followed, if applicable, by
\begin{verbatim}
****** Energy bands
\end{verbatim}
followed by limits for energy bands for superlevels, without any 
comment line.

We will now describe the basic blocks of the atomic data file
in detail.


\subsection{Energy level parameters}
\label{ion_en}

By the term ``level" we mean here either a genuine atomic energy
level, or any reasonably defined group of energy levels, e.g. a
superlevel. As mentioned in Paper~II, \S\,\refsuperl, there are two types of superlevels:
\index{Superlevels}

(i) a genuine superlevel, which is a pre-defined 
group of levels; all the input parameters
have to be specified (used for instance for the iron-peak elements);  

(ii) a merged level, which is a superlevel composed of all merged Rydberg,
partially dissolved, states of an ion. The level parameters, such as its
statistical weight and a mean energy are considered as depth-dependent,
due to a depth-dependence of the occupation probabilities that
are computed by the program.

Each energy level has one input record containing the following
parameters:

\begin{description}

\item[ENION]  -- ionization energy of the level (with respect to the
\index{Input!ionization energies form individual levels}
             ground level of the next ionization state)
             may be given either in erg, eV, cm$^{-1}$,
             or as frequency (s$^{-1}$).
             The atomic data files stored in the {\sc tlusty} website express the
             ionization energy as frequency; this has an advantage that the user
             can immediately recognize which continuum jump in the predicted
             emergent spectrum corresponds to which ion.\\
$\bullet$ if = 0 --  the program assigns the
             hydrogenic value,  $E=Z^2 E_H/n^2$, where $Z$ is the charge
             of the next ion, $E_H$ the ionization energy of hydrogen ($E_H=3.28805\times
             10^{15}$ in the frequency units), and $n$ is the principal quantum number.
             Here it is assumed that the
             principal quantum number is the order number of level
             within the corresponding ion.
\index{Input!statistical weight}
\item[G]     -- statistical weight;\\ 
$\bullet$ if = 0 --  the program assigns the
             hydrogenic statistical weight, $g=2n^2$.
\index{Input!principal quantum number}
\item[NQUANT]  - principal quantum number.\\ 
$\bullet\,=0$  -- the program assigns for NQUANT the serial number of the level;\\
$\bullet\,< 0$ -- indicates that the given level is kept in LTE
                 (even if the model is NLTE);
                 NQUANT is then set to abs(NQUANT)
\item[TYPLEV] -- character{\tt *}10 string -- a spectroscopic identification
     of the level. It 
             appears in outputs, but may also be used as an
             identifier in the case of using the pre-tabulated 
             photoionization cross section data.
\item[IFWOP]  -- a mode of the treatment of level dissolution
\index{Input!setting of occupation probability treatment}
\index{Occupation probability}
(see Paper~II, \S\,\refdissol):\\
$\bullet\,= 0$ -- occupation probability set to 1, i.e no level dissolution;\\
$\bullet\,= 1$ -- occupation probability of the level 
             is calculated, and is used consistently in the rate equations
             and in evaluating the opacities and emissivities;\\
$\bullet\,= 2$ -- depth-dependent statistical weight for the iron-peak superlevels is
calculated using a generalized occupation probability 
(see Hubeny \& Lanz 1995).\\ [2pt]
The occupation probabilities 
are calculated in the hydrogenic approximation, i.e.,
they may be used for any atom/ion, but for non-hydrogenic ions they are
of a limited accuracy.\\
$\bullet\,< 0$ -- signals that the level is a merged level
                 (i.e. the Rydberg states lumped together).
\item[FRODF]   -- dummy (a former option, now obsolete), kept for downward 
compatibility of the input data.
\item[IMODL]  -- mode of treating the linearization of the level population
\index{Input!setting of level group}
\index{Level groups}
There were several options, which all are more or less outdated.
The only practical use of parameter IMODL is to set up the {\em level groups}.\\
$\bullet\,= 0$ -- the level is linearized individually (i.e., it forms its own group);\\
$\bullet\,< -100$ -- the given level is a member of {\em level group}
-- see Paper~II, \S\,\refgroup. A group
is composed by all levels with the same value of IMODL.
For instance, all levels with IMODL=$-101$ will from one group;
the levels with IMODL=$-102$ another group, etc. It is usually advisable
to form the level groups by levels with the same multiplicity
and parity. The actual value of parameter IMODL is inconsequential.

\end{description}


\subsection{Parameters for bound--free transitions}
\label{ion_bf}

The structure is as follows. Each transition 
which is taken into account is specified by one standard record.
If the transition is assumed to be in detailed radiative balance,
there are no other records. Otherwise, there is one or more
additional records for each transition, depending on the actual values of 
some control parameters.

\subsubsection*{The standard record of input parameters for continuum 
\index{Input!data for continuum transitions}
transitions}

\begin{description}
\item[II]      --  relative index of the lower level (i.e. II=1
corresponds to the ground state, etc.)
\item[JJ]      --  relative index of the upper level
\item[MODE]    -- mode of treating treating radiative rates in the
          transition:\\
$\bullet\,=  0$  --  detailed radiative balance (i.e. radiative rates
                 are not evaluated; but collisional rates are);\\
$\bullet\,>  0$  --  primarily linearized transition;\\
$\bullet\,<  0$  --  primarily ALI transition;\\
    Note: the distinction "primarily linearized" or "primarily ALI"
    does not have any meaning for continua;\\
$\bullet\,$ abs(MODE) = 5 or 15  --  signals that the given continuum is
   supplemented by a pseudo-continuum. Pseudocontinuum is the dissolved
   part of a corresponding spectral series converging
                 to the given edge -- see Paper~II, \S\,\refdissol.
                 In this case, there is one additional record
                 immediately following the present one, which
                 specifies the minimum frequency to which the
                 pseudo-continuum is considered (the so-called cutoff frequency).
\item[IBF$\equiv$IFANCY]  --  a mode of evaluation of the
\index{Input!bound-free cross sections}
   photoionization cross sections. All cross sections are in cm${}^{-2}$. 
\begin{itemize}
 \item for {\tt IBF}  = 0  -- simplified hydrogenic (Gaunt factor set to 1)\\
  $\sigma(\nu)= 2.815 \times 10^{-29} Z^4 \nu^{-3} n^{-5}$     
 \item for {\tt IBF}  = 1 -- exact hydrogenic,\\
  $\sigma(\nu)= 2.815 \times 10^{-29} Z^4 \nu^{-3} n^{-5} g^{\rm bf}(n,\nu/Z^2)$
 \item for {\tt IBF}  = 2 -- Peach (1967) formula (obsolete)\\
  $\sigma(\nu)= \alpha [\beta x^s + (1-\beta)x^{s+1}] \times 10^{-18}$
 \item for {\tt IBF}  = 3  -- modified Peach formula after Henry (1970) -- obsolete\\
  $\sigma(\nu)= \alpha [\Gamma x^s + (\beta - 2 \Gamma)x^{s+1}
               +(1+\Gamma-\beta)x^{s+2}] \times 10^{-18}$
 \item for {\tt IBF}  = 4 -- Butler (1990) fit formula of the form,\\
  $\sigma(\nu)= \exp [s+ \alpha \log x +\beta  \log^2 x ] $  
 \end{itemize}
\noindent where $x=\nu_0/\nu$, $s$={\tt S0}, $\alpha$={\tt ALF}, 
$\beta$={\tt BET}, and $\Gamma$={\tt GAM}. For the appropriate options
they are given in the immediately following record.
Further, $n$ is the principal quantum number, and $Z$ the
effective charge (=1 for neutrals, 2 for once ionized, etc.), and $\nu_0$
the threshold frequency.
 \begin{itemize}
  \item for {\tt IBF}  = 5 -- cross section from Verner \& Yakovlev (1995) tables
  \item for {\tt IBF}  = 7 -- hydrogenic cross section with the Gaunt
                       factors from Klaus Werner (priv. comm.)
  \item for {\tt IBF}  = 9 -- Opacity Project data stored in a special file, 
  named RBF.DAT (obsolete option)
  \item for {\tt IBF}  = 11 -- Opacity Project cross section for He I singlet levels  with
                                Fernley et al. (1987)  cubic fits
  \item for {\tt IBF}  = 13 -- analogous for He I triplet levels
  \item for {\tt IBF}  = 21 -- Cross section for ground state of He I from Koester (1985)
                     fit  
\end{itemize}
For the last three options ({\tt IBF} = 11, 13, 21), the data are hardwired in
the code, so no additional input is needed.
\begin{itemize}
  \item for {\tt IBF}  between 50 and 99 -- special cross section for
  a superlevel, pre-calculated. IBF then indicates the input unit number
\index{Superlevels}
  from which the data are read.
  \item for {\tt IBF}  $>$ 100 -- Opacity Project (OP) data, immediately following the given record. There are IBF$-$100 data points. This is the most
  common option for all the transitions for which OP data are available.
  \item for {\tt IBF}  $<$ 0  -- non-standard expression, given by a user
   supplied addition to the subroutine SPSIGK. This option is practically obsolete, 
   but for directions of its use, see \S\,\ref{spsigk}.
\end{itemize}
There is one special value of IBF, namely
\begin{itemize}
\item for {\tt IBF} = 15 -- which indicates that the rest of the input file
has a different form adopted for treating ions for which the inner-shell
ionization is taken into account, called ``X-file''. The structure of this
file is explained in detail in \S\,\ref{xfiles}.
\end{itemize}
\item[ICOL] --  mode of evaluation of the collisional rate
\index{Input!collisional ionization cross sections}
\begin{itemize}
 \item for {\tt ICOL}  = 0  -- standard formulae, different for different species:\\ 
 -- for H, He I, and He II expressions from Mihalas et al;. (1975), \\
 -- for other species by the so-called Seaton formula\\ [-13pt]
 \begin{center}
$\Omega = 1.55 \times 10^{13} T^{-1/2} U_0^{-1} \exp(-U_0) 
\sigma_0 \bar g,$\\ [-2pt]
\end{center}
    where $U_0 =h\nu_0/kT$, and
    $\bar g = 0.1$ for neutral atoms, $\bar g = 0.2$ for singly ionized
    atoms, and $\bar g = 0.3$ otherwise. $\sigma_0$ it the photoionzation cross
    section at the threshold.
    Quantity  $\bar g \sigma_0$ is given by the input parameter {\tt OSC0} -- see below,
 \item for {\tt ICOL}  = 1 -- so-called Allen formula:\\
   $\Omega = c_0 T^{-3/2}U _0^{-2} \exp(-U_0),$
 \item for {\tt ICOL}  = 2 :
   $\Omega = 5.465 \times 10^{-11} c_0 T^{1/2} \exp(-U_0),$
 \item for {\tt ICOL}  = 3 :
   $\Omega = 5.465 \times 10^{-11} c_0 T^{1/2} (1+U_0) \exp(-U_0)$\\
 where in the three above options $c_0$ is given by  {\tt OSC0}, 
\item for {\tt ICOL}  = 4 -- collisional rate evaluated by subroutine {\tt cion}
adapted from program XSTAR by T.~Kallman. This is used only in X-files.
\item for {\tt ICOL}  = 5 -- collisional rate evaluated by subroutine {\tt irc},
adapted from program XSTAR by T.~Kallman.  Also used only for X-files.
\item for {\tt ICOL} = 10 -- radiative charge transfer ionization with protons 
is added to the ordinary collisional ionization rate. It can be used only for the
ground states of neutral atoms. This represents an additional means of including
the charge transfer reaction, in this case selectively for a particular atom.
As described in \S\,\ref{nst2_colh}, one can also switch on this reaction for
all neutral atoms by setting the keyword parameter IFCHTR to a non-zero value.
\item for {\tt ICOL}  = 99 -- collisional rate is set to zero, $\Omega=0$. Again,
used only in X-files for Auger processes.
\item for {\tt ICOL} $<$ 0 -- non-standard expression, given by a user--supplied
                 procedure (CSPEC). Practically obsolete, but for its use see \S\,\ref{cspec}.
\end{itemize}

\item[IFRQ0, IFRQ1]   --  a means of setting some frequency points in the
continuum to the linearized mode in the case of automatic setting of frequency
points. Notice that in such a case the continuum frequency points are 
treated by default in the ALI mode. If IFRQ0 and IFRQ1 are non-zero, then all the
frequency points between the IFRQ0-th and IFRQ1-th point in the continuum are
set to linearization mode. Typically, one sets IFRQ0=1 and IFRQ1=4 for the
hydrogen Lyman continuum, and sometimes analogously for the He II Lyman
continuum.
 
\item[OSC0]  --  collision parameter entering the above expressions, i.e.
$\bar g \sigma_0$ in the case of Seaton formula (ICOL=0); or $c_0$ for 
ICOL=1, 2, 3.
\item[CPARAM]  --  dummy parameter for bound-free transitions.

\end{description}

\subsubsection*{Additional input parameters for continuum transitions}

In most cases, there is one or more additional input records for
continuum transition, which depend on the coded values of the basic
parameters listed in the first record, described above.

\medskip
\noindent %
(1) For MODE = 5 or 15, i.e. when the given continuum is
\index{Input!pseudocontinuum setting}
                 supplemented by a pseudo-continuum (i.e. the dissolved
                 part of a corresponding spectral series converging
                 to the given edge).
                 In this case, there is one additional record
                 immediately following the present one,
                 containing one number:
\begin{description}
\index{Pseudo-continuum cutoff}
\index{CYTLYM keyword parameter}
\index{CYTBAL keyword parameter}
\item[FR0PC] -- the minimum frequency to which the
                 pseudo-continuum is considered; i.e., the cutoff frequency for
                 the pseudo-continuum opacity (see Paper~II, \S\,\refdissol). In the
                 case of the Lyman and Balmer continuum of hydrogen, it may
                 be overwritten through the keyword parameters CUTLYM and
                 CUTBAL - see \S\,\ref{nst2_occup}.
\end{description}

\medskip
\noindent  
(2) For IFANCY = 2, 3, or 4, there is one additional record
\index{Input!bound-free cross sections}
                 containing 4 numbers:
\begin{description}
\item[S0, ALF, BET, GAM] -- parameters for the evaluation of the
      photoionization cross section in the Peach, Henry, and Butler form  --
      see above.
\end{description}

\medskip
\noindent  
(3) For IFANCY $>$ 100, there are two or more additional records, containing
the fit points for the Opacity Project photo-ionization data. The actual
value of IFANCY has the meaning that there are IFANCY$-$100 fit points.
The first (or more, if needed) record(s) then contains IFANCY$-$100 values
of XTOP, followed by the same number of records with values of CTOP,
where

\begin{description}
\item[XTOP] --  the value of $x$, $x = \log_{10}(\nu/\nu_0)$, of a fit point,
where $\nu_0$ is the edge frequency;
\item[CTOP] --  the corresponding value of the cross section, expressed as
${\rm CTOP} = \log_{10}(\sigma_\nu \times 10^{18})$ of a fit point.
\end{description}


\subsection{Parameters for bound-bound transitions}
\label{ion_bb}

The structure is analogous to that for the bound-free transitions, 
\S\,\ref{ion_bf}. 
Each transition 
which is taken into account is specified by one standard record.
If the transition is assumed to be in detailed radiative balance,
there are no other records. Otherwise, there are one or more
additional records for each transition, depending on the actual values of 
some control parameters.

\index{Input!bound-bound transition parameters}
\subsubsection*{The standard record of input parameters for line
transitions}

\begin{description}
\item[II]      --  index of the lower level
\item[JJ]      --  index of the upper level

\item[MODE]    -- mode of treating the radiative rates in the
          transition.\\
$\bullet\,=  0$  --  detailed radiative balance (ie. radiative rates
                 are set to zero but collisional rates are evaluated);\\
$\bullet\,>  0$  --  primarily linearized transition;\\
$\bullet\,<  0$  --  primarily ALI transition;\\
$\bullet\,= 1$ or $-1$ -- an ``ordinary'' line (i.e., any line represented by a symmetric
profile (Doppler, Voigt, or special profiles for hydrogenic lines - see below) \\
$\bullet\,= 2$ or $-2$ -- a ``merged'' line, corresponding to the transition from
     a normal to a ``merged'' level (i.e. all high
     members of a spectral series lumped together). In this case, 
     the corresponding Opacity Distribution Function is calculated by {\sc tlusty}. 
     This option is typically used for H and He II.\\
$\bullet\,= 3$  or $-3$ -- a superline, treated either, (i) in the Opacity Sampling mode
(a standard option), in which 
{\sc tlusty} computes  the detailed cross sections from the basic atomic data, usually
given by Kurucz atomic data files,
or, (ii) In the Opacity Distribution Function (ODF) mode, in which case the ODF 
is calculated by a separate program,
     and is communicated to {\sc tlusty} by means of two additional input files
     As mentioned above, this option is now obsolete.\\
   
\item[IFANCY$\equiv$IPROF]  -- a mode of treatment of the absorption profile. It has a meaning
\index{Input!line profile setting}
for ``normal'' lines only, i.e. with abs(MODE)=1:\\
$\bullet\,= 0$  --  Doppler profile;\\
$\bullet\,= 1$ or $-1$ --  Voigt profile;\\
$\bullet\,= 2$ or $-2$  -- approximate Stark (+ Doppler) profile for hydrogenic lines
after Hubeny et al. (1994);\\
$\bullet\,= 3$ or $-3$  -- hydrogen line profiles given by Lemke's tables. \\
$\bullet\,= 4$ or $-4$  -- hydrogen line profiles given by Tremblay's tables. \\
However, if the
keyword parameter IHYDPR is set to a non-zero value, the Lemke
or Tremblay  profiles
are switched on globally and the program resets the IFANCY for appropriate
transition automatically, so they do not need to be modified by hand.
For details, refer to \S\,\ref{nst2_hyd}.\\
$\bullet\,\geq 10$ -- non-standard expression, given by a user-supplied
                       subroutine (PROFSP);\\
$\bullet\,> 0$  -- the absorption profile at the farthermost frequency point(s) from
          the line center is(are) taken to be 0;\\
$\bullet\,< 0$  -- the absorption profile at the farthermost frequency point(s)
          is(are) evaluated exactly. 
          
\index{Input!collisional excitation cross sections}
\item[ICOL] -  mode of evaluating the collisional rates.
\noindent
The meaning of ICOL is different for H and He, and for other
species.\\ [2pt]
Specifically for H and He:
\begin{itemize}
\item for
{\tt ICOL} = 0  -- standard approximate expression 
taken from Mihalas et al. (1975).
\end{itemize}
For He I bound-bound transitions, the following standard
possibilities are also available:
\begin{itemize} 
\item for
{\tt ICOL} =  1, 2, or 3  -- much more accurate by Berrington \&
Kingston (1987) rates,
                  subroutine written by D.G.Hummer (COLLHE).
                  This procedure can be used only for transitions
                  between states with $n = 1, \ldots 4$.
\item for
{\tt ICOL} =  1  -- means that a given transition is a transition
                  between non-averaged $ls$ states. In this case,
                  labeling of the He I energy levels must agree
                  with that given in subroutine COLLHE, ie. states
                  have to be labeled sequentially in order of
                  increasing frequency.
\item for
{\tt ICOL} =  2  -- means that a given transition is a transition 
                  between a non-averaged $ls$ lower state and an
                  averaged upper state.
\item for
{\tt ICOL} =  3  -- means that a given transition is a transition 
                  between two averaged states.
\end{itemize}

For species other than H or He:
\begin{itemize}
\item for {\tt ICOL}  = 0  -- simplified van Regemorter (1962) formula\\
   $\Omega = 19.7363\, f_{ij} T^{-3/2} U_0^{-1} \exp(-U_0) 
    \max \left[ \bar g, 0.276 \exp(U_0) E_1(U_0) \right],$\\
    with $\bar g = 0.25$
\item for {\tt ICOL}  = 1 : proper van Regermorter formula; the same as above, 
but now $\bar g$ is an input
       parameter ({\tt CPAR}). $\bar g$ should be taken
       $\bar g = 0.2$ for the transitions $nl \rightarrow n^\prime l^\prime$
(where $n \not= n^\prime$), 
   and $\bar g = 0.7$ for the transitions $nl \rightarrow n l^\prime$.
\item for {\tt ICOL}  = 2 :
   $\Omega = 5.465 \times 10^{-11} f_{ij} c_0 T^{1/2} \exp(-U_0),$
\item for {\tt ICOL}  = 3 :
   $\Omega = 5.465 \times 10^{-11} f_{ij} c_0 T^{1/2} (1+U_0) \exp(-U_0),$
\item for {\tt ICOL}  = 4 : Eissner-Seaton (1972) formula:\\
   $\Omega = 8.631 \times 10^{-6} g_i^{-1} T^{-1/2}\,  \exp(-U_0) c_0.$\\
    In the three above options $c_0$ is given by input parameter {\tt CPARAM}
    -- see below
\item  for {\tt ICOL} = 9 : represents the collision rate for a superline, computed
by summing collision strengths for the individual components. 
\item for {\tt ICOL} = -1 :Van Regemorter formula for neutral atoms,\\
  $\Omega =  19.7363\, f_{ij}\,  T^{-3/2}\, U_0^{-1} \exp(-U_0)\,  \Gamma(U_0),\quad$  where\\
  $\quad\quad\Gamma(U_0) = 0.276\,\exp(U_0)\,  E_1(U_0),\quad$ for $U_0 \leq 14,\quad$  and\\
  $\quad\quad\Gamma(U_0) = 0.066\, U_0^{-1/2}\, (1+1.5 U_0),\quad$ for $U_0 > 14$
\end{itemize}
\noindent where $U_0=(E_j - E_i)/(kT)$, $E_i$ and $E_j$ are the energies
of the lower and upper level, $g_i$ is the statistical weight, and $f_{ij}$
the oscillator strength. Function $E_1$ is the first exponential integral
function.

Generally, for any species,
\begin{itemize}
\item for {\tt ICOL} $<0$ -- non-standard expression, given by a user--supplied
addition to the subroutine CSPEC. This option is practically obsolete, but for its use 
see \S\,\ref{cspec}. The value of ICOL= -1 is already used for the Van Regemorter
formula for neutral atoms - see above.
\end{itemize}

\item[IFRQ0, IFRQ1] -- their non-zero values 
                 signal a change for the mode of treatment frequency
                 points (i.e. ALI or linearized) between indices
                 IFRQ0 and IFRQ1 (internal indices for a line, 
                 starting with 1). For instance, if the given
                 transition is primarily an ALI one (MODE$<$0), then the
                 points between IFRQ0 and IFRQ1 will be taken
                 as linearized.
                 
\item[OSC] --   oscillator strength\\
if = 0 -- the program assigns a scaled hydrogenic oscillator strength;

\item[CPARAM]  --  collision parameter, i.e. quantity $\bar g$ for ICOL=1,
or $c_0$ for ICOL= 2, 3, 4.

\end{description}

\subsubsection*{Additional input parameters for line transitions}

In most cases, there is one or more additional input records for
the line transition, which depends on the coded values of the basic
parameters listed in the first record, described above. In case there are more
input records, they should appear in the order in which they are listed
below:

\medskip
\noindent 
(1) modified frequency of the line -- if abs(MODE) $>$ 100.
\begin{description}
\item[FR0INP] -- frequency (or wavelength -- if FR0INP $< 10^{10}$; the value
is understood as wavelength in \AA) of the line, if it is required to be
different from the value computed from the corresponding level energies.
This option is useful, for instance, to avoid a spurious overlap of a 
normal line and a superline represented by an ODF. 
Again, it is obsolete in the Opacity Sampling mode.
\end{description}

\medskip
\noindent 
(2)  Additional input parameters for ``normal'' line transitions,
i.e. those not represented by ODF's -- with abs(MODE)=1.

\begin{description}
\item[LCOMP]   --  a mode of considering the absorption profile:\\
$\bullet$ =  .FALSE. -- depth-independent profile;\\
$\bullet$ =  .TRUE.  -- depth-dependent profile.

\item[INTMOD]  --  a mode of setting the frequency points and weights
                   in the line:\\
$\bullet\,= 0$  -- means that frequency points and weights have
               already been read among  the NJREAD or NFREAD
               frequencies;\\
$\bullet\,\neq 0$  -- frequency points and weights are evaluated, with one of the
following possibilities:\\
$\bullet\,= 1$ -- equidistant frequencies, trapezoidal integration;\\
$\bullet\,= 2$ -- equidistant frequencies, Simpson integration;\\
$\bullet\,= 3$ -- a ``modified Simpson'' integration, which is a set of
               3-point Simpson integrations with each subsequent
               integration interval doubled, until the whole
               integration area is covered;\\
$\bullet\,= 4$ -- frequencies (in units of standard $x$; $x$ being the frequency displacement
from the line center measured in units of fiducial Doppler width) and weights
(for integration over $x$) are read from the record(s) immediately following. This
option is cumbersome to use, and is generally obsolete.

\item[NF]    --  number of frequency points in the line
          (effective only for INTMOD $\neq$ 0)

\item[XMAX]   -- the maximum frequency extent of the line, in units of
fiducial Doppler width, defined as the Doppler width for the given species
evaluated at $T=$ TSTD, and the standard microturbulent velocity VTB-- see below:\\
$\bullet\,=  0$  -- program sets up default XMAX=4.55\\
$\bullet\,> 0$ -- means that the line is assumed symmetric around the
               center; the frequency points are set up between $x=0$ and
               $x=$XMAX, where $x$ is frequency difference from the
               line center in units of the fiducial Doppler width. But see the note below.\\
$\bullet\,< 0$  -- frequency points are set between $x=$XMAX and $x=-$XMAX\\ [2pt]
{\bf Important note}: in the overlapping mode 
(IOVER $>$ 0; which is the default), all lines are
 set by default to the full-profile mode. Therefore, even if XMAX was coded as
  positive, it is reset to  $-$XMAX, and NF is reset to $2\times{\rm NF}-1$.

\item[TSTD]  -- characteristic temperature for evaluating the fiducial
Doppler width:\\ 
$\bullet\,= 0$  -- the program sets  
               TSTD = $(3/4) \, T_{\rm eff}$.

\end{description}

\medskip
\noindent 
(3) If a Voigt profile is assumed, i.e., if abs(IFANCY) = 1, an additional
input record is required which specifies an evaluation of the
relevant damping parameter:

\begin{description}
 \item[GAMR]  -- a natural broadening indicator:\\           
$\bullet\,> 0$  -- has the meaning of natural damping parameter;\\
$\bullet\,= 0$  -- classical natural damping assumed
         $\Gamma = 2.4734 \times 10^{-8} \, \nu_0^2$;\\
$\bullet\,< 0$  -- natural damping is given by a non-standard, user supplied 
expression in  the subroutine GAMSP -- see \S\,\ref{gamsp},

 \item[STARK1] -- Stark broadening indicator:\\
$\bullet\,= 0$  -- Stark broadening is neglected;\\
$\bullet\,< 0$  -- scaled classical expression, i.e.
           $\Gamma = -{\tt STARK1} \times \Gamma^{\rm clas}$, where
           $\Gamma^{\rm clas} = 10^{-8} \, n_{\rm eff}^{5/2}\, n_{\rm e}$,
           where $n_{\rm eff}$ is the effective quantum number of
           the upper level, $n_{\rm eff} \equiv Z_I^2 [E_H/(E_{I}-E_{j)}]^{1/2}$ 
with the  excitation energy $E_{j}$ and the ionization energy $E_I$,
$Z_I$ is the effective  charge ($Z_I=1$ for neutrals) of the ion $I$,  $E_H$ 
is the ionization energy of hydrogen, and $n_{\rm e}$ is the electron density.
\\
$\bullet\,> 0$  -- Stark broadening given by
       $n_{\rm e} \left({\tt STARK1} \times T^{\tt STARK2} + {\tt STARK3}\right)$,
       where {\tt STARK2, STARK3} are the input parameters
 \item[STARK2, STARK3]   --   see above
 \item[VDWH]    -- Van der Waals broadening indicator:\\
$\bullet\,\leq 0$  -- Van der Waals broadening is neglected;\\
$\bullet\,> 0$  -- a scaled classical expression (see Paper~I, Appendix A).
\end{description}

\medskip
\noindent 
(4) Additional input parameters for a ``merged superline'' transition,
 i.e. a transition to a merged level, treated by means of an ODF
 \index{Input!merged superline setting}
    -- i.e. for abs(MODE)=2:

\begin{description}
\item[KDO(${\bf 1,\ldots,4}$), XDO($1,\ldots,3$)] -- the parameters
 which have the following meaning:
 The superline is represented by four frequency intervals.
 Going away from the peak of the corresponding Opacity Distribution 
 Function, the first interval is 
 represented by a KDO(1)--point Simpson integration, with a 
 distance XDO(1) fiducial Doppler widths between the points.
 The same for the second and third interval.
 The rest (the interval between the last point and the 
 corresponding edge) is represented by a KDO(4)-point
 Simpson integration.
 The fiducial Doppler width is taken here as that corresponding to the
 effective temperature.
\end{description}


\subsection{Energy bands for superlevels}
\label{ion_sup}

Finally, for data files of iron-peak elements one must provide the limits of 
energy bands used to
\index{Input!energy band for superlevels}
\index{Superlevels}
build the superlevels, first for even-parity levels, then for odd-parity levels.
This is required only in the Opacity Sampling mode where {\sc tlusty} reads 
the detailed lists of individual energy  levels from the Kurucz files
and constructs the superlevels. This list of energy bands must be 
consistent with the level data given at the top of the file.
The structure of the input is as follows:
\index{Superlevels}
\begin{description}
\item[NEVKU] -- number of even superlevels
\end{description}
and then NEVKU records for i=1,$\ldots,$NEVKU
\begin{description}
\item[XEV(i)]  -- upper energy limit for the I-th even  superlevel [cm${}^{-1}$];
\end{description}
and analogously for odd superlevels
\begin{description}
\item[NODKU] -- number of odd superlevels
\end{description}
and then NODKU records for i=1,$\ldots,$NODKU
\begin{description}
\item [XOD(i)]  -- upper energy limit for the i-th odd  superlevel [cm${}^{-1}$].
\end{description}


\subsection{Examples}
\label{ion_exa}

\subsubsection*{Energy levels}

An input for hydrogen is very simple; for instance the ground state
is specified by the following record
\begin{verbatim}
 0.                  0.    0    ' (N=1)  '   1     0.    0
\end{verbatim}
because the energies and statistical weights are hydrogenic.
The number 1 in the fifth entry signifies that the exact occupation
probability and level dissolution will be taken into account.
A ``merged'' level is specified by
\begin{verbatim}
 0.                  0.    0    ' merged '  -1     0.    0
\end{verbatim}
The He I ground state is specified by 
\begin{verbatim}
 5.94503520D+15      1.    1    '1 sing S'   0     0.    0 
\end{verbatim}
As an example of level data for other species, let us take first 10 levels
of C IV,
\begin{verbatim}
****** Levels
 1.55945583E+16      2.    2  'C IV 2Se 1'  0   0.   0  
 1.36613760E+16      2.    2  'C IV 2Po 1'  0   0.   0  
 1.36581470E+16      4.    2  'C IV 2Po 1'  0   0.   0  
 6.51537202E+15      2.    3  'C IV 2Se 2'  0   0.  -104
 5.99906956E+15      6.    3  'C IV 2Po 2'  0   0.  -104
 5.85471942E+15     10.    3  'C IV 2De 1'  0   0.  -104
 3.56244314E+15      2.    4  'C IV 2Se 3'  0   0.  -104
 3.35343089E+15      6.    4  'C IV 2Po 3'  0   0.  -104
 3.29290270E+15     10.    4  'C IV 2De 2'  0   0.  -104
 3.29005464E+15     14.    4  'C IV 2Fo 1'  0   0.  -104
\end{verbatim}
Notice that both levels of the ${}^2\!P$ doublet (levels No. 2 and 3) are treated 
separately; therefore the two components of the resonance doublet at 1548 and 
1551 \AA\ are treated as individual lines.
This example also illustrates setting level groups. The first three levels are treated
individually, while the higher states are lumped into one group.

\subsubsection*{Bound-free transitions}

The hydrogen Lyman continuum, for a hydrogen model atom composed of
9 levels for H I and one level for H II, is specified, for instance, by
\begin{verbatim}
****** Continuum transitions
 1 10  5  1  0  1  6  0.  0.    ! Lyman cont.+ pseudocontinuum
 2.60D15                        ! Lyman pseudocontinuum cutoff
\end{verbatim}
The 6-th and 7-th entry, IFRQ0 and IFRQ1, signify that the 1st through 6th
frequency points in the Lyman continuum (the points immediately blue-ward
of the discontinuity) will be treated in the linearized (not ALI) mode.
As mentioned earlier, this option is usually a recommended one, since it
usually increases the convergence rate considerably, while the total
computer time remains virtually unchanged (see also an extensive discussion
in Hubeny \& Lanz 1995).

Another example is provided by a specification for the =bound-free transition from
the ground state of C~IV, viz.
\begin{verbatim}
****** Continuum transitions
   1  26   1    116      0      0      0   1.922E-19   0.000E+00
 -0.0149  0.0110  0.1338  0.2761  0.3795  0.5541  0.7093  0.7545
  0.7610  0.7675  0.7739  0.7804  0.7998  0.9744  1.1619  1.3246
 -0.1726 -0.2005 -0.4114 -0.6842 -0.8897 -1.2472 -1.5886 -0.5243 
 -0.4333 -0.3678 -0.3427 -0.3320 -0.3668 -0.8011 -1.2988 -1.7537
\end{verbatim}
which shows how to use a fit point representation of the Opacity Project 
photoionization data. In this case, there are 16 points (IFANCY=116), with relative
frequencies $x$ specified in the second record, and the logarithms of
cross sections (in Mb) in the third. 
Other features: the collisional rate is evaluated by the Seaton formula
(ICOL=0 -- the 5th entry in the first record), with the parameter
 $\bar g\sigma_0 = 1.922 \times 10^{-19}$ -- the 8th entry. Since this model
 atom for C~IV has 25 discrete levels, the index of the upper level of the
 transition -- the ground state of the next ion -- is 26.

\subsubsection*{Line transitions}

The hydrogen L$\alpha$ line is specified, for instance, by
\begin{verbatim}
 1  2 -1    2  0 26 28  0.  0.    ! H I Lyman lines        
  T 3 27 1022. 0.
\end{verbatim}
In this case, the line frequencies are predominantly taken in the ALI mode
(MODE=$-1$), but points 26 -- 28 are set to a different treatment, 
i.e. are linearized. The line is
taken with a depth-dependent (LCOMP=T), 
with approximate Stark + Doppler profile (IFANCY=2).
The  line is assumed to extend to 1022 fiducial Doppler widths.
The number of frequency points is set to 27 (NF=27), but we assume the
standard value of the line-overlap mode switch IOVER=1, in which case the
line is automatically taken with a full profile (extending to both sides 
from the line core), with the actual number of frequency points being
therefore $2 \times 27 - 1=53$. The line frequencies are taken with ALI
except the three frequency points, 26 -- 28, around the center 
which are linearized.

If L$\alpha$ is assumed to be in detailed radiative balance, one codes
a single record:
\begin{verbatim}
 1  2  0  0  0  0  0  0.  0.    
\end{verbatim}
We stress that there is a significant difference between specifying the mode
of treating L$\alpha$ as above, and not specifying the transition at all. In the
later case, neither collisional, nor radiative rates are calculated, i.e. the
transition is assumed to be in both radiative and collisional detailed balance.
The levels are thus forced to be in exact Boltzmann equilibrium within
each other. In the former case, the collisional rates are calculated, but
the levels are not forced to be in equilibrium. Departures from equilibrium
are determined by relative values of the collisional rates and the respective 
photoionization rates.

Another example is the beginning of the block of line transitions in the C~IV
model atom; here we show the transitions from the ground state (1) to the
first 10 levels,
\begin{verbatim} 
   1   2  -1   1   1      0      0   0.952E-01   7.000E-01
  T  3   23 700.  0.
  2.6E+8  7.9E-7   0.  0.  0.
   1   3  -1   1   1      0      0   1.900E-01   7.000E-01
  T  3   23 700.  0.
  2.6E+8  7.9E-7   0.  0.  0.
   1   4   0   0   4      0      0   0.000E+00   5.000E-02
   1   5  -1   1   1      0      0   2.030E-01   2.000E-01
  T  3   21 500.  0.
  4.6E+9  3.1E-6   0.  0.  0.
   1   6   0   0   4      0      0   0.000E+00   5.000E-02
   1   7   0   0   4      0      0   0.000E+00   5.000E-02
   1   8  -1   1   1      0      0   6.100E-02   2.000E-01
  T  3   15 100.  0.
  2.9E+9  9.4E-6   0.  0.  0.
   1   9   0   0   4      0      0   0.000E+00   5.000E-02
   1  10   0   0   4      0      0   0.000E+00   5.000E-02
\end{verbatim}
Here, only the transitions from level 1 to levels 2, 3, and 8, are
dipole allowed; the rest are dipole forbidden. The allowed
transitions are treated with Voigt profile (IFANCY=1); one has therefore
three record for these lines; the last of them specifies the damping parameters,
in this case for the radiative and Stark broadening.
 

\subsection{Optional, non-standard parameters}
\label{new_ion2}

Unlike in the previous part of this chapter, the following parameters are
communicated to the program through the {\em standard input file}.
If the parameter NONSTD of the standard input (see \S\,\ref{new_at})
is coded as positive, the program reads an additional record with the 
following four parameters. These are included for historical reasons and for
possible comparisons with results from the older codes (such as that of Mihalas 
et al. 1975), otherwise they are obsolete and hardly ever used for real calculations.

\begin{description}

\item[IUPSUM] -- mode of evaluation of the total population of higher,
    non-explicit, LTE energy levels of the ion -- the
    so-called upper sum (see Paper~II, \S\,2.3):\\
$= 0$ -- calculated by means of the partition function \\ 
$> 0$ -- calculated as a sum of populations of hydrogenic
        levels starting with the quantum number next to
        the highest explicit level and ending with IUPSUM;\\
$< 0$ -- the occupation probability form. 
The absolute value of IUPSUM specifies the main quantum number of
the highest hydrogenic level considered.\\
DEFAULT: IUPSUM=$-100$ for H; and =0 for other species.

\item[ICUP]  --  mode of considering a ``modified collisional ionization
    rate'', i.e., that allowing for collisional excitation into,
    and collisional de-excitation from, higher,
    non-explicit, LTE energy levels of the ion:\\
$ =  0$ -- this contribution is neglected;\\
$ > 0$ -- calculated as a sum of contributions of
        rates into and from averaged (hydrogenic)
        levels starting with the quantum number next to
        the highest explicit level and ending with ICUP.\\
DEFAULT: ICUP=16 for all ions but He II; ICUP=32 for He II.

\item[MODEFF]   --  mode of evaluating the free-free cross section:\\
 =  0 -- free-free opacity is neglected;\\
 =  1 -- hydrogenic cross section with the Gaunt factor
        set to unity;\\
 =  2 -- hydrogenic cross section with the exact Gaunt
        factor;\\
$<$ 0 -- non-standard expression, given by the user
        supplied subroutine {\tt FFCROS}.\\
DEFAULT: MODEFF=2 for H I and He II; MODEFF=1 for all other ions.

\item[NFF]     --  mode of considering "modified free-free" opacity,
    i.e. allowing for the photoionization from higher,
    non-explicit, LTE energy levels of the ion:\\
 =  0 -- this contribution is neglected;\\
$>$ 0 -- principal quantum number of the first non-explicit
        level.\\
DEFAULT: NFF=0

\end{description}


\subsection{X-files: Atomic data files for inner-shell ionization}
\label{xfiles}

When inner-shell ionization is taken into account, the system of input
explained in the previous section cannot be used, because
(i) one deals with a different form of cross sections, and, 
most importantly,
(ii) when reading data for one ion the indices of energy levels are set
for this particular ion, but the actual index of the level where an atom
ends after an inner-shell ionization (two or more ionization stages higher)
is not yet known.

Therefore, there is a different format of such atomic input files. Also,
the number of explicit energy levels of all ions treated this way
is limited to 1, i.e., these ions are treated as 1-level ions. 
The atoms and ions for which one does
not require a consideration of inner-shell ionization are treated in the
ordinary way, with as many explicit levels as desired. Obviously, the latter
approach is used for H and He, but can be used for other species as
well. The atomic data files for ions with inner-shell ionization,
which is important in particular in the X-ray region, are called here
``X-files."

The first four lines of an X-file is analogous to the ordinary atomic data
file, Here is an example for C III:
\begin{verbatim}
****** Levels
 1.15792157E+16      1.    2  'c3   1Se 1'  0   0.  0
****** Continuum transitions
   1 1004 1 15 4 0 0 0.000E+00 0.000E+00
\end{verbatim}
Data for the one (ground) level is exactly as explained before. It is followed
by a label ``{\tt ****** Continuum transitions}'', followed by the record
containing the introductory data and flags for the first possible ionization 
from this level.The meaning of the individual entries is
similar to that in the ordinary files. Here is a detailed explanation:
\begin{description}
\item[II]   --  relative index of the lower level as before, but it is always 1
because one deals with 1-level ions only.
\item[JJ]  --  here, instead of the relative index of the upper level, this 
number has to be set to be larger than 1000, and JJ-1000 represents the 
core charge of the ion in which the process ends. In this particular case, 
it ends in C IV (or C${}^{+3}$, with a core charge 4 and the actual
charge 3). In this particular case, one has in fact a regular photoionization,
but which is treated analogously, and using analogous fitting formulae, as
for inner-shell photoionizations (see below).
\item[MODE]    -- the number has to be set to 1, which indicates that the 
process is being treated as an explicit transition.
\item[IBF] -- has to be set to 15, which tells the program that the rest of
the file is an X-file, with a different meaning of the subsequent entries.
\item[ICOL] -- a special flag for the corresponding collisional rate
It has to be set to either 4 or 99\\
$\bullet\, = 4$ -- collisional ionization rate evaluated by subroutine
{\tt cion} adapted from program XSTAR by T.~Kallman. The necessary
fitting data are hardwired in {\sc tlusty}, so no additional data are necessary.\\
$\bullet\, = 99$ -- collisional rate is set to zero.
\item[the next three entries] -- all zeros, have no meaning in an X-file.
\end{description} 
Then, there are five lines of input, again for example for the same C III X-file,
\begin{verbatim}
     4.789E+01
      4 3 2
       1.208
      2
       0.0047
\end{verbatim}
which have the following meaning\\
E -- the threshold energy of the transition (in eV)\\
NMAX, IZX, NSH\\
S\\
NA\\
DX\\
which is followed by NA numbers $b_j, j=1,\ldots,N\!A$, and
NA groups of 11 numbers $a_{ij}, i=1,\ldots,11; j=1,\ldots,N\!A$.
In the present case, this block looks is:
\begin{verbatim}
       6.000e-02
       3.000e+02
       -5.258768e+01
       -1.632927e+02
       -1.728917e+02 
       -8.205509e+01
       -1.465739e+01
       0.
       0.
       0.
       0.
       0.
       0.
       +3.220718e+00
       -2.712337e+00
       -2.975667e-01
       +8.581632e-02
       -3.513528e-02
       +8.835004e-03
       0.
       0.
       0.
       0.
       0.
\end{verbatim}
Here, $a$ and $b$, together with the rest of above input parameters,
are the fitting coefficients for evaluating the
corresponding photoionization cross section, using the same approach
as in the XSTAR program by T.~Kallman (subroutine {\tt bkhsgo} there), 
adapted to {\sc tlusty} by O.~Blaes.

The next record is:
\begin{verbatim}
   1 1005 1 15 99 0 0 0.000E+00 0.000E+00
\end{verbatim}
that specifies data for the true inner-shell ionization process that leaves
the C${}^{+2}$ ion in the ground state of C${}^{+4}$. This entry is then
followed by an analogous set of entries to specify the corresponding
photoionization cross sections.

We stress that an X-file for a given ion contains the data for all transitions
that go from this ion to a higher ion. 


\section{Keywords for additional/non-standard physical processes}
\label{nst2_phys}

In this chapter we describe the keyword parameters that switch on,
or provide additional parameters for, additional physical
processes that are not considered by default. For some specific applications,
these may actually be quite important; for instance for white dwarfs, cool stars,
very hot stars, or accretion disks, etc. The users who intend to use {\sc tlusty}
for such applications are encouraged to study the appropriate parts of this chapter
in detail.

\subsection{Hydrogen line profiles}

There are two different types of refinements for
the hydrogen line broadening: \\
(i) using detailed hydrogen Stark broadening tables constructed by
Lemke (1997) or Tremblay \& Bergeron (2009)---hereafter referred to as
Tremblay's tables; and\\
(ii) considering the quasi-molecular satellites
of Lyman $\alpha$, $\beta$, $\gamma$, and H$\alpha$. We shall describe them below.

\subsubsection{Special Stark broadening tables}
\label{nst2_hyd}

\begin{description}
\item[IHYDPR] -- a mode of treatment of the hydrogen line
\index{IHYDPR keyword parameter}
broadening. This keyword can be used to set the treatment of all Lyman
and Balmer lines globally, and its action overwrites the setup specified
in the atomic data file through the parameter IFANCY -- see \S\,\ref{ion_bb}. \\
$\bullet\, = 0$ -- hydrogen line broadening is treated as before (an approximate Stark
broadening after Hubeny, Hummer, \& Lanz 1994), or whichever treatment
was specified individually through parameter IFANCY in the atomic data file;\\
$\bullet\, =1$ -- hydrogen line broadening is computed using Lemke's tables,
for all Lyman and Balmer lines with a principal quantum number of the
upper level $\leq 10$, that are treated explicitly.
The file {\tt lemke.dat} has to be present in, or linked to, the current directory.
The table contains data for higher lines as well, but these are not used because they
turned out to be incorrect.\\
$\bullet\, =2$ -- hydrogen line broadening of all Lyman and Balmer lines
is computed using Tremblay's tables. 
The file {\tt tremblay.dat} has to be present in, or linked to, the current directory.\\
DEFAULT: IHYDPR=0
\end{description}

\subsubsection{Hydrogen quasi-molecular opacities}
\label{nst2_quas}

An interaction of a radiating hydrogen atom with neighboring protons
is usually described  through the pressure broadening of hydrogen lines. 
However, an interaction
with a very close proton leads to the formation of a transient H${}_2^+$
molecule, and the optical properties have to be described through the
wave functions of this molecule. This gives rise to a number of discrete
features, called quasi-molecular satellites of hydrogen lines, usually of
Lyman or Balmer, lines. This effect is important for high densities, so
it may play an important role in the atmospheres of white dwarfs; in some
cases also in the atmospheres of extremely metal-poor stars. 

The process is included in {\sc tlusty} by means of computing the relevant
cross sections using the tables by  Allard \& Koester
(1992). The cross sections were calculated for a single, characteristic,
temperature. One can also use newer, temperature-dependent cross sections
supplied by N. Allard (private comm.).  

When including the quasi-molecular satellites, they are treated as part
of the corresponding hydrogen line profile. The full profile is computed
as specified by the keyword parameter IHYDPR (\S\,\ref{nst2_hyd}). Half
of the cross section, corresponding to the broadening by electrons is kept,
while the other half is replaced by the data for quasi-molecular satellites that
correspond to the broadening by protons (the tables extend to the whole
extent of the profile, not just over the satellite features themselves).

The present version of {\sc tlusty} deals with satellites of Lyman $\alpha$,
$\beta$, $\gamma$, and Balmer $\alpha$,  which are the most important lines. 
Here is a description of the corresponding keyword parameters:
 
\begin{description}

\item[IQUASI] -- a switch for including quasi-molecular 
\index{IQUASI keyword parameter}
satellites\\
$\bullet\,= 0$ -- quasi-molecular satellites are not included;\\
$\bullet\,> 0$ -- quasi-molecular satellites are included. In this case one has
to specify some or all additional parameters NQUALP, NQUBET, NQUGAM,  
and NQUBAL (see below).\\
DEFAULT: IQUASI=0

\item[NQUALP] -- a non-zero absolute value specifies that 
\index{NQUAPL keyword parameter}
the satellites of L$\alpha$ are taken into account. Unlike previous versions
the actual number of the parameter is inconsequential.\\
$\bullet\,>0$ -- standard, temperature-independent cross section.
The corresponding file {\tt laquasi.dat} has to be present in, or linked to, the current directory.\\
$\bullet\,<0$ -- temperature-dependent cross section. This option requires
a different table than that used in the previous option, which is not a part of
the standard distribution of {\sc tlusty}.\\
DEFAULT: NQUALP=0

\item[NQUBET] -- an analogous flag
\index{NQUBET keyword parameter}
for the L$\beta$ satellites.\\
$\bullet\,>0$ -- standard, temperature-independent cross section.
The corresponding file {\tt lbquasi.dat} has to be present in, or linked to, the current directory.\\
$\bullet\,<0$ -- temperature-dependent cross section.\\
DEFAULT: NQUBET=0

\item[NQUGAM] -- an analogous flag
\index{NQUGAM keyword parameter}
for the L$\gamma$ satellites.\\
$\bullet\,>0$ -- standard, temperature-independent cross section.
The corresponding file {\tt lgquasi.dat} has to be present in, or linked to, the current directory.\\
$\bullet\,<0$ -- temperature-dependent cross section.\\
DEFAULT: NQUGAM=0

\item[NQUBAL] -- an analogous flag
\index{NQUBAL keyword parameter}
for the H$\alpha$ satellites.\\
$\bullet\,>0$ -- standard, temperature-independent, cross section.
The corresponding file {\tt lhquasi.dat} has to be present in, or linked to, the current directory.\\
$\bullet\,<0$ -- temperature-dependent cross section.\\
DEFAULT: NQUBAL=0

\item[TQMPRF] -- takes effect only if one considers temperature-dependent
\index{TQMPRF keyword parameter}
cross section(s), where, if set to a non-zero value, TQMPRF specifies the 
temperature for which the cross section will be taken, so that the profile will be
temperature-independent after all. It can be used for testing purposes
to examine the temperature dependence of the profiles.\\
DEFAULT: TQMPRF=0.
\end{description}


\subsection{Additional opacities}
\label{nst2_opadd}

Some additional opacities are already hardwired in {\sc tlusty}, so that
no additional input files are necessary to describe them. 
Some may be switched on by setting an appropriate keyword parameter,
and/or by providing a corresponding additional input file containing their tabular
values.

To the first category opacities contains the hydrogen Rayleigh scattering,
and  the H${}^-$ opacity, although these may also be treated
differently, and the H${}_2^+$ opacity. The second category contains
hydrogen quasi-molecular opacities, described above,  and the 
Collision-Induced Absorption (CIA) opacities. We shall describe them below.

\subsubsection{Rayleigh scattering}
\label{nst2_ray}

The cross section is treated either through an analytic expression
(for hydrogen), or through a corresponding table that may contain
contributions from many species. The former option can be used
with both the standard mode (an evaluation of opacities on
the fly), or with the opacity table mode, while the latter is used only
in conjunction with the opacity table mode (IOPTAB $<$ 0).

The analytical expression for hydrogen is from Kurucz (1970):
\begin{equation}
\label{rays_dal}
\sigma(\nu)=5.799\times 10^{-13} x^{-4} + 1.422\times 10^{-6} x^{-6}
+ 2.784 x^{-8},
\end{equation}
where $x=2.997925 \times 10^{18}/\min(\nu, 2.922 \times 10^{15})$.
\begin{description}
\item[IRSCT] -- a switch for including Rayleigh scattering 
\index{IRSCT keyword parameter}
on hydrogen;\\
$\bullet\,=0$ -- Rayleigh scattering is neglected.\\
$\bullet\,>0$ -- Rayleigh scattering is included, using eq. (\ref{rays_dal})\\
DEFAULT: IRSCT=0
\item[IFRAYL] -- a switch for a treatment of Rayleigh scattering in the
\index{IFRAYL keyword parameter}
\index{Rayleigh scattering}
case of using the opacity table (IOPTAB $<$ 0);\\
$\bullet\,<0$ -- the scattering coefficient, possibly for many species,  is given by 
the special table called {\tt rayleigh.tab} which has to be given in the
same grid of temperatures, densities, and frequencies as the global
opacity table {\tt absopac.dat};\\
$\bullet\,>0$ -- the scattering coefficient, only on hydrogen,  is still evaluated
by an analytical formula using eq. (\ref{rays_dal}).\\
$\bullet\,=0$ -- Rayleigh scattering is neglected.\\
DEFAULT: IFRAYL=0 

\end{description}

\subsubsection{Hardwired optional H and He opacities}
\label{nst2_hhe}

The following keyword parameters, if set to a non-zero value, switch on the individual opacity sources described below. If switched on, their evaluation does not require
any additional input data.

\begin{description}

\item[IOPHMI] --  a switch for including the
\index{IOPHMI keyword parameter}
H${}^-$ opacity (both bound-free and free-free), assuming LTE.
Note that H${}^-$ can be considered as one of the explicit ions,
then the opacity is calculated automatically (and moreover in NLTE if
such a model is being calculated). In this case one must code
IOPHMI=0 in order not to include the H${}^-$ opacity twice.
In both cases, selecting H${}^-$ as a one-level separate ion, or
using this option, IOPHMI $ >0$, the bound-free as well as free-free
opacity due to H${}^-$ is evaluated using the table (for bound-free) 
or an analytic expression  (for free-free)  from Kurucz (1970). \\
DEFAULT: IOPHMI=0

\item[IOPH2P]  --  a switch for considering the
\index{IOPH2P keyword parameter}
H$_2^+$ opacity, bound bound-free and free-free, using the analytic 
expression from Kurucz (1970).\\
DEFAULT: IOPH2P=0

\item[IOPHE1] --  a switch for considering the
\index{IOPHE1 keyword parameter}
approximate hydrogenic opacity of neutral helium
given as a sum of bound-free transitions from averaged
levels with principal quantum numbers between 
the highest level considered explicitly and an averaged level
corresponding to the principal quantum number equal to IOPHE1
(after Mihalas et al. 1975).
Outdated; included for historical reasons only.\\
DEFAULT: IOPHE1=0

\item[IOPHE2] --  a switch for considering the
\index{IOPHE2 keyword parameter}
approximate hydrogenic opacity of ionized helium
given as a sum of bound-free transitions from averaged
levels with principal quantum numbers between 
the highest level considered explicitly and an averaged level
corresponding to the principal quantum number equal to IOPHE2
(again after Mihalas et al. 1975).
Outdated; included for historical reasons only.\\
DEFAULT: IOPHE2=0

\end{description}

\subsubsection{CIA opacities}
\label{nst2_cia}

The Collision-Induced Opacity (CIA) for H${}_2$ -- H${}_2$ interactions,
using data from Borysow \& Frommhold (1990) can be included by 
setting the keyword parameter IFCIA. In this case, one needs an additional file
{\tt CIA-H2H2.Yi}. The corresponding subroutine was written by M. Montgomery.
\begin{description}
\item[IFCIA] -- switch for including the CIA 
\index{IFCIA keyword parameter}
opacity;\\
$\bullet\,=0$ -- the CIA opacity is neglected;\\
$\bullet\,>0$ -- the CIA opacity is included.\\
DEFAULT: IFCIA=0
\end{description}

\subsection{Collisions}
\label{nst2_colis}

Although the evaluation of collisional rates is controlled by the corresponding
parameters in the atomic data files (see \S\,\ref{ion_bf} and \S\,\ref{ion_bb}),
{\sc tlusty} offers several keyword parameters that are designed to set up
a modification of such an evaluation for a whole group of transitions without 
the necessity to modify the standard atomic data files.

\subsubsection{Modified electron collisional rates for hydrogen}
\label{nst2_colm}

\begin{description}
\item[ICOLHN] -- a switch for setting up the default calculation of the
\index{ICOLHN keyword parameter}
hydrogen collisional excitation rates. We stress that setting this parameter
to 1 or 2 will overwrite the mode of evaluation of collisional rates specified
by input parameter ICOL (coded in the input atomic data file  -- see \S\,\ref{ion_bb}).\\
$\bullet\,=0$ -- hydrogen collisional excitation rates are evaluated according 
to the values of ICOL.\\
$\bullet\,=1$ -- hydrogen collisional excitation rates are evaluated using the data by
Przybilla \& Butler (2004) (for transitions with the upper level up to $n=7$);\\
$\bullet\,=2$ -- hydrogen collisional excitation rates are evaluated using the data by
Giovanardi et al. (1987)  (for transitions with the upper level up to $n=15$).\\
DEFAULT: ICOLHN=1
\end{description}

\subsubsection{Charge transfer reactions}
\label{nst2_colh}

\begin{description}
\item[IFCHTR] -- a global switch indicating that all the charge 
\index{IFCHTR keyword parameter}
transfer reactions of neutral or ionized atoms (with atomic numbers 2 to 30) with 
hydrogen are switched on. The charge transfer reaction is treated as a
collisional transition from the ground state of an atom (neutral or ionized)
to the ground state of the next ionization stage. We employ 
an appropriately modified routine written originally 
Kingdon and Ferland (1996). The corresponding subroutine was written by
C. Allende-Prieto.
One can also set up charge transfer
for individual species separately by appropriately modifying parameter
ICOL (see \S\,\ref{ion_bf}).\\
$\bullet\,= 0$ -- charge transfer is not set globally; the process can be switched
on for individual ions through ICOL\\
$\bullet\,\not= 0$ -- charge transfer is set globally for all species with atomic
number between 2 and 30 (He to Zn). In this case, charge transfer
reactions are considered up to IFCHTR-1 times ionized atoms;  that is
IFCHTR=1 means that the charge transfer reactions are switched on for neutral
atoms only, IFCHTR=2 switches charge transfer for neutrals and
once-ionized atoms, etc.\\
DEFAULT: IFCHTR=0 

\end{description}

\subsubsection{Dielectronic recombination}
\label{nst2_diel}

As explained in Paper~II, \S\,\reftrans, dielectronic recombination is included automatically
if the corresponding photoionization cross section is included with all
the resonances. 

If the photoionization cross section is treated as a smooth function of
frequency, there is still a possibility to take dielectronic recombination
into account, using the data from Aldrovandi \& Pequignot (1973), 
Nussbaumer \&  Storey (1983), and Arnaud \& Raymond (1992).  
Numerically, the
procedure described by Hubeny et al. (2001, Appendix B) is used 
that treats dielectronic recombination by introducing an artificial modification 
of the photoionization cross sections. The necessary data are hardwired in the code.
\begin{description}
\item[IFDIEL] -- a global switch for treating the dielectronic 
\index{IFDIEL keyword parameter}
recombination.\\
$\bullet\,=0$ -- dielectronic recombination is neglected (for bound-free 
transitions with smooth cross actions), or is included automatically 
(for transitions with cross sections that include resonances);\\
$\bullet\,\not= 0$ -- dielectronic recombination is considered using the procedure
and data mentioned above.\\
DEFAULT: IFDIEL=0
\end{description}

\subsection{Occupation probabilities and pseudocontinua}
\label{nst2_occup}

Calculation of the occupation probabilities and related pseudocontinuum opacities is
controlled by the positional parameter IFWOP imported in the atomic data
file -- see \S\,\ref{ion_en} (switching on occupation probabilities), and
positional parameters MODE and FR0PC -- see \S\,\ref{ion_bf} (pseudocontiua). 
The following keyword parameters only provide additional options that
can be used mostly for testing.

\begin{description}
\item[BERGFC] -- a switch for using the so-called Bergeron \
\index{BERGFC keyword parameter}
empirical factor (see Paper~II, \S\,\refdissol)\\
$\bullet\,=2$ -- this value represents the Bergeron suggestion, which is now 
obsolete in the context of current {\sc tlusty}.\\
DEFAULT: BERGFC=1

\item[CUTLYM] -- a switch for resetting the frequency cutoff
\index{CUTLYM keyword parameter}
for the Lyman pseudocontinum\\
$\bullet\,=0$ - no resetting, the cutoff remains as specified by the hydrogen 
atomic data file;\\
$\bullet\,>0$ - the cutoff is reset to the value of CUTLYM [in s${}^{-1}$].\\
DEFAULT: CUTLYM=0.

\item[CUTBAL] -- a switch for resetting the frequency cutoff
\index{CUTBAL keyword parameter}
for the Balmer pseudocontinum\\
$\bullet\,=0$ - no resetting, the cutoff remains as specified by the hydrogen 
atomic data file;\\
$\bullet\,>0$ - the cutoff is reset to the value of CUTBAL [in s${}^{-1}$].\\
DEFAULT: CUTBAL=0.

\end{description}

\subsection{Compton scattering}
\label{nst2_compt}
\index{Compton scattering}

We employ a numerical procedure outlined by Hubeny et al. (2001),
and described in detail in Paper~II, Appendix A6 (formal solution) and
C3 (linearization).
First, we describe keyword parameters that define a general setup:
\begin{description}

\item[ICOMPT] -- the basic switch for including Compton scattering. \\
$\bullet\,= 0$ -- Compton scattering is not included; electron scattering is 
\index{ICOMPT keyword parameter}
treated as coherent, Thomson, scattering;\\
$\bullet\,>0$ -- Compton scattering is included.
In this case, several other keyword parameters take effect -- see below.\\
DEFAULT: ICOMPT=0
\item[ICOMRT] -- a flag for setting the method of solution of the explicit
\index{ICOMRT keyword parameter}
angle-dependent transfer equation with Compton scattering\\
$\bullet\, = 0$ -- formal solution done by the Feautrier method.\\
$\bullet\, \not= 0$ -- formal solution done by the Discontinuous Finite
Element  (DFE) method.\\
DEFAULT: ICOMRT=0
\item[ICHCOO] -- a switch for selecting the treatment of frequency derivatives 
\index{ICHCOO keyword parameter}
and boundary conditions in frequency:\\
$\bullet\, = 0$ -- frequency derivatives and boundary conditions in frequency
are treated as described in Hubeny et al. (2011)\\
$\bullet\, > 0$ -- frequency derivatives and boundary conditions in frequency
are treated as described in Chang \& Cooper (1970)\\
DEFAULT: ICHCOO=0
\item[FRLCOM] -- takes effect only if ICOMPT $> 0$ (Compton scattering
is included), and then it
sets the critical frequency above which all the frequency
\index{FRLCOM keyword parameter}
points are set to the linearization mode, regardless of the original setting
specified by keyword parameter IJALI.\\
DEFAULT: FRLCOM=8.2E14
\item[KNISH] -- if set to a non-zero value, switches on the Klein-Nishina
\index{KNISH keyword parameter}
form of the scattering cross section, however still within the framework
of the Kompaneets approximation. The cross section has the form 
(e.g., Rybicki \& Lightman 1979)
\begin{equation}
\label{kjnsih}
\frac{4}{3}
\frac{\sigma(x)}{\sigma_{\mathrm{e}}} =
\frac{1+x}{x^3}\!\left[\!\frac{2x(1+x)}{1+2x}-\ln(1+2x)\!\right]\!+\frac{\ln(1+2x)}{2x}-\frac{1+3x\ }{(1+2x)^2},
\end{equation}
where $\sigma_{\rm e}$ is the Thomson cross section.
This approach is not fully consistent, and
is used for testing purposes only.\\
DEFAULT: KNISH=0
\end{description}
The following parameters switch on various, mostly unphysical, approximations,
that may sometimes be used to avoid numerical problems, or for various testing
purposes.
\begin{description}
\item[ICOMST] -- a switch to specify the treatment of stimulated emission part of 
\index{IMOMST keyword parameter}
the source function with Compton scattering.\\
$\bullet\, = 0$ -- the stimulated emission term in the
Compton scattering source function is neglected. This is useful only for testing
and comparison purposes.\\
$\bullet\, > 0$ -- stimulated emission is included; moreover the numerical value
sets the number of internal iterations of solving the fully coupled implicit transfer
equation (see Paper~II, Appendix A6).\\
DEFAULT: ICOMST=1
\item[ICOMDE] -- if set to 0, the frequency derivative terms in the
\index{ICOMDE keyword parameter}
Compton source function are neglected. This is an unphysical option,
but sometimes useful for testing purposes.\\
DEFAULT: ICOMDE=1
\item[ICOMVE] -- a switch for inhibiting an update of the Eddington factors
\index{ICOMVE keyword parameter}
\index{Eddington factor}
as a result of a solution of the angle-dependent transfer equation with
Compton scattering.\\
$\bullet\, = 0$ -- Eddington factor is not updated, and is kept to the value
determined from the previous solution.\\
$\bullet\, > 0$ -- Eddington factor is updated.\\
DEFAULT: ICOMVE=0
\item[ICMDRA] -- a switch for linearizing the Compton scattering terms
\index{ICMDRA keyword parameter}
in the radiative equilibrium equation with respect to the mean intensities.\\
$\bullet\, = 0$ -- the tern is not linearized (which is sometimes more stable)\\
$\bullet\, \not= 0$ -- the term is linearized.\\
DEFAULT: ICMDRA=0
\end{description}
As explained in Paper~II, Appendix A6, 
the overall formal solution  can be done in several different ways. 
To provide a flexible scheme, we have introduced
five keyword parameters, NCFOR1, NCFOR2, NCFULL, NCITOT, and
NCCOUP, described below, to set up the procedure of the formal solution.

The following parameters control the subroutine {\tt RTECOM} which
is the basic routine for performing the formal solution of the transfer
equation for all frequencies at the current values of the state parameters.
The procedure is composed of several steps:
\begin{itemize}
\item[(i)] An (optional) preparatory loop consisting of several (NCFOR1) iterations
\index{NCFOR1 keyword parameter}
of solving the angle-dependent explicit transfer equation, Eq. (320) or
(321) of Paper~II, depending on whether the Feautrier of the DFE scheme is used
(which is determined by parameter ICOMRT). This is done by subroutine {\tt RTECF1}.
\item[(ii)] A general iteration loop that is composed of NCFULL iterations.  
\index{NCFULL keyword parameter}
This loop consists of two parts:
\begin{itemize}
\item a solution of the fully coupled (in frequency and depth)  implicit transfer
equation; possibly with ICOMST internal 
iterations to update the stimulated emission term (subroutine {\tt rtecmc}),
\item a nested iteration loop, with NCITOT iterations, which again is 
\index{NCITOT keyword parameter}
composed of
\begin{itemize}
\item another (optional) nested iteration loop of NCCOUP iterations, 
\index{NCCOUP keyword parameter}
each performing an
explicit solution of the transfer equation [Eq. (318) of Paper~II] to update the mean
intensities of radiation with a fixed Eddington factor,
\item a loop of NCFOR2 iterations
\index{NCFOR2 keyword parameter}
of explicit angle-dependent solutions of the transfer equation, Eq. (320) or (321)
of Paper~II.
\end{itemize}
\end{itemize}
\end{itemize}

\begin{description}
\item[NCFOR1] -- see above\\
DEFAULT: NCFOR1=0
\item[NCFOR2] -- see above\\
DEFAULT: NCFOR2=1
\item[NCCOUP] -- see above\\
DEFAULT: NCCOUP=0
\item[NCITOT] -- see above\\
DEFAULT: NCITOT=1
\item[NCFULL] -- see above\\
DEFAULT: NCFULL=1
\end{description}

\subsection{Additional convection parameters}
\label{nst2_conv}

Here we describe several less important, or rarely used keyword parameters
for treating convection.

\begin{description}

\item[ILGDER] -- a switch for the numerical representation
\index{ILGDER keyword parameter}
of the mid-grid point values  and for evaluating the logarithmic gradient $\nabla$:\\ [2pt]
$\bullet\,$ $=0$ -- the mid-point values between depth points 
$d$ and $d-1$ (for instance of the temperature), and the logarithmic
gradient are computed ass
\begin{equation}
T_{d-1/2} = (T_d +T_{d-1})/2, \quad {\rm and} \quad
\nabla_d \equiv \nabla_{d-1/2} =
\frac{T_d - T_{d-1}}{P_d-P_{d-1}} \frac{P_d+P_{d-1}}{T_d + T_{d-1}}.
\end{equation}
$\bullet\,$ $\not= 0$
\begin{equation}
T_{d-1/2} = \sqrt{T_d T_{d-1}}, \quad {\rm and} \quad
\nabla_d  = \ln (T_d /T_{d-1}) / \ln (P_d /P_{d-1})
\end{equation}
DEFAULT: ILGDER=0

\item[IPRESS] --  a flag for treating the total pressure in the convection
\index{IPRESS keyword parameter}
zone:\\
$\bullet\,= 0$ -- total pressure is held fixed when evaluating the
derivatives of the convective flux;\\ 
$\bullet\,= 1$ -- the derivatives w.r.t. the total pressure are calculated.\\
DEFAULT: IPRESS=0

\item[IPRINT] --  a flag that controls a diagnostic output for models with 
\index{IPRINT keyword parameter}
convection:\\
$\bullet\,= 0$ -- no additional output, only the final model is printed;\\
$\bullet\,= 1$ -- the convective flux and the results of routine 
CONCOR are printed after each iteration.\\
DEFAULT: IPRINT=0

\item[ITMCOR] -- if set to a non-zero value, it switches on an old procedure 
\index{ITMCOR keyword parameter}
(now essentially obsolete) 
that correct the temperature in the convection zone if the convective flux, 
corresponding to the newly determined temperature and the logarithmic gradient
$\nabla$,  is larger than the total flux. If so, the temperature is modified
by an iterative procedure for determining new temperature that yields
the convective flux that does not exceed $\sigma T_{\rm eff}^4$.\\
DEFAULT: ITMCOR=0

\item[IDCONZ] -- a parameter for artificially switching off convection
\index{IDCONZ keyword parameter}
for upper layers, where the radiation pressure contribution may
lead to spurious convection. IDCONZ indicates the depth index above
which (i.e., for depth index $d \leq$ IDCONZ), the convection is switched 
off regardless of the values of the actual and the adiabatic gradient.\\
DEFAULT: IDCONZ=31

\item[DERT] -- the value of $\Delta T/T$ for a numerical evaluation of
\index{IDERT keyword parameter}
derivatives of the convective flux with respect to temperature. It takes
effect only in the Rybicki scheme (if IFRYB $>0$). (Notice that when
a standard scheme is used, the value of $\Delta T/T$ is hardwired at
0.001.)\\
DEFAULT: DERT=0.01

\end{description}


\subsection{State equation and thermodynamic quantities}
\label{nst2_eos}

Here is a summary of keyword parameters, useful mostly for
testing purposes:

\begin{description}
\item[IIRWIN] -- a mode of using the Irwin partition functions tables.
\index{IIRWIN keyword parameter}
In any case, they are used only for $T\leq 16000$ K.\\
$\bullet\,=0$ -- Irwin tables are only used when molecules are taken
into account in the equation of state (IFMOL $>0$);\\
$\bullet\,>0$ -- Irwin tables are used by default (even for IFMOL=0);\\
DEFAULT: IIRWIN=0 

\item[IFTENE] -- a mode of evaluating the internal 
\index{IFTENE keyword parameter}
energy -- see Paper~II, Appendix B3;\\
$\bullet\,\leq 1$ -- the internal energy only contains the ionization energies;\\
$\bullet\,>1$ -- the internal energy also contains contributions from the
partition functions, $d\ln U/d\ln T$.\\
DEFAULT: IFTENE=0

\item[IFENTR] -- a mode of evaluation of the the specific heat and
\index{IFENTR keyword parameter}
the adiabatic gradient\\
$\bullet\,=0$ -- standard evaluation through the $T, P$, and internal energy;\\
$\bullet\,>0$ -- an evaluation through the $T,P$, and entropy.\\
DEFAULT: IFENTR=0

\end{description}


\subsection{Details of accretion disks}
\label{nst2_disk}

As shown in Paper~II, \S\,2.2.2,
the (generally depth-dependent) kinematic viscosity $w$ is allowed to vary as a step-wise 
power law of the column mass density, viz. 
\begin{equation}
\label{visc1}
w(m) = w_0 \left( {m/m_0} \right)^{\zeta_0}\, , \quad m>m_{\rm d} \, ,
\end{equation}
\begin{equation}
\label{visc2}
w(m) = w_1 \left( {m/m_0} \right)^{\zeta_1}\, , \quad m<m_{\rm d} \, ,
\end{equation}
where $m_{0}$ is the column mass at the mid-plane, and $m_{\rm d}$ is the 
column mass at the so-called division point, which is an input parameter.
With this parametrization, we allow for a different power-law exponent for inner and
outer layers. This represents a generalization
of an approach we used previously, based on a single power-law representation

There are thus four independent parameters: the power law
exponents $\zeta_0$ and $\zeta_1$, the division point, $m_{\rm d}$; and $f$,
the fraction of energy dissipated in deep layers where $m>m_{\rm d}$. 
The coefficients
$w_0$ and $w_1$ are derived from the condition on the vertically averaged
viscosity, 
$ \int_0^{m_0} w(m) dm/m_0 = \lbar w $, and
$ \int_{m_{\rm d}}^{m_0} w(m) dm/m_0 = f \, \lbar w $. We obtain
\begin{equation}
\label{w0}
w_0 = {f \, \lbar w (\zeta_0 + 1) \over 1 -(m_{\rm d}/{m_0})^{\zeta_0 + 1} }\, ,
\end{equation}
\begin{equation}
\label{w1}
w_1 = {(1-f)\, \lbar w (\zeta_1 + 1) \over 
(m_{\rm d}/{m_0})^{\zeta_1 + 1}}\, .
\end{equation}
Generally, $w(m)$ does not have to be continuous at the division point
$m_{\rm d}$. If we require a continuity, then $f$ and $m_{\rm d}$ are no 
longer two independent parameters; instead, they are related through
\begin{equation}
\label{mdiv}
{m_{\rm d}\over m_0} = \left( 1 + {\zeta_0 + 1 \over \zeta_1 + 1 } 
{f \over 1 - f} \right)^{-{1 \over \zeta_0 + 1}} \, .
\end{equation}
Typically, the deep-layer power law exponent $\zeta_0$ is set to 0 
(constant viscosity), while the ``surface'' power law exponent $\zeta_1$ is 
sometimes set to a value larger than zero.

In {\sc tlusty}, parameter $\zeta_0$ is set by the optional parameter ZETA0;
parameter $\zeta_1$ is set by the optional parameter ZETA1; 
parameter $f$ is set by the optional parameter FRACTV; 
and the division point $m_d$ is set by the optional parameter DMVISC;.
\begin{description}

\item[ZETA0] -- the viscosity parameter $\zeta_0$\\
\index{ZETA0 keyword parameter}
DEFAULT: ZETA0=0 (viscosity is constant in the inner layers).

\item[ZETA1] -- the viscosity parameter $\zeta_1$\\
\index{ZETA1 keyword parameter}
DEFAULT: ZETA1=0 (viscosity is constant in the outer layers).

\item[FRACTV] -- the viscosity parameter $f$; i.e. the fraction of energy
\index{FRACTV keyword parameter}
dissipated in inner layers (with the power-law exponent $\zeta_0$);\\
$\bullet\,> 0$ -- the value of $f$ (must obviously be between 0 and 1)\\ 
$\bullet\,$  $<$ 0 -- the kinematic viscosity is assumed to be a continuous function
of depth, and FRACTV is computed through the division mass DMVISC;\\
DEFAULT: FRACTV=-1

\item[DMVISC] -- the division mass $m_0$ in the viscosity prescription.
\index{DMVISC keyword parameter}
(expressed as $m_0/m_{\rm tot}$, where $m_{\rm tot}$ is the column mass 
at the disk midplane).\\
DEFAULT: DMVISC=0.01 (i.e. 1\%\ of the total column mass is considered
as ``outer layers'').

\item[IZSCAL] -- a switch determining whether the basic depth scale
\index{IZSCAL keyword parameter}
is given by the column mass, $m$, or a geometrical distance, $z$.\\
$\bullet\,= 0$ -- the basic scale is the $m$-scale. In this case, the total
column mass at the mid-plane is computed in the LTE-grey model (or 
read from the input model), and subsequently held fixed.\\
$\bullet\,= 1$ -- the basic scale is the $z$-scale. It is not a recommended
option.\\
DEFAULT: IZSCAL=0

\item[IFZ0] -- this keyword has two functions: (i) it is a switch for treating 
\index{IFZ0 keyword parameter}
the lower boundary condition for the radiative transfer equation for disks; 
and (ii) it indicates a number of global iterations for which the $z$-scale
is being recalculated in the formal solution. This number is determined
by the absolute value of IFZ0.\\
$\bullet\,\geq 0$ -- lower boundary condition is a symmetry boundary condition
given by Eq. (52) of Paper~II;\\
$\bullet\,<0$ -- lower boundary condition is represented by the diffusion
approximation, Eq. (5) of Paper~II.\\
DEFAULT: IFZ0=9 (but reset to -1 for atmospheres)

\item[IBCHE] -- a mode of treating the upper boundary condition of the 
\index{IBCHE keyword parameter}
vertical hydrostatic equilibrium equation:\\
$\bullet\,=0$ -- the boundary condition is the same as in the case of stellar
atmospheres;\\
$\bullet\,=1$ -- the boundary condition is in the form derived specifically for
disks, expressed as -- see Paper~II, \S\,\refgraydisk,
\begin{equation}
m_1 = H_g \rho(z_1) f[(z-H_r)/H_g],
\end{equation}
where $f(x)=(\sqrt\pi/2)\exp(x^2)\, {\rm erfc}(x)$, and $H_g$ and $H_r$ are the
gas pressure scale height and radiation pressure scale height, respectively.\\
$\bullet\,=2$ -- the same expression as above, but a different (older) variant of 
its linearization, namely with the scale heights $H_g$ and $H_r$ held fixed.
This is an outdated option.\\
DEFAULT: IBCHE=1

\item[ICOMGR] -- a switch for a treatment of Compton scattering in the
\index{ICOMGR keyword parameter}
LTE-gray starting model\\
$\bullet\,=0$ -- the Compton scattering is neglected in the LTE-gray model;\\
$\bullet\,=1$ -- the Compton scattering is taken into account in evaluating the
LTE-gray model, using the procedure described in Paper~II, \S\,\refgraydisk, 
Eqs. (185) -- (186).\\
DEFAULT:  ICOMGR=0

\end{description}


\subsection{Additional, approximately described physical processes}
\label{nst2_addphys}

Here we describe several processes that are not a main emphasis of {\sc tlusty},
but can be included in an approximate way.

\subsubsection{Approximate partial frequency redistribution}
\label{nst2_prd}

There are two keywords that control the adopted treatment of partial redistribution:
\begin{description}
\item[IFPRD] -- switch for treating the four lines with approximate partial
\index{IFPRD keyword parameter}
redistribution;\\
$\bullet\,=0$ -- no partial redistribution, all lines are treated with complete redistribution;\\
$\bullet\,>0$ -- partial redistribution in all four lines considered by {\sc tlusty} 
(L$\alpha$, Mg I and Mg II resonance lines) is switched on.\\
DEFAULT: IFPRD=0
\item[XPDIV] -- division frequency (in units of Doppler width) for the
\index{XPDIV keyword parameter}
partial coherent scattering approximation.\\
DEFAULT: XPDIV=3.
\end{description}

\subsubsection{Wind blanketing}
\label{nst2_winbl}

This approximation, introduced by
Abbott and Hummer (1985), and slightly modified by Voels et al. (1988),
mimics an influence of atmospheric layers outside the uppermost depth point, presumably
a stellar wind, by a frequency-dependent albedo that specifies the portion of
radiation reflected back due to the wind. It attains values between 0 and 1.

This approach is relatively obsolete and is kept mostly for historical reasons.
Here are the parameters that sets the treatment of the wind albedo:
\index{IWINBL keyword parameter}
\begin{description}
\item[IWINBL] -- a switch indicating whether the wind-blanketing albedo is considered.\\
$\bullet\,= 0$  --  wind blanketing is not considered;\\
$\bullet\,> 0$  --  wind blanketing is considered, basically as in
   Abbott and Hummer (1985), slightly modified after Voels et al. (1988) 
   to treat properly the angle-averaged albedos;\\
$\bullet\,= -1$ -- wind blanketing is not considered \\ 
DEFAULT: IWINBL=$-1$ 

\item[ALBAVE] -- frequency-integrated wind blanketing albedo for constructing
\index{ALBAVE keyword parameter}
a starting LTE-gray model.\\
DEFAULT: ALBAVE=0 (no wind blanketing)

\end{description}


\section{Keywords for additional numerical options}
\label{nst2_num}

In this chapter we describe a number of keyword parameters that
control various aspects of the numerical setup of model atmosphere
calculations.


\subsection{Working with opacity tables}
\label{nst2_optab}

As stated in \S\,\ref{nonst_glob}, using the opacity table is switched on by the keyword
IOPTAB. The table has the following structure:\\
$\bullet\,$ 1st record: number of frequencies (NUMFREQ), temperatures (NUMTEMP),  and densities (NUMRHO)\\
$\bullet\,$ 2nd record: vector of logarithms of temperatures (TEMPVEC)\\
$\bullet\,$ 3rd record: vector of logarithms of densities (RHOVEC)\\
$\bullet\,$ NUMFREQ blocks containing: the frequency, and subsequently\
NUMRHO sub-blocks of NUMTEMP values of log(opacity).

The maximum number of temperatures and densities is given by parameters
MTABT and MTABR, set in the INCLUDE file {\tt BASICS.FOR} to 50.
This is also the number of actual temperatures and densities in the current
table {\tt absopac.dat}. The maximum number of frequencies, MFRTAB, is also
specified in {\tt BASICS.FOR}. In the standard setup, where the opacity
table is not used, MFRTAB is set to 1 to save memory, so when the 
opacity table is being used, this number should be so to at least the current
number of frequencies in the table, such as MFRTAB=30000 (which
is the statement commented out in the distributed file {\tt BASICS.FOR};
the actual distributed {\tt absopac.dat} has NUMFREQ=28706.

The opacity table is distributed as an ASCII file, but there is a simple
program ({\tt optabtr.f}) that transfers it to the binary format. Reading 
the binary opacity table is much faster, so when computing many models
it is advantageous to use the binary format. The format is
controlled by the keyword IBINOP.

The opacity table is read at the beginning of calculations. The first step
is to interpolate the opacity from the internal frequencies of the table to
the frequency points actually used, for each pair of temperature and density
given by TEMPVEC and RHOVEC. This avoids the necessity of repeated
interpolations in frequency during the calculations, so one interpolates only
in temperature and density. \\ [-8pt]

There are several additional options:
\begin{description}
\item[IBINOP]  -- a switch for the format of the opacity table.\\
$\bullet\,=0$ -- opacity table is in the ASCII format;\\
$\bullet\,=1$ -- opacity table is in the binary format.\\
DEFAULT: IBINOP=1

\item[IFRSET] -- a switch for resetting the frequency points
used in the linearization (which are otherwise given by the positional
parameter NFREAD and several keyword parameters such as FRCMAX,
described in \S\,\ref{nonst_freq}) when using the opacity table.\\
$\bullet\,=0$ -- frequency points are not reset, and are given as specified
independently of the opacity table;\\
$\bullet\,>0$ -- the frequency points used in the linearization are reset to the 
tabular frequencies. No interpolation in frequency is thus needed.\\
DEFAULT: IFRSET=0

\end{description}

 
\subsection{Details of the formal solution}
\label{nst2_for}
 
The {\em formal solution}, as discussed in physical terms in Paper~II, \S\,\refformal,
and Appendix B,
is a set of calculations between two consecutive iterations of the global 
linearization scheme. It can influence the performance of the code
significantly. In the code, it is driven by subroutine {\tt RESOLV}, which
calls a number of additional subroutines. The formal solution has several
basic steps:

(i) Computing new components of the state vector after completing
of a previous global linearization iteration (subroutine {\tt INITIA}); 

(ii) update of temperature (in particular when convection is taken into 
account);

(iii) update of density and pressure (and the vertical distance from the
midplane in case of disks);

(iv) update of level populations and radiation field.\\ [2pt]
\noindent
Steps (ii) - (iv) may be iterated several times, and steps (ii) - (iii) may
be missing altogether for certain models.

Below, we describe these steps and the corresponding keyword parameters that
control a detailed setup of the global formal solution.

\subsubsection*{(i)  Computing new components of the state vector}
By default, this is done, as explained in Paper~II, \S\,\refcl,
\begin{equation}
\label{clcorr}
\psi_{di}^{(n)} = \psi_{di}^{(n-1)} + \delta\psi_{di}^{(n)},
\end{equation}
where the superscript denotes the iteration number. This can be done for all
components of the vector $\psi$; however, it was found already in the early
days of computing NLTE model atmospheres that in some cases it is
better to use a different approach. This is controlled by the following
keyword parameters:

\begin{description}
\item[IFPOPR] -- a switch for treating the calculation of new populations just
\index{IFPOPR keyword parameter}
after completing an iteration of the linearization scheme.\\
$\bullet\,= 0$ -- the original Auer-Mihalas scheme: after a completed linearization
iteration, new populations (i.e. those obtained as 
$n^{\rm new} = n^{\rm old} + \Delta n$) are not used; instead one uses
a new radiation field to compute new radiative rates, and the populations
are determined by solving the rate equations. The option is kept for
historical reasons; it is only useful for the pure complete linearization
scheme.\\
$\bullet\,>$ 0 -- the populations directly coming from linearization are used.
The individual values of IFPOPR switch on a different setup; again, these
have only historical meaning; there is virtually no practical reason to 
change the default value.\\
DEFAULT: IFPOPR=4
\item[ITEMP] --  a flag for evaluating the  ``new'' temperature when 
\index{ITEMP keyword parameter}
convection is taken into account:\\
$\bullet\,= 0$ --  new temperature is calculated as
  $T^{\rm new}=T^{\rm old}+\delta T$ (i.e., exactly as it is done without convection);\\
$\bullet\,= 1$ -- new temperature is calculated through the logarithmic gradient
$\nabla$ in the convection zone;\\
$\bullet\,= 2$ -- new temperature is calculated through $\nabla$ everywhere.\\
DEFAULT: ITEMP=0
\end{description}

\subsubsection*{(ii) Update of temperature} 

It should be kept in mind that the temperature is the critical quantity of a
model, on which the other state parameters depend sensitively. 
If the temperature is seriously wrong at some points in the atmosphere,
it is likely that the whole procedure based on a linearization scheme will diverge,
or at least converge slowly.
A care should therefore be taken to assure that the ``new'' temperature is
as consistent with other state parameters, and behaves as smoothly, as possible.
There are two types of correction procedures, (i) numerical smoothing, and
(ii) physical corrections, based on the solution of an energy balance equation.
The first category is controlled by the following flag:
\begin{description}
\item[IOSCOR] -- if set to a non-zero value, it invokes a procedure for 
\index{IOSCOR keyword parameter}
smoothing a possible oscillatory behavior of the new temperature after a completed 
iteration of the linearization scheme (subroutine {\tt OSCCOR}).
The absolute value of IOSCOR specifies the index of the deepest depth point
at which is the correction procedure performed.
The routine finds the limiting depth indices of the region where the oscillations
occur, $d_{\rm min}$ and $d_{\rm max}$, and sets the new temperature
in this region by interpolating in logarithms,
\begin{equation}
\log T_{d}^{\rm new} = \log T_{d_{\rm min}} + 
\frac{\log(T_{d_{\rm max}}/T_{d_{\rm min}})}
{\log(m_{d_{\rm max}}/m_{d_{\rm min}})}\, \log(m_{d}/m_{d_{\rm min}}),
\end{equation}
$\bullet\,$ if $<0$, then in addition the program finds the minimum temperature,
and resets the local new temperature for all depths above the minimum (lower
depth indices) to the minimum temperature.\\
DEFAULT: IOSCOR=0
\end{description}
There is a number of flags controlling the update of temperature in the
convection zone; they are described in detail in \S\,\ref{nonst_conv} and
\S\,\ref{nst2_conv}.

\subsubsection*{(iii) Update of density and pressure}

There are several flags in this category; their action differs depending on
whether one computes a model atmosphere or a vertical structure of a disk.
\begin{description}
\item[IHECOR] -- a mode of treating the total particle density $N$ during
\index{IHECOR keyword parameter}
the global formal solution:\\
$\bullet\,= 0$ -- the total particle density is held fixed;\\
$\bullet\,> 0$ -- the total particle density is recalculated by solving the
hydrostatic equilibrium equation with the current values of the other state
parameters ($T$, radiation intensities).\\
DEFAULT: IHECOR=0
\item[IHESO6] -- if set to a non-zero value, it activates a special
\index{IHESO6 keyword parameter}
procedure (subroutine {\tt HESOL6}) in the formal solution for a simultaneous
solution for six variables -- $P$, $P_{\rm gas}$, $\rho$, $N$, $n_{\rm e}$,
and $z$ by solving the hydrostatic equilibrium equation, the definitions
of $P$, $P_{\rm gas}$, and $\rho$, and the $z$-$m$ relation -- see Paper~II, 
Appendix B4. It operates only for disks.\\
DEFAULT: IHESO6=0
\end{description}

\subsubsection*{(iv) Update of level populations and radiation field}

As discussed in Paper~II, \S\,\refformal, this is the most important step of the
global formal solution that lies at the very heart of the NLTE approach.
This step is sometimes called the ``restricted NLTE problem'', which
represents a simultaneous solution of the kinetic equilibrium and the
radiative transfer equations to determine the level populations and the radiation
intensities, with other state parameters held fixed. In the original
approach one used a simple Lambda iteration procedure, while
in the present version one uses a more efficient procedure based on
the ALI scheme with preconditioning -- see Paper~II, Appendix B1.

We summarize here the corresponding keyword parameters that
control this procedure. Parameter NLAMBD was already described in
\S\,\ref{nonst_discr}.
\begin{description}
\item[NLAMBD] -- number of  iterations for solving the coupled radiative
\index{NLAMBD keyword parameter}
transfer and kinetic equilibrium equation of the global formal solution.
Historically, these were ordinary Lambda iterations, hence the name NLAMBD.
The current default procedure is instead the ALI scheme with preconditioning.\\ 
DEFAULT: NLAMBD=2 for NLTE models; NLAMBD=1 for LTE models

\item[CHMAXT] --  a parameter that enables to change the number of
\index{CHMAXT keyword parameter}
iterations of the global formal solution (given by NLAMBD) when the model 
is almost converged. If the maximum of the absolute values of the relative changes 
of temperature at all depths decreases below CHMAXT,
the number of iterations is set to NLAMT. \\
DEFAULT: CHMAXT=0.01

\index{NLAMT keyword parameter}
\item[NLAMT] --  the reset number of ``Lambda" iterations -- see above.\\
DEFAULT: NLAMT=1

\item[IFPREC] -- a flag for treating the preconditioning of the statistical
\index{IFPREC keyword parameter}
equilibrium equations (after Rybicki \& Hummer 1991):\\
$\bullet\,= 0$ -- no preconditioning;\\
$\bullet\,= 1$ -- a diagonal (local) preconditioning, as described
in Paper~II, Appendix B1, is switched on;\\
$\bullet\,> 1$ -- a tri-diagonal (non-local) preconditioning is switched on
(this option is not supported in version 205, and thus should not
be used).\\
DEFAULT: IFPREC=1

\item[IACPP] -- a switch for invoking the Ng acceleration of the preconditioned formal
\index{IACPP keyword parameter}
solution:\\
$\bullet\,= 0$ -- no acceleration;\\
$\bullet\,> 0$ -- acceleration is done first in the IACCP-th iteration
         of the formal solution, and is repeated every IACDP iterations;\\
 Notice that if IACCP $>$ NLAMBD (total number of iterations of
  the formal solution), then no acceleration is performed.\\
DEFAULT: IACPP=7 

\item[IACDP] --  step for the Ng acceleration of the formal solution 
(see above).\\
\index{IACDP keyword parameter}
DEFAULT: IACDP=4

\item[IELCOR] -- a flag for turning off an iterative update of the
\index{IELCOR keyword parameter}
electron density by solving iteratively a non-linear system of the kinetic equilibrium
equations and the charge conservation equation.
IELCOR has the meaning of the serial number of the global iteration
till which is this procedure performed.\\
DEFAULT: IELCOR=100 (i.e. ELCOR is called always)

\end{description}


\subsection{Additional parameters for LTE-gray models}
\label{nst2_ltegr}

These are mostly obsolete parameters, kept for downward
compatibility, and for possible comparisons with other modeling
approaches and codes.
\begin{description}
\item[TSURF] -- a mode of evaluating the surface 
\index{TSURF keyword parameter}
temperature:\\
$\bullet\,= 0$  -- the surface temperature ($T_0$) and the Hopf function 
($q_0$) are evaluated 
exactly, that is, $T_0 = [(3/4) q_0)]^{1/4} T_{\rm eff}$, with $q_0= 1/\sqrt{3}$,
so $T_0=0.8112\, T_{\rm eff}$;\\
$\bullet\,> 0$  -- the value of surface temperature is set to TSURF, and
 the Hopf function is assumed to be constant, corresponding to TSURF
 (this has only a pedagogical significance).\\
DEFAULT: TSURF=0.

\item[ALBAVE] -- frequency-integrated wind blanketing 
\index{ALBAVE keyword parameter}
albedo;\\
DEFAULT: ALBAVE=0 (no wind blanketing)

\item[DION0] -- the initial estimate of the degree of ionization at
\index{DION0 keyword parameter}
the first depth point (=1 for completely ionized; =1/2 for completely 
neutral).\\
DEFAULT: DION0=1.

\item[NDGREY] -- the number of depth points for evaluating 
\index{NDGREY keyword parameter}
the LTE-grey model.\\
$\bullet\,$ if = 0 -- NDGREY is set to ND\\
DEFAULT: NDGREY=0

\item[IDGREY] -- a mode of determining the mass-depth scale to be used
\index{IDGREY keyword parameter}
in the subsequent linearization:\\
$\bullet\, =  0$ -- the depth grid $m_d$ (in g cm$^{-2}$) is evaluated as a column mass
 corresponding to the Rosseland optical depths which
 are equidistantly spaced in logarithms between
 the first point TAUFIR and the last point TAULAS[
 the last-but-one point is, however, set to TAULAS$-1$.\\
$\bullet\,=  1$ -- similar, but now $m_d$ is evaluated as the mass
 corresponding to the Rosseland optical depths which
 are equidistantly spaced in logarithms between
 the first point TAU1 and the last-but-one point
 TAU2; the last point is TAUL [with TAU1, TAU2, and TAUL are additional
 input parameters, read at the end of the standard input (unit 5) file.].
 This option is similar to IDGREY=0, but now TAU1 and TAUL may
  be different from TAUFIR and TAULAS. This option is now obsolete.\\
DEFAULT: IDGREY=0

\item[IHM] -- if non-zero, the negative hydrogen ion is considered in the particle 
\index{IHM keyword parameter}
and charge conservation when constructing an LTE-gray model;\\
DEFAULT: IHM=0

\item[IH2] -- if non-zero, the hydrogen molecule is considered in the particle 
\index{IH2 keyword parameter}
and charge conservation when constructing an LTE-gray model;\\ 
DEFAULT: IH2=0

\item[IH2P] -- if non-zero, the ionized hydrogen molecule is considered in the
\index{IH2P keyword parameter}
particle and charge conservation when constructing an LTE-gray model \\ 
DEFAULT: IH2P=0

\end{description}


\subsection{Setup of the radiative transfer equation
and $\Lambda^\ast$ operator}
\label{nonst_rte2}

Here we describe some additional parameters for the setup of the transfer
equation and the evaluation of the approximate $\Lambda^\ast$ operator.
These are rarely used, and if so then mostly for testing and comparison
purposes. Their default values usually yield the best results.

\begin{description}

\item[NELSC] -- a  mode of treating the electron scattering by the Feautrier
\index{NELSC keyword parameter}
scheme (it has no meaning for the DFE scheme, ISPLIN$=5$):\\
$\bullet\,= 0$ -- the electron scattering source function is treated
exactly; i.e. an single-dependent transfer equation contains an explicit
angular coupling due to the $J_\nu$-dependence of the
electron scattering source function;\\
$\bullet\,> 0$ -- the electron scattering source function is treated as
a thermal source function, i.e. it is given through
the current mean intensity $J_\nu$.
It is included for pedagogical and testing purposes only.\\
DEFAULT: NELSC=0

\item[DJMAX] -- the maximum relative change of the mean intensity in
\index{DJMAX keyword parameter}
the internal ALI iteration loop for treating electron scattering in
the case of the DFE formal solution. It has an effect only if ISPLIN=5.\\
DEFAULT: DJMAX=0.001

\item[NTRALI] -- the maximum number of iterations of 
\index{NTRALI keyword parameter}
the internal ALI iteration loop for treating electron scattering in
the case of the DFE formal solution. It has an effect only if ISPLIN=5.\\
DEFAULT: NTRALI=3

\item[ILMCOR] -- a mode of including the electron scattering contribution
\index{ILMCOR keyword parameter}
to the approximate Lambda operator $\Lambda^\ast$:\\
$\bullet\,= 0$ --  the $\Lambda$ operator is defined to act on the source function 
$S=\eta/(\kappa +\sigma)$, i.e. $J = \Lambda[\eta/(\kappa +\sigma)]$;
here $\eta$ and $\kappa$ are the thermal emission and absorption coefficients,
and $\sigma$ is the scattering coefficient, $\sigma=n_{\rm e}\sigma_{\rm e}$,
with $\sigma_{\rm e}$ being the electron scattering cross section.\\
$\bullet\,= 1$ --  the $\Lambda$ operator is defined to act
on the thermal source function 
$S^{\rm th}=\eta/\kappa$, i.e. $J = \Lambda[\eta/\kappa]$. \\
DEFAULT: ILMCOR=1

\item[ILPSCT] --  a mode of including the electron scattering correction
\index{ILPSCT keyword parameter}
in the preconditioning scheme:\\
$\bullet\,= 0$ --  the $\Lambda$ operator is defined to act on the 
thermal source function;\\
$\bullet\,= 1$ --  the $\Lambda$ operator is defined to act on 
the source function of the form $S=\eta/(\kappa +\sigma)$.\\
DEFAULT: ILPSCT=0

\item[ILASCT] -- a mode of including the electron scattering correction
\index{ILASCT keyword parameter}
in the evaluation of the derivative of the source function with respect
to the state parameters ($T, n_{\rm e}$, and populations -- subroutine
{\tt ALIFR1}):\\
$\bullet\,= 0$ --  the $\Lambda$ operator is defined to act on the 
thermal source function;\\
$\bullet\,= 1$ --  the $\Lambda$ operator is defined to act on the 
source function of the form $S=\eta/(\kappa +\sigma)$.\\
DEFAULT: ILASCT=0

\item[IBC] --  mode of the treatment of the $\Lambda^\ast$ operator at the 
\index{IBC keyword parameter}
lower boundary:\\ 
$\bullet\,= 0$ --  $\Lambda^\ast$  at depth points ND and ND$-$1 is given by 
$J_\nu/S_\nu$\\
$\bullet\,> 0$ --  $\Lambda^\ast$  at depth points ND and ND$-$1 is computed
exactly;\\
$\bullet\,= 3$ --  in addition, all appropriate derivatives in the linearization are 
calculated exactly.\\
DEFAULT: IBC=3

\end{description}


\subsection{Setup of the kinetic equilibrium equations}
\label{nonst_ese2}

Similarly to \S\,\ref{nonst_rte2}, we list here several
less used keyword parameters that control the setup of the kinetic
equilibrium equation; the more important parameters were already described in 
\S\,\ref{nonst_ese}.

\begin{description}

\item[MODREF] -- a flag for setting up the reference levels
\index{MODREF keyword parameter}
of the individual explicit species.
Recall that the reference level is the energy level for each species
for which the kinetic equilibrium equation is replaced by
the abundance definition equation.
The indices of the reference levels are stored in the array 
NREF(IAT), IAT=1,NATOM.\\
$\bullet\,= 0$ --
  NREF(IAT) is set to NKA(IAT), i.e. the highest
  ionization state of the species IAT.\\
$\bullet\,= 1$ --
  NREF(IAT) is determined by the program to be the
  index of ground level of the most populated ion of the
  species IAT.\\
$\bullet\,= 2$ --
  NREF(IAT) is set to index of the ground state of the second highest
  ionization state of the species IAT.\\
DEFAULT: MODREF=1

\item[IFLEV] -- a switch for globally changing the mode of treating the
\index{IFLEV keyword parameter}
 linearization of atomic level populations:\\
 $\bullet\,= 0$ -- the mode is specified by the positional parameter IMODL
(\S\,\ref{ion_en}), and is not changed;\\
$\bullet\,> 0$ -- the mode is reset, for all levels except the highest
 ionization stage, to IMODL=1, i.e. the updated LTE
 mode.\\
DEFAULT: IFLEV=0 for NLTE; IFLEV=1 for LTE models

\item[ICHC] -- switch for selecting the closing equation for the set of 
\index{ICHC keyword parameter}
kinetic equilibrium equations:\\
$\bullet\,= 0$ -- the closing equation is the abundance definition equation;\\
$\bullet\,= 1$ -- the closing equation is the charge conservation equation.\\
DEFAULT: ICHC=0

\item[IRSPLT] -- a switch for the mode of solution of the global system 
\index{IRSPLT keyword parameter}
of the kinetic equilibrium equations:\\
$\bullet\,= 0$ -- kinetic equilibrium equations for all species are solved
simultaneously (with one big rate matrix);\\
$\bullet\,= 1$ -- kinetic equation is solved for one species at a time
(i.e. the big rate matrix is split into partial rate matrices for the
individual chemical species).\\
DEFAULT: IRSPLT=1

\item[POPZR2] -- a secondary parameter for setting up a level 
\index{POPZR2 keyword parameter}
zeroing.  \\
DEFAULT: POPZR2=1.e-20

\end{description}


\subsection{Auxiliary, mostly obsolete, parameters}
\label{nst2_aux}

\begin{description}

\item[IFMOFF] -- a flag for considering the modified free-free
\index{IFMOFF keyword parameter}
cross section for all explicit ions even if the atomic data files 
do not specify it. If set
to a non-zero value, the characteristic quantum number is set
to the main quantum number of the highest level of an ion + 1.\\
DEFAULT:  IFMOFF=0
 
\item[IOVER] -- a flag for turning on the "line-overlapping" 
\index{IOVER keyword parameter}
mode:\\
$\bullet\,= 0$ -- no overlapping is allowed for (only one line may
contribute to the opacity at any single frequency), This option is kept
for historical reasons only;\\
$\bullet\,> 0$ -- a general line overlap is allowed.\\
DEFAULT: IOVER=1

\item[ITLAS] -- a flag for turning off laser lines, i.e. those for which 
\index{ITLAS keyword parameter}
 the absorption coefficient (= true absorption minus                    
 stimulated emission) becomes negative.
 Turning off laser lines means that the line absorption and
 emission coefficient are set to zero at depths where the
 absorption coefficient would be negative.
 ITLAS has the meaning of the global iteration number 
 starting from which the laser lines are turned off
 (ITLAS=0 turns off laser lines from the very beginning).\\
DEFAULT: ITLAS=100

\item[IBFINT] --   a mode of storing the photoionization 
\index{IBFINT keyword parameter}
cross sections:\\
$\bullet\,= 0$ -- means that the cross sections are stored for all frequency points;\\
$\bullet\,= 1$ -- means that the photoionization cross sections are 
 stored only for continuum frequencies, and are interpolated 
 for line frequencies;\\
DEFAULT: IBFINT=1

\item[IRDER] --  a mode of treatment of linearization of the kinetic
\index{IRDER keyword parameter}
equilibrium equations in the hybrid CL/ALI scheme: \\
$\bullet\,= 0$ -- the rate equations in the CL/ALI scheme are not linearized;\\
$\bullet\,> 0$ -- the rate equations in the CL/ALI scheme are linearized; there 
are several variants of treating specific derivatives, which are of historical
significance as they were used for testing purposes;\\
$\bullet\,= 3$ -- full linearization; all derivatives are calculated exactly.\\
DEFAULT: IRDER=3

\item[ILDER] --  a flag for controlling the evaluation of derivatives of
\index{ILDER keyword parameter}
recombination rates with respect to temperature.
Introduced for testing purposes only.\\
$\bullet\,= 0$ -- derivatives are calculated;\\
$\bullet\,> 0$ -- derivatives are set to zero.\\
DEFAULT: ILDER=0

\item[IBPOPE] -- a flag for controlling the derivatives of the rate equations.
Introduced for testing purposes only.  \\
$\bullet\,= 0$ -- derivatives of the rows of rate equations with
\index{IBPOPE keyword parameter}
respect to the mean intensity in the linearized frequency points are set to zero.\\
$\bullet\,> 0$ -- derivatives are calculated.\\
DEFAULT: IBPOPE=1

\item[DPSILG] -- during linearization, the relative changes of all
\index{DPSILG keyword parameter}
state parameters are artificially limited not to exceed certain values.
DPSILG sets up a general limit for all quantities, in a sense that
$\delta(\psi_i)/\psi_i = \max[1/{\rm DPSILG}-1, \delta(\psi_i)/\psi_i]$, and
$\delta(\psi_i)/\psi_i = \min[{\rm DPSILG}-1, \delta(\psi_i)/\psi_i]$;\\
DEFAULT: DPSILG=10. (i.e. all the relative changes are truncated to
have values between $-0.9$ and 9.)

\item[DPSILT] -- analogous, but limits specifically the relative
changes in temperature. If DPSILT $<$ DPSILG, DPSILT overwrites
DPSILG.\\
DEFAULT: DPSILT=1.25
\index{DPSILT keyword parameter}

\item[DPSILN] -- analogous, but for the relative changes of electron
density.\\
DEFAULT: DPSILN=10.
\index{DPSILN keyword parameter}

\item[DPSILD] -- analogous, but for the logarithmic gradient $\nabla$
(effective only if convection is switched on, ICONV $\neq$ 0.)\\
DEFAULT: DPSILD=1.25
\index{DPSILD keyword parameter}

\end{description}


\subsection{Collisional-radiative switching}
\label{nonst_crsw}

This is an obsolete procedure suggested originally by Hummer \& Voels (1988),
somewhat modified by considering a depth-dependent switching parameter.
It consists in artificially multiplying the radiative rates by a factor CRSW, which
satisfies CRSW$\,<1$, and gradually increase this factor until it reaches unity. 
This procedure may lead to a greater stability of the kinetic equilibrium equations.
The actual setup is controlled by the following keywords:
\begin{description}
\item[ICRSW] -- switch for turning on the collisional-radiative 
\index{ICRSW keyword parameter}
switching;\\
$\bullet\,=0$ -- collisional-radiative switching is not considered;\\
$\bullet\,>0$ -- collisional-radiative switching is considered.\\
DEFAULT: ICRSW=0
\item[SWPFAC] sets the 
\index{SWPFAC keyword parameter}
initial CRSW=SWPFAC $\times$ min(collis.rate/rad.rate)\\
DEFAULT: SWPFAC=0.1
\item[SWPLIM] -- a limiting factor for CRSW; in the sense 
\index{SWPLIM keyword parameter}
that\\
if CRSW $>$ SWPLIM, then CRSW = 1,\\
DEFAULT: SWPLIM=0.001
\item[SWPINC] -- an increment factor for determining new CRSW, 
\index{SWPINC keyword parameter}
namely\\
CRSW(actual) = CRSW(previous) $\times$ SWPINC\\
DEFAULT: SWPINC=3.

\end{description}


\subsection{Parameters determining the amount of additional output}
\label{nonst_print}

\begin{description}

\item[IPRIND] -- if set to a non-zero value, a condensed model atmosphere
\index{IPRIND keyword parameter}
is stored after each iteration of complete linearization (Unit 17)
-- see \S\,\ref{out_aux}\\
DEFAULT: IPRIND=0

\item[ICHCKP] -- if set to a non-zero value, an additional output
\index{ICHCKP keyword parameter}
showing the total transition rates in and out of all explicit levels
is generated (Unit 16) -- see \S\,\ref{out_aux}.\\
DEFAULT: ICHCKP=0

\item[IPRINP] -- if set to a non-zero value, the explicit level
\index{IPRINP keyword parameter}
populations are stored in output unit 7 (condensed model atmosphere)
even for LTE models (the populations are always stored for a NLTE
model, so in NLTE this parameter has no effect).\\
DEFAULT: IPRINP=1

\item[IPOPAC] -- if set to a non-zero value, an additional output
\index{IPOPC keyword parameter}
of useful quantities for all continuum frequencies is generated.\\
$\bullet\,=1$ -- stores the opacity for all continuum frequencies as a function of depth
(Unit 85)\\
$\bullet\,=2$ -- stores the absorption, scattering, emission coefficients, and the
mean intensity for all continuum frequencies, as a function of depth -- see \S\,\ref{out_aux}.\\
DEFAULT: IPOPAC=0

\item[ICOOLP] -- if set to a non-zero value, the net total cooling
\index{ICOOLP keyword parameter}
rate is stored as a function of depth (Unit 87); 
if moreover it is set to a vale $> 10$, also the individual
cooling rates for all explicit ions are also stored (Unit 88)
 -- see \S\,\ref{out_aux}.\\
DEFAULT: ICOOLP=0

\end{description}


\section{User-supplied routines}
\label{userdef}

To help the user to implement his/her own expressions for various atomic 
parameters, we have set up a  number of dummy routines that are prepared
to accept the user supplied expressions. In this way the user does not have 
to worry about where exactly to place the desired expression, and about 
possible indirect effects of the new piece of code. Also, the user does not
have to understand the detailed structure of the code.

There are four such routines which are described below.
\subsection{{\tt SPSIGK} -- non-standard photoionization cross sections}
\label{spsigk}

The routine is called with the following parameters:
\begin{verbatim}
     SUBROUTINE SPSIGK(IB,FR,SIGSP)
\end{verbatim}
The input parameters are: 
\begin{itemize}
\item{\tt IB} is the switch given by the
input parameter IFANCY (renamed to IBF), which has to be negative to
activate this routine  -- see \S\,\ref{ion_bf},. The actual value of IB invokes 
one particular expression or a hardwired table. We have already used the
following values to invoke some special formulae or tables:
\begin{itemize}
\item IB $=-602$ -- an outdated formula for the He I ground state;
\item IB $=-202$ -- another outdated formula for the He I ground state;
\item IB $=-602$ -- special table of cross section for C I $2p^2\,{}^1\!D$ level
(provided by G.B. Taylor, priv. communication);
\item IB $=-603$ -- special table of cross section for C I $2p^2\,{}^1\!S$ level
(G.B. Taylor, priv. communication);
\item between $-101$ and $-137$ -- cross section after Hidalgo (1968).
In this case, $-$IB$-100$ is the Hidalgo's index;
\item between $-301$ and $-337$ -- cross section after Reilman \& Manson (1979).
Again, $-$IB$-300$ is their index.
\end{itemize}
Any other values are free and available for the user.
\item {\tt FR} is the frequency [in s${}^{-1}$],
\end{itemize}
The output is {\tt SIGSP} [in cm ${}^{-2}$].

\subsection{{\tt GAMSP} -- non-standard Voigt damping parameters}
\label{gamsp}

The routine is called with the following parameters:
\begin{verbatim}
      SUBROUTINE GAMSP(ITR,T,ANE,AGAM)
\end{verbatim}
The subroutine is called if the input parameter GAMAR (see \S\,\ref{ion_bb})
is coded negative. The input parameters are: 
\begin{itemize}
\item {\tt ITR} is the index of the transition,
\item $T$ is the temperature [K], 
\item {\tt ANE} is the electron density [cm ${}^{-3}$].
\end{itemize}
The output is {\tt AGAM} the Voigt parameter, $a$, where
$a\equiv\Gamma/(4\pi \Delta\nu_D)$, where $\Gamma$ is the (physical)
damping parameter, and $\Delta\nu_D$ is the Doppler width.

\subsection{{\tt CSPEC} -- non-standard collisional rates}
\label{cspec}

The routine is called with the following parameters:
\begin{verbatim}
      SUBROUTINE CSPEC(I,J,IC,OS,CP,U0,T,CS)
\end{verbatim}
The input parameters are:
\begin{itemize}
\item {\tt I,J} -- indices of the lower and the upper level of the transition;
\item {\tt IC} -- a collisional switch, given by the input parameter ICOL,
which must be set negative to invoke this routine -- see \S\,\ref{ion_bb}.
The following values are  already used:
\begin{itemize}
\item IC$=-1$, already mentioned in \S\,\ref{ion_bb}, is reserved for
the van Regemorter formula for neutral atoms;
\item IC$=-2$ -- a usual van Regemorter formula for ions. This is
a redundant option because this formula is invoked by setting ICOL=0
without a need to call this special routine.
\end{itemize}
\item {\tt OS} -- oscillator strength; given by the input parameter OSC0;
\item {\tt CP} -- collisional parameter, given by the input parameter CPARAM --
see \S\,\ref{ion_bb};
\item {\tt U0} -- the value of $U_0=h\nu_0/kT$;
\item {\tt T} -- temperature [K].
\end{itemize}
The output is {\tt CS}, the collisional rate $\Omega$ -- see \S\,\ref{ion_bb}.

\subsection{{\tt PFSPEC} -- non-standard partition functions}
\label{pfspec}

The routine is called with the following parameters:
\begin{verbatim}
      SUBROUTINE PFSPEC(IAT,IZI,T,ANE,U,DUT,DUN)
\end{verbatim}
The subroutine is called if the input parameter MODPF (see \S\,\ref{new_at})
is set to a negative value.
The input parameters are:
\begin{itemize}
\item {\tt IAT} -- atomic number;
\item {\tt IZI} -- ionization stage (IZI=1 for neutrals, etc.);
\item {\tt T} -- temperature [K];
\item {\tt ANE} -- electron density [cm ${}^{-3}$].
\end{itemize}
The output is {\tt U}, the partition function, and its derivatives with respect
to temperature and electron density, respectively ({\tt DUT} and {\tt DUN});
the derivatives are usually set to 0.


\section{Summary of useful tricks; troubleshooting}
\label{tricks}

{\sc tlusty} can employ a variety of numerical and physical tricks that
aim at speeding up calculations, avoiding convergence 
problems, or introducing minor physical simplifications that do not
deteriorate the quality of the resulting model. These approaches 
prove useful in some cases, but in some other cases they may harm the
convergence or even lead to additional numerical problems.
Consequently, they should not be used blindly.

The keyword parameters that switch on/off such options
are set to their default values and may be changed by setting an appropriate
value in the keyword parameter file. The choice of defaults was
made based on experience gained from many different models, and typically
a default value was found successful in most cases. However, there
are cases where those default values have to be changed to improve the
convergence properties of the run. 

We will summarize the tricks and their appropriate control keyword(s) 
together with a brief discussion below.  We stress that we describe 
here only computational options. Obviously, the model atmosphere
itself as well as its convergence properties may be significantly
influenced by the global choice of explicit species, energy levels,
and general atomic data. The topic of atomic data and their choice is
discussed in detail in Chap. \ref{ions}

This chapter will present an explanation of many optional
keyword parameters introduced above. The previous chapters presented
a formal explanation; the present one will concentrate more on
practical aspects, in particular on the role of the individual
parameters in troubleshooting and dealing with convergence problems.

There are essentially four categories of such parameters;
(i) flags for invoking purely numerical tricks; (ii) parameters
that influence the accuracy of a model; (iii) parameters that introduce
minor, but helpful physical approximations; and (iv) parameters
that switch on/off a potentially important, but still optional,
physics. We will discuss these categories below.


\subsection{Purely numerical tricks}
\label{tricks_num}

These procedures are defined as that their use does not
change the accuracy of the resulting model. In other words, using or not using
an option, provided that both options converge, would lead to
the same numerical results (that is, essentially the same results;
the rounding errors and other computer-related inaccuracies may
lead to differences at 4th and higher decimal place of the computed
model parameters).
\begin{itemize}

\item {\em Setup of the hybrid CL/ALI scheme}\\
This is partly set up by the atomic data files, where one uses default
settings  -- the first six frequencies in the hydrogen Lyman continuum;
the three central frequencies in the L$\alpha$ line, and the central
frequency in L$\beta$ and L$\gamma$ lines, together with the first three
points in the He II Lyman continuum are set to linearized mode;
the rest of the frequencies are in the ALI mode.\\
There is also a useful keyword IFRALI (see \S\,\ref{nonst_rte}) 
\index{IFRALI keyword parameter}
which can change the setup globally.

\item {\em Rybicki scheme}\\
Setting this scheme is in a large majority cases preferable for LTE
models. Surprisingly, it can also be used cautiously for NLTE models
in case of convergence problems. As discussed in Paper~II, \S\,\refryb, the
scheme is not designed for computing full NLTE models, but can
serve as a means to compute intermediate NLTE model from which the final
model may be converged more easily. An actual example of using the Rybicki 
scheme in the context of NLTE models was shown in \S\,\ref{examp_cwd}.

\item {\em Accelerations}\\
There are two basic acceleration schemes: Ng and Kantorovich,
see Paper~II, \S\,\refaccel. 
The corresponding keywords are described in \S\,\ref{nonst_acc}. 
Both accelerations are helpful tools to speed up calculations
considerably, or even sometimes to improve convergence properties,
but on the other hand they may cause problems in some cases. Therefore,
they should be used with caution.

As demonstrated in \S\,\ref{trouble_conv},
one should be careful about the parameter IACC, which represents 
\index{IACC keyword parameter}
the iteration
number where the Ng acceleration is done for the first time. If the
value is too low and the acceleration is done too soon, the consequence
is often a harmful or disastrous effect upon the convergence. If it is done
too late it does not do what it could, namely speed up the
calculations. In some cases the Ng acceleration is crucial even for
the success of the run; it is when the normal linearization produces
undamped oscillations, so that only the Ng acceleration is able to find the
way out of otherwise non-convergent situation. The best strategy is
to examine carefully the convergence log (file {\tt fort.9}); if
the convergence pattern deteriorates drastically exactly at the iteration
number when the Ng acceleration was switched on, a simple cure is to increase
the value of IACC. If, in contrast, one sees an oscillatory
behavior for many iterations, and just at the onset of Ng acceleration
the maximum change suddenly drops, it is advisable to try to decrease
the value of IACC.

In the case of convection, Ng acceleration may often be harmful;
in this case one may skip the acceleration completely, i.e. to set
IACC to a value larger than NITER.

An analogous warning can be expressed about the Kantorovich scheme,
whose onset is controlled by the keyword parameter ITEK. Turning it on too
soon may slow down the convergence; or lead to divergence,
turning it on too late would not
do any harm to the convergence rate, but increases the execution time
significantly. Again, it is very dangerous to use it for convective models,
unless starting it late in the iteration process when the position of the
convection zone does not change any more, but in such a case it does
not lead to a significant reduction of the computer time. Therefore, 
a safe strategy is to turn it off completely for convective models.

\item {\em Level grouping}\\
This numerical trick may significantly reduce the number
of level populations to be linearized -- see Paper~II, \S\,\refgroup. We remind that
a level group is a set of several levels whose populations are 
assumed to vary in a coordinated way in the linearization. More precisely, 
instead of linearizing the individual level populations, one
linearizes the total populations of the groups, assuming that the
ratios of the individual level populations within the group to the
total population of the group is unchanged in the linearization. In the
formal solution step, one solves for all the individual level
populations exactly. 
The grouping is set up by the atomic data files (parameter IMODL;
\index{IMODL positional parameter}
see Sect. \ref{ion_en}). The standard files,
for instance those distributed through the {\sc tlusty} website, contain 
a default level grouping. This grouping works well in most cases. 
However, the user should be aware that under some circumstances
(e.g., low temperature), this grouping may cause convergence problems,
and should be reset by changing parameter IMODL, typically by letting 
more low-lying levels to form a single-level group.

\item {\em Choice of the radiative transfer formal solver}\\
This is driven by keyword ISPLIN (see \S\,\ref{nonst_rte}). 
\index{ISPLIN keyword parameter}
Usually, the original
2nd-order Feautrier scheme (ISPLIN=0; the default) works very well.
One exception is the case of a strong external irradiation;
in this case setting ISPLIN=5 (the DFE scheme) may be preferable.

\item {\em Number of iterations of the global formal
solution (``Lambda" iterations)}\\
This is driven by keyword NLAMBD. The default value (NLAMBD=2) usually works
\index{NLAMBD keyword parameter}
well; in some cases one may increase this number, which increases the
computer time, but may provide a more stable scheme. An actual example
is shown in \S\,\ref{trouble_conv_nlte}.

\item {\em Choice of the approximate $\Lambda^\ast$-operator}\\
This is driven by keywords IFALI (see \S\,\ref{nonst_rte}) -- 
\index{IFALI keyword parameter}
diagonal or tridiagonal operator,
and JALI (Rybicki-Hummer or Olson-Kunasz evaluation). While the
\index{JALI keyword parameter}
value of JALI usually does not matter, in some cases a tridiagonal
operator (set up with IFALI=6 - notice that the default is IFALI=5, 
corresponding to the
diagonal operator) may provide a more stable solution. However,
at present, this option is not yet fully supported, and may not work
properly in some cases.

\index{IATREF keyword parameter}
\index{MODREF keyword parameter}
\item {\em Choice of the so-called ``reference atom'' and 
the reference levels}\\
This is driven by keyword IATREF and MODREF (see \S\,\ref{nonst_ese}
and \S\,\ref{nonst_ese2}). 
Usually, the default values work well. However, for highly unusual 
chemical compositions, in particular for extremely hydrogen-deficient
objects, it is advisable to set IATREF=2 (so that He will be the 
reference species).

\item {\em Limiting the values of the relative changes to be used to the obtain
next iterates of the structural parameters}\\
This is driven by keywords DPSILT, DPSILG, DPSILN, DPSILD -- see
Sect. \ref{nst2_aux}. Usually, the default values work well.

\end{itemize}


\subsection{Changing the accuracy of a model}
\label{tricks_accu}

There are parameters In this category that are originally meant
to influence the accuracy of a model without significantly changing
its general properties (e.g., the number of depth points). However,
these parameters may in certain cases significantly influence the
global convergence properties.

\begin{itemize}

\index{ND keyword parameter}
\item {\em Number of depth points}\\
The default value (ND=70) is usually satisfactory. In some cases 
(for instance if some quantities vary very sharply with depth), 
it may be helpful to increase this value.
On the other hand, for hot stars it is recommended to decrease
the value to ND=50, which saves computer time and memory without 
a decrease of the accuracy of the model. In fact, this was used in our recent
O-star and B-star model grid calculations -- Lanz \& Hubeny (2003, 2007).

\item {\em Setting up the minimum and maximum Rosseland optical
depth for a model} (only if the model is computed from scratch)\\
This is driven by the keywords TAUFIR and TAULAS (see \S\,\ref{nonst_gr}).
\index{TAUFIR keyword parameter}
\index{TAULAS keyword parameter}
The default values usually work well; in special cases one may need
to increase TAULAS or to decrease TAUFIR. For instance, to produce
X-ray spectra of, say, a DA white dwarf, one has to set TAULAS to
a very large value ($\approx 10^4$ or more) because hydrogen provides
very little opacity in the X-ray region and thus the atmosphere
is very transparent there. Another example are strongly irradiated companions
of hot stars, where the external irradiation may penetrate deep into the atmospheres
of the companions. Again, one may need to set TAULAS to a larger value.

\item {\em Convergence criterion}\\
It is driven by the keywords NITER and CHMAX (see \S\,\ref{nonst_discr}). 
\index{NITER keyword parameter}
\index{CHMAX keyword parameter}
We recall that NITER represents
the maximum number of the global linearization iterations.
If set to zero, the program only computes the radiation field and prints
the output information. This is useful for instance if some output files
of the previously computed model (but not the resulting model file
on unit 7) were accidentally deleted; or for some exploratory
calculations to test a basic setup without the goal to produce a converged
model.

\item {\em Treatment of the iron-peak lines}\\
This is controlled by keywords ISPODF, DDNU, STRLX, CNU1, CNU2,
CHMAXT, and JIDS - see \S\,\ref{nonst_blank}. These parameters
influence typically the accuracy of a model (and, obviously, 
the computer time needed to generate a model), but usually not
its convergence properties.

\end{itemize}


\subsection{Introducing some helpful minor physical approximations}
\label{tricks_appr}

Here, we consider physical approximations that either shorten
the computer time, or help to prevent convergence problems.
In some cases, these tricks lead to negligible deficiencies in the
computed model; in some cases (e.g., setting detailed
radiative balance for too many lines) they would provide a model
of insufficient accuracy. In such a case, the option is useful
to compute an intermediate model, starting from which one may compute
a more realistic model that would not converge otherwise.

From the point of view of troubleshooting, the following parameters
are the most important ones:
\begin{itemize}

\item {\em Division point between the differential and integral form of
the radiative equilibrium equation}\\
This is driven by the keywords TAUDIV, IDLST, and NDRE -- see 
\S\,\ref{nonst_re}. The most important parameter is TAUDIV --
\index{TAUDIV keyword parameter}
\index{IDLST keyword parameter}
\index{NDRE keyword parameter}
the division point for treating energy balance
as a linear combination of the differential and integral form (see
Hubeny \& Lanz 1995). In the past, changing this value
is often {\em the} decisive trick to make LTE models converge. 

It should be stressed that when the Rybicki scheme is used, it usually
eliminates most of the problems with the radiative (or radiative/convective) 
equilibrium equation, and the value
of TAUDIV is not so critical; in fact, it can be set to a small value
(say 0.01). Also, in this case it is mandatory to set IDLST=0.

If the Rybicki scheme is not used, there is
no a priori recipe to choose the best value of TAUDIV.
The most reasonable strategy is to try two values of both
sides of the default (which is 0.5), say 0.05 and 5 or so. Experience
shows that for cool stars ($T_{\rm eff} < 10,000$ K) setting
TAUDIV=5 usually helps. In contrast, for accretions disks the
value TAUDIV=0.05 is sometimes more advantageous.
However, the user should be aware of the fact that this trick may
lead to numerical inaccuracies. When a linear combination
of the differential and the integral forms of the radiative equilibrium
equation is satisfied, it
does not necessarily mean that both are satisfied individually.
Specifically, if Eq. (16) of Paper~II is satisfied exactly, 
\begin{equation}
\label{relc}
\alpha\bigg[\int_0^\infty\!\!\left(\chi_\nu J_\nu - \eta_\nu^{\rm tot}\right)d\nu\bigg] +
\beta \bigg[\int_0^\infty \frac{d(f_\nu J_\nu)}{ d\tau_\nu}\,d\nu -
\frac{\sigma_R }{ 4\pi}\,T_{\mathrm{eff}}^4\bigg] = 0,
\end{equation}
one can still have
\begin{equation}
\label{rela}
\alpha\bigg[\int_0^\infty\!\!\left(\chi_\nu J_\nu - \eta_\nu^{\rm tot}\right)d\nu\bigg] 
= -\epsilon,
\end{equation}
and
\begin{equation}
\label{relb}
\beta \bigg[\int_0^\infty \frac{d(f_\nu J_\nu)}{ d\tau_\nu}\,d\nu -
\frac{\sigma_R }{ 4\pi}\,T_{\mathrm{eff}}^4\bigg] = \epsilon,
\end{equation}
so that radiative equilibrium is in fact {\em not satisfied}.
The higher the value is set for TAUDIV, the larger the errors could be. 
If the user sets a value larger than the default, it is advisable
to recompute the converged model with the default value of TAUDIV.
In other words, increasing TAUDIV to obtain an intermediate model 
may enable to converge a model that would not converge directly with
the default value of TAUDIV. To verify that the model is properly converged,
one has to inspect the last table of the standard output (Unit 6) for the
conservation of the total flux.

\item {\em Artificial lowering of the total radiative pressure at the
surface}\\
Radiative acceleration increases in the superficial layers 
($\tau_{\rm Ross} < 10^{-5}$), and may overcome the gravitational 
acceleration. The model may become numerically unstable in these 
layers, although physically stable at larger optical depths.
The option XGRAD=0 (see \S\,\ref{nonst_misc}) allows to auto-limit 
\index{XGRAD keyword parameter}
the radiative acceleration to some fraction
of the gravity in the ten uppermost layers. This is the default approach.
The resulting model is not exact at the surface layers (in fact, it cannot be
because one does not include a wind anyway), but is very reasonably
accurate at deeper layers where most of the observed spectral features form.
This option is therefore extremely useful.

Setting XGRAD=-1., or -2., allows to impose a more stringent cut-off in the 20 
uppermost layers, which may critically help the convergence near the 
Eddington limit.
XGRAD $> 0$ auto-limits the radiative acceleration to fraction XGRAD 
of the gravity acceleration $g$ {\em everywhere}
in the atmosphere. Potentially, this may lead to an incorrect
atmospheric structure, so this option should be used only with
extreme caution.

\item {\em Fixed temperature at a surface layer}\\
This is set up by the parameter NRETC (see \S\,\ref{nonst_misc}).
\index{NRETC keyword parameter}
This option is useful if the model does not converge at 
the upper layers,
while the convergence is reasonable elsewhere. In this case one may
set up, for instance, NRETC=10 (that is, the temperature at the first 10
depth points will be held fixed to the values given by the input
model). Once one obtains a converged solution, it is recommended
to re-converge the model with NRETC=0 (the default), i.e. with
the radiative equilibrium solved exactly at all depths. 

\item {\em Setting up the minimum and maximum frequency}\\
This is driven by the parameters FRCMIN and FRCMAX -- see 
\S\,\ref{nonst_freq}. In some cases it may be advantageous to obtain
\index{FRCMIN keyword parameter}
\index{FRCMAX keyword parameter}
an intermediate model by lowering the maximum frequency FRCMAX,
because high frequencies may cause numerical problems due to
high sensitivity of mean intensities to the state parameters,
in particular to temperature.

\item {\em Setting up frequency points in the high-frequency tail}\\
This is driven by parameters NFTAIL, DFTAIL, and CFRMAX - see
\S\,\ref{nonst_freq}. The default values usually work well;
in some cases, typically with Comptonization (i.e. for very hot
atmospheres or, in particular, disks), one may consider more frequencies
in the high-energy tail, that is, to increase NFTAIL.

\item {\em Setting a detailed radiative balance in lines}\\
In many cases, this trick helps to provide an intermediate model
if other tricks fail. This topic was discussed at length in Chap.\,\ref{trouble}.

\item {\em Level zeroing}\\
This is a useful trick that may avoid some floating-point
problems. The idea is as follows (see Paper~II, \S\,\refgroup):
Whenever a local population of a level becomes lower
than a prescribed fraction of the total population of the species
(given by the parameter POPZER; with a default value $10^{-20}$), 
\index{POPZER keyword parameter}
the population is set exactly
to zero, and instead of considering an appropriate rate equation for 
such a level, one replaces it by a simple condition $n_i=0$. 
This does not decrease
the number of state parameters, but improves the numerical stability
without compromising the final solution.

However, a level may also be super-zeroed, if it is zeroed in all depth
points. In that case it is permanently removed from the set of explicit
levels and thus the number of state parameters is indeed reduced.
This option is very useful for instance for hot accretion disks, where one can
originally set all ions of a species, for example Fe I to Fe XXVII, 
and let the code decide which are going to be removed based on the actual
physical conditions.

\item {\em Zeroing of the mean intensity}\\
Analogously to level zeroing, one may set the mean intensity of radiation to zero
if the value of the mean intensity decreases below a prescribed
fraction of the maximum mean intensity. This fraction is
given by the optional parameter RADZER, with a default value $10^{-20}$.

\item {\em Keeping some selected structural parameters fixed}\\
This is driven by the keywords described in \S\,\ref{nonst_lin}.
Any of these options may be useful to compute an intermediate
model in the case when convergence cannot be achieved with all structural
parameters being linearized. Usually, keeping just the temperature fixed 
(INRE=0) will provide an intermediate converged model; in some cases
it is better to also keep the electron density and the total particle
density fixed; in this case one may set the shortcut parameter
IFIXMO=1 -- see also \S\,\ref{trouble_conv_nlte}.
\index{IFIXMO keyword parameter}

\end{itemize}


\subsection{Optional, but potentially important physical processes}
\label{tricks_phys}

This category of tricks allows one to remove or include some major
physical mechanism (e.g. convection, Compton scattering, etc.) that
should be either present always, or should be uniquely given
by the considered model parameters, but are left to the discretion
and judgment of the user whether they are included or not.
The reason is that in an intermediate case between a completely
negligible and almost negligible influence of a given mechanism
its inclusion or rejection may significantly influence the convergence
properties of a model.

\begin{itemize}

\item {\em Convection}\\
The parameters for switching on convection are described
in \S\,\ref{nonst_phys} (HMIX0, MLTYPE); and the parameters for 
controlling numerical details are in \S\,\ref{nonst_conv}, and \S\,\ref{nst2_conv}.
If the convection is weak, it may be advantageous
to converge first a model without convection; 
however usually it is better to include convection from the outset.
There may still be problems for very cool stars; we are working
on a more robust scheme for treating convection.

\item {\em Additional opacity sources}\\
This is driven by the parameters IRSCT, IOPHMI, IOPH2P --
see \S\,\ref{nst2_ray} and \S\,\ref{nst2_hhe}.
\index{IRSCT keyword parameter}
This option was useful in the past where only a limited amount
\index{IOPHMI keyword parameter}
of explicit atoms, ions, and levels could be considered. The utility
\index{IOPHMI keyword parameter}
of the so-called additional opacities is diminished now, but there are
still cases where it is useful. For instance, one may simply switch on
a H$^-$ opacity even if the H$^-$ ion is not an explicit ion (however,
treating H$^-$ as an explicit ion is preferable because this approach
can account for NLTE effects in H$^-$). Another examples are the Rayleigh
\index{Rayleigh scattering}
scattering opacity (IRSCT),  H$_2^+$ opacity (IOPH2P), and the Collision-Induced
Absorption (CIA) opacity (parameter IFCIA). In the two
latter cases, setting IOPH2P or IFCIA to non-zero values is the
only available mechanism in {\sc tlusty} how to switch on these opacities.

\index{Compton scattering}
\item {\em Compton scattering}\\
Compton scattering  is important for very hot atmospheres and,
in particular, for accretion disks. It is switched on by setting the
parameter ICOMPT to a non-zero value. The equations, procedures,
\index{ICOMPT keyword parameter}
and corresponding keyword parameters are described in Paper~II,
Appendix A6 and C3, and here in \S\,\ref{nst2_compt}.
Computing a model with Compton scattering is typically slower than
without it, but hot models are obviously more accurate when it is included, 
and in some cases it can even improve the convergence properties and the stability 
of the numerical scheme. For very hot models, where all species
are essentially fully ionized, it is even mandatory to include the Compton
scattering because it becomes the dominant agent for establishing the
radiative equilibrium.

\item {\em Microturbulence and turbulent pressure}\\
Their treatment is driven by the parameters VTB and IPTURB -- 
see \S\,\ref{nonst_phys}. The value of VTB does not usually influence
\index{VTB keyword parameter}
\index{IPTURB keyword parameter}
the convergence properties of the model. On the other hand, parameter
IPTURB, which controls whether the assumed microturbulence 
will be associated with a corresponding turbulent pressure, may
influence the hydrostatic equilibrium and thus the convergence
properties. In case of a high
turbulence (close to, or larger than, the thermal velocity),
a safer and perhaps
more physical option is to set IPTURB=0, in which case there
is no turbulent pressure associated with the assumed turbulent 
velocity.
\end{itemize}


\section{List of keyword parameters}
\label{list}

Here we give the list of all keyword parameters in alphabetic order,
including their default values, and the reference to the section where they
are described in detail.

\begin{tabbing}
keyword \ \ \ \ \ \  \= default value \ \ \ \ \ \ \ \=  described in \\ [3pt]
ABPLA0   \>   0.3   \>   \ref{nonst_gr}\\
ABPMIN   \>   $10^{-5}$    \>  \ref{nonst_gr}  \\
ABROS0   \>   0.4    \>  \ref{nonst_gr}   \\
ADIST   \>   0.  \>  \ref{exter}   \\
ALBAVE   \>   0    \>  \ref{nst2_winbl}  \\
ALPHAV   \>   0.1     \>  \ref{nonst_accr}  \\
BERGFC   \>   0.   \>  \ref{nst2_occup}  \\
CFRMAX   \>   2 (0 for disks)   \> \ref{nonst_freq}  \\
CHMAX    \>   $10^{-3}$    \>  \ref{nonst_discr}  \\
CHMAXT   \>   0    \>  \ref{nst2_for}  \\
CNU1     \>   4.5    \>  \ref{nonst_blank}  \\
CNU2     \>   3    \>  \ref{nonst_blank}  \\
CRFLIM   \>   0.7   \>  \ref{nonst_conv}  \\
CUTBAL   \>   0.   \>  \ref{nst2_occup}  \\
CUTLYM   \>   0.   \>  \ref{nst2_occup}  \\
DDNU     \>   0.75   \>  \ref{nonst_blank}  \\
DERT    \>   0.01   \>  \ref{nst2_conv}  \\
DFTAIL   \>   0.25   \>  \ref{nonst_freq}  \\
DION0    \>   1.   \>  \ref{nst2_ltegr}  \\
DJMAX    \>   0.001   \>  \ref{nonst_rte2}  \\
DM1      \>   $10^{-3}$  \>   \ref{nonst_gr}  \\
DMVISC   \>   0.01   \>  \ref{nst2_disk}  \\
DPSILD   \>   1.25   \>  \ref{nst2_aux}  \\
DPSILG   \>   10.   \>  \ref{nst2_aux}  \\
DPSILD   \>   1.25   \>  \ref{nst2_aux}  \\
DPSILN   \>   10.   \>  \ref{nst2_aux}  \\
FRLCOM  \>  8.2E14   \>   \ref{nst2_compt}  \\
FRACTV   \>   -1   \>  \ref{nst2_disk}  \\
FRCMAX   \>   0.   \>  \ref{nonst_freq}  \\
FRCMIN   \>   $10^{12}$   \>  \ref{nonst_freq}  \\
FRLMAX   \>   0.   \>  \ref{nonst_freq}  \\
FRLMIN   \>   $10^{13}$   \>  \ref{nonst_freq}  \\
HCMASS   \>   0   \>  \ref{nonst_wd}  \\
HMIX0    \>   -1.   \>  \ref{nonst_phys}  \\
IACC     \>   7   \>  \ref{nonst_acc}  \\
IACD     \>   4   \>  \ref{nonst_acc}  \\
IACDP    \>   4   \>  \ref{nst2_for}  \\
IACPP    \>   7    \>  \ref{nst2_for}  \\
IATREF   \>   1   \>  \ref{nonst_ese}  \\
IBC      \>   3   \>  \ref{nonst_rte2}  \\
IBCHE      \>   1   \>  \ref{nst2_disk}  \\
IBFINT   \>   1   \>  \ref{nst2_aux}  \\
IBPOPE   \>   1   \>  \ref{nonst_glob}  \\
ICHANG   \>   0   \>  \ref{nonst_mod}  \\
ICHANM   \>   1   \>  \ref{nonst_gr}  \\
ICHC     \>   0   \>  \ref{nonst_ese2}  \\
ICHCKP   \>   0   \>  \ref{nonst_print}  \\
ICHCOO  \>   0   \>  \ref{nst2_compt}  \\
ICOLHN   \>   1   \>  \ref{nst2_colm}  \\
ICMDRA   \>   0   \>  \ref{nst2_compt}  \\
ICOMDE   \>   1   \>  \ref{nst2_compt}  \\
ICOMGR     \>   0   \>  \ref{nst2_disk}  \\
ICOMPT   \>   0   \>  \ref{nonst_glob}, \ref{nst2_compt}  \\
ICOMRT   \>   0   \>  \ref{nst2_compt}  \\
ICOMST   \>   1   \>  \ref{nst2_compt}  \\
CONRE    \>   0   \>  \ref{nonst_conv}  \\
ICONV    \>   0   \>  \ref{nonst_conv}  \\
ICOOLP   \>   0   \>  \ref{nonst_print}  \\
ICRSW    \>   0    \>   \ref{nonst_crsw}   \\
IDCONZ    \>   31   \>  \ref{nst2_conv}  \\
IDEEPC   \>   2   \>  \ref{nonst_conv}  \\
IDGREY   \>   0   \>  \ref{nst2_ltegr}  \\
IDLST    \>   5   \>  \ref{nonst_re}  \\
IDLTE    \>   1000   \>  \ref{nonst_ese}  \\
IDMFIX   \>   1   \>  \ref{nonst_gr}  \\
IELCOR   \>   100   \>  \ref{nst2_for}  \\
IFALI    \>   5   \>  \ref{nonst_rte}  \\
IFCHTR   \>   0   \>  \ref{nst2_colh}  \\
IFCIA   \>   0   \>  \ref{nst2_cia}  \\
IFDIEL   \>   0   \>  \ref{nst2_diel}  \\
IFENTR  \>  0   \>   \ref{nst2_eos}  \\
IFIMXO   \>   0   \>  \ref{nonst_lin}  \\
IFLEV    \>   0 for NLTE   \>  \ref{nonst_ese2}  \\
              \>   1 for LTE   \>  \ref{nonst_ese2}  \\
IFMOFF   \>   0   \>   \ref{nst2_aux} \\
IFMOL    \>   0   \>   \ref{nonst_glob}  \\
IFPOPR   \>   4   \>  \ref{nst2_for}  \\
IFPREC   \>   1   \>  \ref{nst2_for}  \\
IFPRAD    \>   0   \>  \ref{nonst_misc}  \\
IFPRD    \>   0   \>  \ref{nst2_prd}  \\
IFRALI   \>   0   \>  \ref{nonst_rte}  \\
IFRAYL    \>   0   \>  \ref{nst2_ray}  \\
IFRYB    \>   0   \>   \ref{nonst_glob}  \\
IFTENE  \>  0   \>   \ref{nst2_eos}  \\
IFZ0     \>   -1  (9 for disks)   \>  \ref{nst2_disk}  \\
IH2      \>   0   \>  \ref{nst2_ltegr}  \\
IH2P     \>   0   \>  \ref{nst2_ltegr}  \\
IHECOR   \>   0   \>  \ref{nst2_for}  \\
IHESO6   \>   0   \>  \ref{nst2_for}  \\
IHM      \>   0   \>  \ref{nst2_ltegr}  \\
IHYDPR   \>   0   \>  \ref{nst2_hyd}  \\
IIRWIN    \>   0    \>   \ref{nst2_eos}    \\
ILASCT   \>   0   \>  \ref{nonst_rte2}  \\
ILDER    \>   0   \>  \ref{nst2_aux}  \\
ILGDER    \>   0   \>  \ref{nst2_conv}  \\
ILMCOR   \>   1   \>  \ref{nonst_rte2}  \\
ILPSCT   \>   0    \> \ref{nonst_rte2}  \\
IMUCON    \>   0   \>  \ref{nonst_conv}  \\
INDL     \>   0    \>  \ref{nonst_lin}  \\
             \>   3 with convection   \>  \ref{nonst_lin}  \\
INHE     \>   1   \>  \ref{nonst_lin}  \\
INMP     \>   0   \>  \ref{nonst_lin}  \\
INPC     \>   3    \>  \ref{nonst_lin}  \\
             \>   4 with convection   \>  \ref{nonst_lin}  \\
INRE     \>   2   \>  \ref{nonst_lin}  \\
INSE     \>   4    \>  \ref{nonst_lin}  \\
              \>   5 with convection   \>  \ref{nonst_lin}  \\
INTRPL   \>   0   \>  \ref{nonst_mod}  \\
IOPH2P   \>   0   \>  \ref{nst2_hhe}  \\
IOPHE1   \>   0   \>  \ref{nst2_hhe}  \\
IOPHE2   \>   0   \>  \ref{nst2_hhe}  \\
IOPHMI   \>   0   \>  \ref{nst2_hhe}  \\
IOFTAB    \>   0   \>   \ref{nonst_glob}  \\
IOSCOR    \>   0   \>  \ref{nst2_for}  \\
IOVER    \>   1   \>  \ref{nst2_aux}  \\
IPOPAC   \>   0   \>  \ref{nonst_print}  \\
IPRESS   \>   0   \>  \ref{nonst_conv}  \\
IPRIND   \>   0   \>  \ref{nonst_print}  \\
IPRING   \>   0   \>  \ref{nonst_gr}  \\
IPRINP   \>   0   \>  \ref{nonst_print}  \\
IPRINT   \>   0   \>  \ref{nonst_conv}  \\
IPTURB   \>   1   \>  \ref{nonst_phys}  \\
IQUASI   \>   0   \>  \ref{nst2_quas}  \\
IRDER    \>   3   \>  \ref{nst2_aux}  \\
IRSCT    \>   0   \>  \ref{nst2_ray}  \\
IRSPLT   \>   1   \>  \ref{nst2_for}  \\
ISPLIN   \>   0   \>  \ref{nonst_rte}  \\
ISPODF   \>   0   \>  \ref{nonst_glob}, \ref{nonst_blank}  \\
ITEK     \>   4   \>  \ref{nonst_acc}  \\
ITEMP    \>   0   \>  \ref{nst2_for}  \\
ITGMAX   \>  10   \>  \ref{nonst_gr}  \\
ITLAS    \>   100   \>  \ref{nst2_aux}  \\
ITMCOR    \>   0   \>  \ref{nst2_conv}  \\
IVISC    \>   0   \>  \ref{nonst_accr}  \\
IWINBL   \>   $-1$   \>  \ref{nst2_winbl}  \\
IZSCAL   \>   0   \>  \ref{nonst_phys}  \\
JALI     \>   1   \>  \ref{nonst_rte}  \\
JIDS     \>   0   \>  \ref{nonst_blank}  \\
KNISH  \>   0   \>  \ref{nst2_compt}  \\
KSNG     \>   7   \>  \ref{nonst_acc}  \\
MLTYPE   \>   0   \>  \ref{nonst_phys}  \\
MODREF   \>   1   \>  \ref{nonst_ese2}  \\s
NCCOUP   \>   0   \>  \ref{nst2_compt}  \\
NCFOR1   \>   0   \>  \ref{nst2_compt}  \\
NCFOR2   \>   1   \>  \ref{nst2_compt}  \\
NCFULL   \>   1   \>  \ref{nst2_compt}  \\
NCITOT   \>   1   \>  \ref{nst2_compt}  \\
NCONIT   \>   10   \>  \ref{nonst_gr}  \\
ND       \>   70   \>  \ref{nonst_discr}  \\
NDCGAP   \>   2   \>  \ref{nonst_conv}  \\
NDGREY   \>   0   \>  \ref{nst2_ltegr}  \\
NDRE     \>   0   \>  \ref{nonst_re}  \\
NELSC    \>   0   \>  \ref{nonst_rte2}  \\
NFTAIL   \>   21   \>  \ref{nonst_freq}  \\
NITER    \>   30   \>  \ref{nonst_discr}  \\
NITZER   \>   3   \>  \ref{nonst_ese}  \\
NLAMBD   \>   2 (1 for LTE)   \>  \ref{nonst_discr}, \ref{nst2_for}  \\
NLAMT    \>   1   \>  \ref{nst2_for}  \\
NMU      \>   3   \>  \ref{nonst_discr}  \\
NNEWD    \>   0   \>  \ref{nonst_gr}  \\
NQUALP   \>   3   \>  \ref{nst2_quas}  \\
NQUBAL   \>   0   \>  \ref{nst2_quas}  \\
NQUBET   \>   0   \>  \ref{nst2_quas}  \\
NQUGAM   \>   0   \>  \ref{nst2_quas}  \\
NRETC    \>   0   \>  \ref{nonst_re}  \\
NTRALI   \>   3   \>  \ref{nonst_rte2}  \\
NQUALP   \>   3   \>  \ref{nst2_quas}  \\
NQUBET   \>   0   \>  \ref{nst2_quas}  \\
ORELAX   \>   1.   \>  \ref{nonst_acc}  \\
POPZER   \>   $10^{-20}$   \>  \ref{nonst_ese}  \\
POPZR2   \>   $10^{-20}$   \>  \ref{nonst_ese2}  \\
RADSTR   \>   0   \>  \ref{nonst_wd}  \\
RADZER   \>   $10^{-20}$   \>  \ref{nonst_rte}  \\
REYNUM    \>   0.   \>  \ref{nonst_accr}  \\
RSOURC  \>    0.   \>  \ref{exter} \\
SPRFAC  \>  0.667  \>  \ref{exter}  \\
STRL1    \>   0.001   \>  \ref{nonst_blank}  \\
STRL2    \>   0.02   \>  \ref{nonst_blank}  \\
STRLX    \>   $10^{-10}$   \>  \ref{nonst_blank}  \\
SWPFAC    \>   0.1    \>   \ref{nonst_crsw}   \\
SWPINC    \>   1.    \>   \ref{nonst_crsw}   \\
SWPLIM    \>   0.001    \>   \ref{nonst_crsw}   \\
TAUDIV   \>   0.5   \>  \ref{nonst_re}  \\
TAUFIR   \>   $10^{-7}$   \>  \ref{nonst_gr}  \\
TAULAS   \>   316.   \>  \ref{nonst_gr}  \\
TDISK    \>   0    \>  \ref{nonst_gr}  \\
TQMPRF   \>   0   \>  \ref{nst2_quas}  \\
TRAD     \>   0.   \>  \ref{exter}  \\
TSURF    \>   0.   \>  \ref{nst2_ltegr}  \\
VTB      \>   0.   \>  \ref{nonst_phys}  \\
XGRAD    \>   0.   \>  \ref{nonst_misc}  \\
XPDIV    \>   0   \>  \ref{nst2_prd}  \\
WDIL     \>   1.   \>  \ref{exter}  \\
ZETA0    \>   0.   \>  \ref{nonst_accr}, \ref{nst2_disk}  \\
ZETA1    \>   0.   \>  \ref{nonst_accr}, \ref{nst2_disk}  \\
\end{tabbing}


\section{Outlook}
\label{outlook}

{\sc tlusty} has been under a continuous development for over
three decades, and there is no reason to stop it now. We plan to
do various upgrades on different levels to make the code even more 
flexible, efficient, and possibly also more user-friendly.

We will first describe the upgrades or additions as far as the physical 
processes  are concerned, going roughly from low temperatures to high.
\begin{itemize}
\item Including opacity and scattering due to condensates (clouds; dust).
Since these are included in a special variant Cool{\sc tlusty}, this upgrade
is quite straightforward and needs only transporting the corresponding
routines. The necessary tables were already constructed
by Budaj et al. (2014) and made publicly available. 
\item Considering departures from chemical equilibrium for carbon and
nitrogen chemistry, following the approach described in Hubeny \& Burrows
(2007). It is also implemented in Cool{\sc tlusty}, so transporting the 
corresponding routines is straightforward.
\item Developing additional and more sophisticated approaches for treating
convection. Some are already included in Cool{\sc tlusty}, but more are needed
to cope with convergence and stability problems in some cases.
\item There are several new promising approaches to treat convection beyond the
framework of the mixing-length theory. We will explore them, and possibly
implement and test them in {\sc tlusty}.
\item The treatment of occupation probabilities will be extended to treat
also neutral perturbers.
\item There is a theoretical development in progress (Gomez, in prep.)
to replace the present treatment of pseudocontiua with a better and physically
more realistic approach. That would provide a significant improvement
in the accuracy of white dwarfs models.
\end{itemize}
On the high-energy side:
\begin{itemize}
\item Implementing a better treatment of inner-shell transitions.
\item Implementing a treatment of K lines
\item Upgrading the treatment of the Compton scattering to a fully
relativistic case with Klein-Nishina cross section and the actual redistribution
function to go beyond the Kompaneets approximation.
\end{itemize}

On an algorithmic level, we envisage several upgrades to help the
efficiency and stability of the model construction. These include
\begin{itemize}
\item Modifying the Rybicki scheme to linearize with respect to $T$ and $N$,
not just $T$ as it is at present.
\item To extend the current rudimentary treatment of the over-relaxation
acceleration to a more sophisticated approach.
\item To provide an improved flexibility in treating partial opacity tables
in a hybrid approach so that some species are treated as explicit and some
through the opacity tables.
\end{itemize}

As to the global geometrical framework is concerned, we do not plan
to further upgrade the current {\sc tlusty} to treat 3-D effects and/or
dynamical structures, because it requires a completely different philosophy
of modeling. However, we plan to upgrade our 3-D radiation transfer
solver program called IRIS (Ibgui et al. 2013) by transporting all routines from
{\sc tlusty} that describe local physics; in particular the evaluation of opacities,
and the solution of the kinetic
equilibrium equation with ALI and preconditioning, which would  provide
a means to construct NLTE models for snapshots of independent
hydrodynamical calculations, and improved estimate of radiation moments
to better describe the radiation energy density, flux, and pressure in
hydrodynamical calculations.


\section*{Acknowledgements}

Although we have coded the large majority of the code ourselves, we
took and modified several routines from other codes, namely subroutines
written by Larry Auer, Bob Kurucz, late David Hummer, Mats Carlsson, 
Detlev Koester, Gary Ferland and Tim Kallman. 
Several colleagues have written some specific routines
or parts of the routines specifically for {\sc tlusty}. We gratefully acknowledge
contributions from Carlos Allende-Prieto, Martin Barstow, Omer Blaes,
Veronika Hubeny, Alex de Koter, and Mike Montgomery. 
We are indebted to Knox Long for his contribution in polishing the
code and removing outdated features in order to adhere to modern 
FORTRAN standards.

We also acknowledge a contribution from a large number of
users who have pointed out some bugs or encouraged us to
develop various upgrades and improvements. The following list is far
from being complete, but we would like to acknowledge contributions from
Eric Agol, Nicolle Allard, Pierre Bergeron, 
Jano Budaj, Adam Burrows, Pierre Chayer, Katia Cunha, Shane Davis, 
Marcos Diaz, Paul Dobie, 
Stefan Dreizler, Boris G\"ansicke, Thomas Gomez, Stefan Haas, 
Sally Heap, John Hillier, late David Hummer, Detlev Koester, 
Lars Koesterke, Julian Krolik, Chela Kunasz, late Al Linnell, Knox Long, 
late Dimitri Mihalas, late Mirek Plavec, Simon Preval, 
Robert Ryans, Ed Sion, Chantal Stehl\'e, David Sudarsky, Steve Voels, 
Richard Wade, Klaus Werner, and Dayal Wickramasinghe.

Our special thanks go to Knox Long, Yeisson Ossorio, Simon Preval, Klaus Werner,
and, in particular, to Peter Nemeth,
for their careful reading of the manuscript and many useful comments and 
suggestions to the draft of this Guide.

I.H. also gratefully acknowledges the support from the Alexander von Humboldt Foundation,
and wishes to thank especially to Klaus Werner for his hospitality at the Institute of Astronomy 
and Astrophysics of the University of T\"ubingen, where a part of the work on this paper was done. 

Last, but not least, we would like to acknowledge the encouragement
and help of late Fran\c coise Praderie in the early stages of the {\sc tlusty}
development; it may well be that without her influence and encouragement {\sc tlusty}
would not have survived even its early childhood.


\section*{References}
\addcontentsline{toc}{section}{References} 

\def\reference{\par \leftskip20pt \parindent-20pt\parskip4pt}
\noindent
\reference Abbott, D.C., \& Hummer, D.G. 1985, ApJ, 294, 286.
\reference Aldrovandi, S.M.V., \& Pequignot, 1973, A\& A, 25, 137.
\reference Allende Prieto, C., Lambert, D., Hubeny, I., \& Lanz, T. 2003,
ApJS, 147,. 363.
\reference Allard,  N., \& Koester, D. 1992, A\&A, 258, 464.
\reference Arnaud, M., \& Raymond, J., 1992, ApJ, 398, 394.
\reference Auer, L.H., \& Mihalas, D. 1969, ApJ 158, 641
\reference Avrett, E.H., \& Loeser, R. 1982, ASP Conf. Ser.  26, 489.
\reference Bergeron, P., Wesemael, F., \& Fontaine, G. 1991, ApJ, 367, 253.
\reference Bergeron, P., Wesemael, F., \& Fontaine, G. 1992, ApJ 387, 288.
\reference Berrington, K. \& Kingston 1987, J. Phys. B, 20, 6631
\reference Borysow, A., \& Frommhold, L. 1990, ApJ, 348, 41.
\reference Budaj, J., Kocifaj, M., Salmeron, R., \& Hubeny, I. 2015, MNRAS, 454, 2.
\reference Butler, K. 1990, in Accuracy of element abundances from stellar 
atmospheres, Springer, Berlin and New York, p.19.
\reference Carlsson, M. 1986, Uppsala Astron. Obs. Tech. Rep. 33.
\reference Castor, J.I., Dykema, P., \& Klein, R.I., 1992, ApJ 387, 561.
\reference Crandall, D.H., et al., 1974, ApJ 191, 789.
\reference Daeppen, W., Anderson, L.S., \& Mihalas, D.  1987, ApJ, 319, 195.
\reference Eissner, W.; Seaton, M. J. 1972, J. Phys. B, 5, 2187.
\reference Feautrier, P. 1964, C. R. Acad. Sci. Paris., Ser. B, 258, 3189.
\reference Fernley, J. A.; Seaton, M. J.; Taylor, K. T. 1987, J. Phys. B, 32, 5507.
\reference Fontaine, G., Villeneuve, B., \& Wilson, J. 1981, ApJ 343, 550.
\reference Frank, J., King, A. \& Raine, D. 1992, {\it Accretion Power in 
Astrophysics}, 2nd ed., Cambridge Univ. Press, Cambridge.
\reference Giovanardi, C., Natta, A., \& Palla, F., 1987, A\& AS,70, 269.
\reference Grevesse, N., \& Sauval, A. 1998, Space Sci. Rev., 85, 161.
\reference Henry, R.J.W. 1970, ApJ, 161, 1153.
\reference Hidalgo, M.B. 1968, ApJ, 153, 981.
\reference Hubeny, I. 1988, Computer Physics Comm. 52, 103.
\reference Hubeny, I. 1990, ApJ, 351, 632.
\reference  Hubeny, I., Blaes, O., Agol, E., \& Krolik, J.H., 2001,
     ApJ, 559, 680.
\reference Hubeny, I., Burrows, A., \& Sudarsky, D. 2003, ApJ,   594, 1011.   
\reference Hubeny, I., \& Hubeny, V., 1998, ApJ,  505, 558.
\reference Hubeny, I., Hummer, D.G., \& Lanz, T. 1994, A\&A, 282, 151.
\reference Hubeny, I., \& Lanz, T. 1992, A\&A 262, 501.
\reference Hubeny, I., \& Lanz, T. 1995, ApJ, 439, 875.
\reference Hubeny, I., \& Lanz, T. 2011, Synspec: General Spectrum Synthesis Program, Astrophysics Source Code Library, ASC1109.022
\reference Hubeny, I., Lanz, T., \& Jeffery, C.S. 1994, in Newsletter on
   Analysis of Astronomical Spectra No. 20, ed. C.S. Jeffery,
   St. Andrews Univ., p.30.
\reference Hubeny, I. \& Mihalas, D. 2014, {\it Theory of Stellar Atmospheres},
Princeton Univ. Press, Princeton.
\reference Hummer, D.G., \& Mihalas, D. 1988, ApJ  331, 794.
\reference Hummer, D.G., \& Voels, S.A. 1988, A\&A, 192, 279.
\reference Ibgui, L., Hubeny, I., Lanz, T.; \& Stehl\'e, C. 2013 A\&A, 549, 126.
\reference Irwin, A.W. 1981, ApJS, 45, 621.
\reference Kallman, T.R, 2000, {\it XSTAR: A Spectral Analysis Tool}, NASA
Goddard Space Flight Center.
\reference Kingdon, J., \& Ferland, G. 1996, ApJS, 106, 205.
\reference Koester, D. 1985, A\&A 149, 423.
\reference K\v r\'\i\v z, S., \& Hubeny, I., 1986, 
Bull. Astron. Inst. Czechosl., 37, 129.
\reference Krolik, J.H. 1999, {\it Active Galactic Nuclei}, Princeton Univ. Press, Princeton.
\reference Kurucz, R.L. 1970, SAO Spec. Rep. 309.
\reference Kurucz, R.L. 1979, ApJS, 40, 1.
\reference Lanz, T., \& Hubeny, I. 2003, ApJS, 146, 417.
\reference Lanz, T., \& Hubeny, I. 2007, ApJS, 169, 83.
\reference Lemke, M., 1997, A\&AS, 122, 285.
\reference Lynden-Bell, D. \& Pringle, J.E., 1974, MNRAS, 168, 603.
\reference Mihalas, D., 1978, {\it Stellar Atmospheres}, 2nd ed., Freeman,
    San Francisco.
\reference Mihalas, D., Heasley, J.N., \& Auer, L.H., 1975, 
    A Non-LTE Model Stellar Atmospheres Computer Program, NCAR-TN/STR 104.
\reference Mihalas, D., \& Hummer, D.G. 1974, ApJS, 28. 343. 
\reference Novikov, I.D. \& Thorne K.S 1973, in {\it Black Holes}, ed. by.
C. de Witt \& B. de Witt, Gordon \& Breach, New York.
\reference Nussbaumer, H., \& Storey, P.J., 1983, A\& A, 126, 75.
\reference Olson, G.,  \& Kunasz, P. 1987, JQSRT, 38, 325.
\reference Peach, G. 1967, Mem. Roy. Astron, Soc., 71, 13.
\reference Przybilla, N., \& Butler, K., 2004, ApJ, 609, 1181. 
\reference Reilman, R.F., \& Manson, S.T. 1979, ApJS, 40, 815.
\reference Riffert, H., \& Harold, H. 1995, ApJ, 450, 508.
\reference Rybicki, G.B. 1969, JQSRT, 11, 589.
\reference Rybicki, G.B., \& Hummer, D.G., 1991, A\&A 245, 171.
\reference Rybicki, G.B., \& Hummer, D.G., 1992, A\&A 262, 209.
\reference Rybicki, G.B., \& Lightman, A.P. 1979, {\it Radiative Processes in
Astrophysics}, Willey \& Sons, New York.
\reference Seaton, M., editor, 1995, {\it The Opacity Project, Vol. I}, Inst. of Physics
Publishing, Bristol.
\reference Shakura, N. I., \& Sunyaev, R. A. 1973, A\&A  24, 337.
\reference Sharp, C.S., \& Burrows, A. 2007, ApJS, 168, 140.
\reference Sparks \& Fischel 1971, NASA Spec. Rep. 3066.
\reference Sudarsky, D., Burrows, A., Hubeny, I., \& Li, A. 2005, ApJ, 627, 520.
\reference Traving, G., Baschek, B., \& Holweger, H. 1966, Abhand. Hamburg.
C     Sternwarte. Band VIII, Nr. 1.
\reference Tremblay, P.-E., \& Bergeron, P. 2009, ApJ, 696,1755.
\reference Trujillo Bueno, J., \& Fabiani Bendicho, P. 1995, ApJ, 455, 646.
\reference van Regemorter, H. 1962, ApJ, 136, 906.
\reference Vennes, S., Pelletier, C., Fontaine, G., Wesemael, F. 1988, ApJ 331, 876.
\reference Verner, D.A. \& Yakovlev,  D.G. 1995, A\&AS, 109, 125.
\reference Voels, S.A., Bohannan, B., Abbott, D.C., \& Hummer, D.G. 1989, ApJ, 340,
  1073.
\reference Werner, K., 1984, A\&A, 139, 237.

\newpage

\def\noreference{\par \leftskip0pt \parindent0pt\parskip4pt}

\noreference
\section*{List of acronyms}
\addcontentsline{toc}{section}{List of acronyms}

We tried to keep the number of used acronyms at a minimum
because a heavy usage of them is often annoying for a reader.
We only used common acronyms, which we list below.

\begin{tabbing}
ALI  \ \ \ \ \ \   \= Accelerated Lambda Iteration   \\
CFR  \> Complete Frequency Redistribution   \\
CIA  \> Collision-Induced Absorption   \\
CL  \> Complete Linearization   \\
DFE  \> Discontinuous Finite Element (method)   \\
LTE   \>    Local Thermodynamic Equilibrium   \\
ML2  \> modified Mixing Length prescription for convective flux    \\
NLTE  \>  non-LTE (any departure from LTE)   \\
ODF  \> Opacity Distribution Function   \\
OP  \> Opacity Project   \\
OS   \> Opacity Sampling   \\
PFR  \> Partial Frequency Redistribution    \\
RAP \> Resonance-Averaged Profile   \\
SOR  \> Successive Over-Relaxation   \\
\end{tabbing}

\newpage
\addcontentsline{toc}{section}{index}
\printindex

\end{document}